\documentclass[aps, prd, superscriptaddress, nofootinbib, preprintnumbers, twocolumn, showpacs]{revtex4}
\usepackage{graphicx}
\usepackage{amsmath}
\def\fsl#1{\setbox0=\hbox{$#1$}                 
   \dimen0=\wd0                                 
   \setbox1=\hbox{/} \dimen1=\wd1               
   \ifdim\dimen0>\dimen1                        
      \rlap{\hbox to \dimen0{\hfil/\hfil}}      
      #1                                        
   \else                                        
      \rlap{\hbox to \dimen1{\hfil$#1$\hfil}}   
      /                                         
   \fi}                                         %
\newcommand{\tr}{\mbox{tr}}
\newcommand{\Tr}{\mbox{Tr}}
\newcommand{\VEV}[1]{\langle #1 \rangle}

\newcommand{\NDA}{\Omega_{\rm NDA}}
\newcommand{\DxSB}{D$\chi$SB}
\newcommand{\hyper}[4]{F\left(#1,#2,#3; #4\right)}
\newcommand{\conf}[3]{{}_1F_1\left(#1,#2; #3\right)}

\makeatletter
\@addtoreset{equation}{section}
\makeatother

\begin{document}
\title{Gauged Nambu-Jona-Lasinio model with extra dimensions}
\date{\today}

\preprint{DPNU-04-11}
\preprint{TU-723}
\preprint{PNUTP-04-A04}
\preprint{hep-ph/0406194}


\author{Valery P. Gusynin}
\email[E-mail: ]{vgusynin@bitp.kiev.ua}
\affiliation{
       Bogolyubov Institute for Theoretical Physics,
       03143, Kiev, Ukraine}
\author{Michio Hashimoto}
\email[E-mail: ]{michioh@charm.phys.pusan.ac.kr}
\affiliation{Department of Physics, Pusan National University, 
             Pusan 609-735, Korea}
\author{Masaharu Tanabashi}
\email[E-mail: ]{tanabash@tuhep.phys.tohoku.ac.jp}
\affiliation{Department of Physics, Tohoku University, Sendai 980-8578, Japan}
\author{Koichi Yamawaki}
\email[E-mail: ]{yamawaki@eken.phys.nagoya-u.ac.jp}
\affiliation{Department of Physics, Nagoya University, Nagoya 464-8602, Japan}

\begin{abstract}
We investigate phase structure of the 
$D (>4)$-dimensional 
gauged Nambu-Jona-Lasinio
(NJL) model with $\delta(=D-4)$ extra dimensions 
compactified on TeV scale, based on the
improved ladder Schwinger-Dyson (SD) equation in the bulk. We assume that  
the bulk (dimensionless) running gauge coupling in the SD equation 
for the $SU(N_c)$ gauge theory with $N_f$ massless flavors 
is given by the truncated Kaluza-Klein effective theory and hence has
a nontrivial ultraviolet fixed point (UVFP), resulting in the walking 
coupling.
We find the critical line in the parameter space of 
two couplings, the gauge coupling (value fixed
at the UVFP) and the (dimensionless) four-fermion coupling, which is 
similar to that of the gauged NJL model with fixed 
(walking) gauge coupling in four dimensions. It is shown that 
{\it in the presence of such walking gauge interactions} the
four-fermion interactions 
become ``nontrivial'' and ``renormalizable'' 
even in higher dimensions, similarly to the four-dimensional gauged NJL model.
Such a renormalizability/nontriviality  holds 
only in the restricted region of the critical line 
(``nontrivial window'') with the gauge coupling 
larger than a {\it non-vanishing} value
(``marginal triviality (MT)'' point), 
in
contrast to the four-dimensional case where
such a renormalizability holds for all regions of the critical line
except for the pure NJL point (without gauge coupling).
In the nontrivial window the renormalized effective potential yields a 
nontrivial interaction which is conformal invariant. 
The existence of the nontrivial window implies ``cutoff insensitivity'' 
of the physics prediction in spite of the ultraviolet dominance of the dynamics. In the formal limit $D\to 4$, the nontrivial window 
shrinks to the pure NJL point but with a nontrivial condition which
coincides
with the known 
condition of the renormalizability/nontriviality of the four-dimensional gauged
NJL model, $\frac{9}{2}\frac{1}{N_c}<N_f-N_c <\frac{9}{2}N_c$.
\end{abstract}

\maketitle

\section{Introduction}

Growing attention has recently been paid to the model buildings
based on the scenario of extra dimensions compactified with TeV 
scale~\cite{antoni90,Dienes:1998vh}. Along such a scenario 
the
top quark condensate
(Top-Mode Standard Model: TMSM)~\cite{MTY89,Nambu89,Marciano89,BHL90} was
reconsidered by
several 
authors~\cite{Dobrescu,Cheng:1999bg,Arkani-Hamed:2000hv,Hashimoto:2000uk,Gusynin:2002cu,Hashimoto:2003ve}  
in the versions
with compactified extra dimensions, where 
the ad hoc four-fermion couplings in the original TMSM
may be replaced by the strong attractive forces due to 
Kaluza-Klein (KK) modes of the bulk Standard Model (SM)
gauge bosons. 
In particular, Arkani-Hamed, Cheng, Dobrescu and
Hall (ACDH)~\cite{Arkani-Hamed:2000hv} proposed a version of the TMSM 
where the third generation quarks/leptons as well as  the 
SM gauge bosons are put in the bulk,
 while other generations in the brane. In a series of papers~\cite{Hashimoto:2000uk,Gusynin:2002cu} we
 investigated the full gauge dynamics ({\it without four-fermion interactions})
 in the bulk, 
based on the ladder Schwinger-Dyson (SD) equation.
We further studied~\cite{Hashimoto:2003ve} 
phenomenological implications on the ACDH scenario and found that $D=8$ (
four extra dimensions) will be a viable case which predicts
a correct top mass, $m_t=172-175$ GeV,  and a characteristic Higgs mass, 
$m_H=176-188$ GeV, from the requirement that
only the top can condense (``Topped MAC'' or tMAC 
requirement).

Actually, the most prominent feature of the gauge theories with extra dimensions
is that the theories become strongly coupled,
with the coupling growing
by the power running in the ultraviolet region beyond the
compactification scale. Let us take a class of
$SU(N_c)$ gauge theories with massless $N_f$ flavors. 
It was pointed out~\cite{Hashimoto:2000uk} (see also
\cite{Agashe:2000nk, Kazakov, Dienes:2002bg})
that the bulk gauge coupling of such gauge theories can have a (nontrivial) 
ultraviolet fixed point (UVFP) within the
truncated KK effective theory~\cite{Dienes:1998vh}.
Although
existence of such a UVFP is somewhat subtle in the 
lattice studies~\cite{EKM, kawai, Farakos:2002zb} and other nonperturbative
approach~\cite{Gies:2003ic}, unless it is an artifact of
the truncated KK effective theory, it implies existence of 
a renormalizable interacting gauge theory even in dimensions higher than four.
It is in sharp contrast to the conventional view (based on the perturbation)
that higher dimensional gauge theories
are nonrenormalizable and trivial.
Once we assume that such a nontrivial UVFP does exist beyond 
the truncated KK effective theory, the theory behaves near the UVFP 
as a {\it walking gauge theory}~\cite{Holdom,YBM86,BMSY87,AY86,AKW86} 
having a large anomalous dimension
$\gamma_m=D/2-1$~\cite{Hashimoto:2000uk,Gusynin:2002cu}.
In the four-dimensional case, as was pointed out by Bardeen 
et al.~\cite{BLL86}, the four-fermion interaction becomes a 
marginal operator due to the large 
anomalous dimension, $\gamma_m \simeq 1$, of the walking gauge 
theory, and hence can be mixed with the gauge interaction. 
Then {\it the phase structure of the walking gauge theories
should be analyzed in a larger parameter space including the four-fermion
interactions as well as the gauge interactions}. 
 
The gauged Nambu-Jona-Lasinio (NJL) model, a model of 
gauge theory plus four-fermion
interaction, was first studied by Bardeen, Leung and Love~\cite{BLL86} within the ladder SD equation for the fixed gauge coupling in four dimensions.
The critical line of the four-dimensional model for the fixed gauge coupling 
was discovered by Kondo, Mino and 
Yamawaki~\cite{KMY89} 
and independently by Appelquist, Soldate, Takeuchi and Wijewardhana~\cite{ASTW}. Since then the model  has been offering
many interesting applications for physics of dynamical symmetry breaking,
such as the walking technicolor~\cite{Holdom,YBM86,BMSY87,AY86,AKW86}, the
strong ETC technicolor~\cite{MY89,ATEW89}/topcolor-assisted 
technicolor~\cite{Hill}, and the TMSM~\cite{MTY89,Nambu89,Marciano89,BHL90} or
topcolor~\cite{Hill2}, etc. (See for reviews Refs.~\cite{Yamawaki:1996vr,Miransky:vk,Hill:2002ap}). 

The characteristic dynamical feature of the four-dimensional
gauged NJL model is the large anomalous
dimension~\cite{MY89}, $1\le \gamma_m <2$, 
for the fixed (walking) gauge 
coupling, and the finiteness of all (renormalized) couplings
 at composite level as well as 
induced Yukawa coupling of the fermion to the composite bosons
in the continuum limit (infinite cutoff limit). 
This implies the (nonperturbative) ``renormalizability'' or 
``nontriviality'' of the
model, as was pointed out in Ref.\cite{Kondo:1991yk,Krasnikov:1992zn}
 for the 
(``moderately'')  walking gauge coupling, with $A>1$
~\footnote{
$A\equiv 6C_2/(-b)$, where $C_F=(N_c^2-1)/2N_c$ and $-b =(11N_c -2 N_f)/3$.
The parameter $A$ measures the speed of running of the running gauge coupling~\cite{BMSY87,Yamawaki:1996vr}: $A<1$ is the normal running as in the QCD, while
$A \gg 1$ is the walking coupling and  
$A\to \infty$ corresponds to the fixed (``standing'' limit
of the walking) gauge coupling.
Combined with the asymptotic freedom, $A>1$ reads
$\frac{9}{2}\frac{1}{N_c}<N_f-N_c <\frac{9}{2}N_c$.
}, 
and was subsequently shown more
systematically for the fixed (walking) gauge coupling ($A \to \infty$)
in Ref.~\cite{Kondo:1992sq}
as well as for the moderately walking one in Ref.\cite{Kondo:1992sq,Harada:1994wy,Kubota:1999jf}.
For the fixed gauge coupling, such a phenomenon
takes place along the whole critical line 
except for the pure NJL point (without gauge coupling 
and ``$\gamma_m=2$'')
where the above couplings 
vanish logarithmically in the continuum limit and hence
the theory becomes nonrenormalizable or trivial.   
 
Inclusion of four-fermion interactions into 
the gauge theories has not been done
so far in the context of the extra dimension scenario~\footnote{
The pure NJL model with extra dimensions has been already studied, 
e.g. in Refs.~\cite{Hashimoto:2000uk, Andrianov:2003hx}.
See also Ref.~\cite{Abe:2000ny}.}. 
The analysis of the phase 
structure of the gauged NJL model 
with extra dimensions for such walking gauge theories may be
useful not only for the TMSM but for other strong
coupling theories such as the bulk technicolor, various composite 
models, etc.. 
 
In this paper we investigate phase structure of the gauged NJL
model in the $D (>4)$-dimensional bulk, based on the improved 
ladder SD equation for the bulk fermion propagator, assuming that 
the (dimensionless) bulk running gauge coupling 
is given by the truncated KK effective theory 
and hence almost fixed at the nontrivial UVFP
in a wide energy range above the compactified energy scale.
As far as the cutoff energy $\Lambda$ is large, the effect of 
the compactified scale is 
negligible and the SD equation is well approximated by that with the fixed 
coupling~\cite{Hashimoto:2000uk,Gusynin:2002cu}.  
Thus we take the gauge coupling be fixed at the value of UVFP
for simplicity. Here we should stress that our results in this paper crucially 
depend on the existence of such a UVFP.

By solving the SD equation we find the critical line 
in the two-dimensional parameter space ($\kappa_D$,$g$) of the
coupling strength $\kappa_D$ (essentially the squared gauge coupling 
fixed at the UVFP value) and the (dimensionless) four-fermion
coupling $g$:
\begin{equation}
g^{\rm crit} = \frac{\frac{D}{2} -1}{4}
 \left( 1+ \sqrt{1-\kappa_D/\kappa_D^{\rm crit} } \right)^2\, ,
 \label{criticalline1}
\end{equation}
which exists only for $0\le \kappa_D \le \kappa_D^{\rm crit}$,
where $\kappa_D^{\rm crit}$ is the critical coupling strength
of the pure gauge theory without four-fermion interaction~\cite{Hashimoto:2000uk,Gusynin:2002cu}.
 The critical line separating the spontaneously broken 
phase (with nonzero dynamical mass $M (\ne 0)$ of the fermion) 
and the unbroken one (with 
zero dynamical mass) resembles that in the four-dimensional gauged NJL
model with fixed gauge coupling~\cite{KMY89,ASTW}
except for the prefactor $D/2-1$ instead of $1$.
Note, however, that we actually consider in this paper a class of 
$SU(N_c)$ gauge theories (with $N_f$ flavors) which are asymptotically free
having  a trivial UVFP in four dimensions, $\kappa_D \to 0 \, (D\to 4)$, but do have the nontrivial one 
in higher dimensions, $\kappa_D \ne 0 \, (D>4)$ whose value 
is  uniquely determined once 
the model parameters such as $N_c$ and $N_f$
are specified. Each point on the above 
critical line thus corresponds to a single gauged
NJL model with a different gauge theory with different $N_c, N_f$. 
We cannot freely adjust its value. 

We further discuss the phase structure of the model, following the
manner in the four-dimensional case~\cite{Kondo:1992sq}. 
We find that 
the decay constant $F_\pi$ of the composite Nambu-Goldstone (NG) boson $\pi$
becomes finite in the continuum limit ($\Lambda/M \rightarrow \infty$), 
if $N_c$ and $N_f$ are arranged so as to satisfy the condition 
\begin{equation}
  \kappa_D^{\rm MT} < \kappa_D < \kappa_D^{\rm crit}, \quad
  \mbox{(nontrivial window)},
\end{equation}
where $\kappa_D^{\rm MT} (\ne 0)$ is given by
\begin{equation}
\kappa_D^{\rm MT} = \left( 1-\frac{1}{(D/2-1)^2}\right) \kappa_D^{\rm crit}\,.
\end{equation}
We may call $\kappa_D^{\rm MT}$ as a ``marginal triviality (MT)'' point, 
where the decay constant $F_\pi$ of the composite NG boson $\pi$
diverges logarithmically, $F_\pi^2 \sim \ln \Lambda$,
in the continuum limit $\Lambda/M \rightarrow \infty$ 
and hence 
resembles the pure NJL point (zero gauge coupling) 
in the four-dimensional case.
Then in the nontrivial window we find a surprising result 
that even in higher dimensions 
the four-fermion theory {\it in the presence of gauge interactions}
becomes ``renormalizable'' and ``nontrivial'' 
in the sense that 
the induced (renormalized) Yukawa coupling of the fermion to 
the NG boson and the (renormalized) couplings 
among $\pi$ and composite scalar $\sigma$ all remain finite 
in the continuum limit ($\Lambda/M \rightarrow \infty$). 
This is analogous to the renormalizability and nontriviality of 
the four-dimensional 
case~\cite{Kondo:1991yk,Krasnikov:1992zn,Kondo:1992sq,Harada:1994wy,Kubota:1999jf}
except that 
the MT point in four-dimensional case corresponds to 
the pure NJL point, $\kappa_D^{\rm MT} \to 0$ for $D \to 4$. 
We demonstrate these facts by explicit
computation of $F_\pi$, the Yukawa vertex $\Gamma_s$, 
the scalar propagator $D_\sigma$, 
and the effective potential $V(\sigma,\pi)$.

Here we should emphasize one distinct feature of the
inclusion of the four-fermion interaction
into the gauge theory in the bulk with compactified
extra dimensions: 
In the case of pure gauge theories (without four-fermion interactions),
the ``nontrivial'' theory defined at
the critical point $\kappa_D^{\rm crit}$ is only formal, 
since the gauge coupling $\kappa_D$ 
at the UVFP is not a continuous parameter
but a discrete one depending on 
$N_c$ and $N_f$ and hence cannot be 
fine-tuned arbitrarily close to the critical value~\cite{Hashimoto:2000uk,Gusynin:2002cu}. It means that even if 
we assume existence of the UVFP, {\it the pure gauge 
theory 
in higher dimensions can only be renormalizable in a formal sense}.  
Thus there is no finite theory for pure gauge theories. 
On the other hand, in the gauged NJL model 
we do have a continuous parameter, the four-fermion
coupling $g$, and hence {\it the nontrivial theory
can be defined by fine-tuning $g$ arbitrarily close to the criticality
of the four-fermion coupling} at the critical 
line Eq.~(\ref{criticalline1}). Note that for $\kappa_D > \kappa_D^{\rm crit}$ 
where no criticality of the four-fermion coupling exists, there is 
no finite theory even in the gauged NJL model.

For the theories in the nontrivial window,
$\kappa_D^{\rm MT} <\kappa_D <\kappa_D^{\rm crit}$,
we perform explicitly 
the renormalization of the effective potential in a way similar to 
the four-dimensional case given in Ref.\cite{Kondo:1992sq}. 
The renormalized four-fermion coupling $g_R$ also has a UVFP 
$g_R(\infty)=g^{\rm crit}$.
We then find that the 
theory has a large anomalous dimension at the UVFP $g_R(\infty)$:
\begin{equation}
\gamma_m = \left(\frac{D}{2} -1\right) \left(1+ \sqrt{1-\kappa_D/\kappa_D^{\rm crit}}\right)\, ,
\end{equation} 
which takes the same form 
as that of the four-dimensional model~\cite{MY89,Kondo:1992sq} when $D$ 
in the prefactor 
is set to be $D=4$. 
Since $D/2> \gamma_m > D/2-1$,
the dynamical dimension of the 
four-fermion interaction, ${\rm dim} (\bar \psi \psi)^2= 2[(D-1) - \gamma_m]$,
reads $D-2<{\rm dim} (\bar \psi \psi)^2 < D$ for 
$\kappa_D^{\rm MT} <\kappa_D < \kappa_D^{\rm crit}$, 
and such an operator is a relevant operator similarly to the
four-dimensional gauged NJL model~\cite{Kondo:1992sq}. 
Thus, as we were motivated by  the four-dimensional
case, we conclude that {\it the phase structure of the 
gauge theory 
with extra dimensions which behaves as a walking
theory should also be studied with inclusion of the 
four-fermion interaction}.  As expected, in the renormalized effective 
potential there exist interactions of $\sigma_R$ and $\pi_R$
for $\kappa_D^{\rm MT}< \kappa_D < \kappa_D^{\rm crit}$ (nontriviality),
which are {\it conformal invariant}.
The renormalizability/nontriviality implies a remarkable fact
that the physics prediction
is {\it cutoff insensitive}, notwithstanding the ultraviolet
dominance of the dynamics of both the walking gauge and the four-fermion interactions
strongly coupled in the ultraviolet region, which may be useful for model
buildings.   

We further study RG flows of each theory in the nontrivial window
governed by a UVFP $(\kappa_D,\;g_R(\infty))$ corresponding to
a set of $(N_c,N_f)$ 
(each point lying on the critical line $g^{\rm crit}$).
Then we find that 
our renormalization is consistently performed even off the UVFP.

For $ 0\le\kappa_D \le \kappa_D^{\rm MT}$, on the other hand,
 we find that the gauged NJL model
becomes trivial and nonrenormalizable in spite of the 
fact that we can fine-tune
the four-fermion coupling to make finite the dynamical mass of
the fermion. 
The ``renormalized effective potential'' 
actually has no interaction terms of $\sigma_R$ and $\pi_R$.
This renormalization would lead to
the ``anomalous dimension''
$\gamma_m = \frac{D}{2}$ and 
${\rm dim} (\bar \psi \psi)^2=D-2$ for $0\le \kappa_D \le\kappa_D^{\rm MT}$. 
However, our renormalization is incomplete
for $0\le \kappa_D \le\kappa_D^{\rm MT}$:
There exists another marginal operator 
$\partial_M (\bar\psi\psi)\partial^M (\bar\psi\psi)$ whose dynamical
dimension
is given by 
dim($\partial_M (\bar\psi\psi)\partial^M (\bar\psi\psi))=D$ 
(marginal operator!) 
in our renormalization scheme. 

We further discuss in Sec.~\ref{summary} the formal limit $D \to 4$, in which 
$\kappa_D^{\rm MT}\to 0$ and $\kappa_D\to 0$ and hence the UVFP's for both 
regions of 
the nontrivial window $\kappa_D^{\rm MT} < \kappa_D < \kappa_D^{\rm crit}$
and  trivial region $0\leq \kappa_D\leq \kappa_D^{\rm MT}$
shrink to a single point of the pure NJL point. Nonetheless,
the nontrivial window for $D\to 4$ coincides with the condition
of the renormalizability/nontriviality of the
four-dimensional gauged NJL model with (moderately) walking gauge coupling
characterized by
$A>1$~\cite{Kondo:1991yk,Krasnikov:1992zn,Kondo:1992sq,Harada:1994wy,Kubota:1999jf},
which is clearly distinguished from $A\le 1$ corresponding to 
the trivial region  $0\leq \kappa_D\leq \kappa_D^{\rm MT}$.
Thus the nontrivial window 
is not a peculiarity of the extra dimensions but actually  exists already in 
the four-dimensional gauged NJL model in a rather sophisticated manner.

The paper is organized as follows:
In Sec.~\ref{sec2} we give our model setting, Lagrangian and the running 
gauge coupling constant within the truncated KK effective theory.
In Sec.~\ref{sec3} we study phase structure of the model, based on the
simplest approximation, the bifurcation technique~\cite{kn:Atki87},
to the SD equation. We find the critical line,
and calculate simplified scaling relation, scalar boson propagator
and the decay constant  $F_\pi$ of the NG boson $\pi$. 
We also find the MT point where $F_\pi^2$ diverges logarithmically
in the continuum limit. In Sec.~\ref{sec-moreabout}  we incorporate 
subleading terms in the asymptotic region by more involved approximations 
than the bifurcation, ``linearized approximation''~\cite{Fomin,Miransky:vk}
and ``power expansion method''~\cite{Bardeen:sv}. 
The sub-dominant term is required for the effective potential.
In Sec.~\ref{eff_pot} we calculate the effective potential.
In Sec.~\ref{sec-RG} we 
discuss the renormalization making finite the induced Yukawa 
coupling, the effective potential and the
scalar boson propagator. We give detailed discussions on 
the RG flow in the nontrivial window.
Sec.~\ref{summary} is devoted to Summary and Discussion: In particular 
we discuss the $D\to 4$ limit. 
In Appendix~\ref{app1}, we give another detailed calculations of the
composite scalar propagator. 
Appendix~\ref{sec-appendix2} and~\ref{sec-appendix3} are for details of 
the ``linearized approximation'' and ``power expansion method'', 
respectively.
Appendix~\ref{app-D} is for more discussions on the nonrenormalizability of 
the region $0 \leq \kappa_D \leq \kappa_D^{\rm MT}$.

\section{The Model}
\label{sec2}

Let us 
start with the renormalization group properties of the
gauge interaction in the $D=4+\delta$ dimensional bulk.  
Extra $\delta=(D-4)$ spatial dimensions are assumed to be compactified
at scale $R^{-1}$. 
The negative mass dimension of the bulk gauge coupling implies strong 
interaction at high energy.
Thus the higher loop effects somehow need to be taken into account.
The task, however, immediately meets serious trouble with the
nonrenormalizability of the bulk gauge theory $D>4$.
We therefore {\em define} the bulk gauge theory by using the ``truncated
Kaluza-Klein (KK)'' effective theory~\cite{Dienes:1998vh}, which
allows us to calculate the loop effects within 4D effective field theory. 
A remarkable feature of the truncated KK effective theory is the
existence of the non-trivial ultraviolet fixed point (UVFP) at least
for its $\overline{\rm MS}$
coupling~\cite{Hashimoto:2000uk,Agashe:2000nk, Kazakov, Dienes:2002bg}.
We briefly review the properties of the renormalization group equation
(RGE) in this model. 

After the decomposition of the bulk gauge field into its KK-modes, the 
running of the four-dimensional gauge coupling $g_{4D}$ can be described by the 
RGE,
\begin{equation}
  (4\pi)^2 \mu \frac{d g_{4D}^{}}{d \mu} = N_{\rm KK}(\mu) \, b' \,g_{4D}^3,
  \label{rge-ED1}
\end{equation}
where we assumed the renormalization scale $\mu$ is sufficiently
larger than the compactification scale $R^{-1}$, $\mu \gg R^{-1}$.
The RGE coefficient $b'$ is given by
\begin{equation}
  b' = -\frac{22-\delta}{6} N_c + \frac{2}{3} \cdot 2^{\delta/2} \, N_f ,
\label{bprime}
\end{equation}
for $SU(N_c)$ gauge theory with $N_f$ bulk fermions.
$N_{\rm KK}(\mu)$ stands for the number of KK modes below the
renormalization scale $\mu$.
We use an approximation
\begin{equation}
N_{\rm KK}(\mu) = \frac{1}{2^n}
                  \frac{\pi^{\delta/2}}{\Gamma(1+\delta/2)}(\mu R)^\delta.
  \label{nkk_app}
\end{equation}
The factor $1/2^n$ in Eq.~(\ref{nkk_app}) arises from 
the orbifold compactification on 
$T^\delta/Z_2^n$. (See Refs.~\cite{Hashimoto:2000uk,Hashimoto:2003ve}.)
The four-dimensional gauge coupling $g_{4D}^{}$ can be matched with 
the {\it dimensionful} bulk gauge coupling $g_{(4+\delta)D}^{}$,
\begin{displaymath}
   g_{(4+\delta)D}^2=(2\pi R)^\delta g_{4D}^2/2^n .
\end{displaymath}
It is convenient to define the {\it dimensionless} bulk gauge coupling
$\hat g$, 
\begin{displaymath}
  \hat g^2 \equiv g_{(4+\delta)D}^2 \mu^\delta, 
\end{displaymath}
and thus
\begin{equation}
  \hat g^2(\mu) = \frac{(2\pi R \mu)^\delta}{2^n}g_{4D}^2 (\mu).
  \label{hat-g}
\end{equation}
Combining Eqs.~(\ref{rge-ED1}), (\ref{nkk_app}), and (\ref{hat-g}),
we obtain RGE for $\hat g$~\cite{Hashimoto:2000uk} 
\begin{equation}
 \mu \frac{d}{d \mu} \hat g = \frac{\delta}{2}\,\hat g
 + \left(1+\frac{\delta}{2}\right) \NDA \, b'\, \hat g^3, 
 \quad (\mu \gg R^{-1}), \label{rge_ED3}
\end{equation}
with $\NDA$ being the loop factor of naive dimensional analysis (NDA)
in $D$ dimensions,
\begin{equation}
  \NDA \equiv \frac{1}{(4\pi)^{D/2}\Gamma(D/2)} .
\end{equation}
The RGE (\ref{rge_ED3}) leads to 
a UVFP $g_{*}$~\cite{Hashimoto:2000uk},
\begin{equation}
  g_{*}^2 \NDA = \frac{1}{-(1+2/\delta)\,b'}, \label{UV-FP}
\end{equation}
for $b' < 0$. 

So far we have shown the UVFP only within the truncated KK effective
theory at one-loop level.
Does such a UVFP really exist beyond the approximation we adopted?
This question is, of course, a highly non-perturbative 
problem~\cite{EKM, kawai, Farakos:2002zb, Gies:2003ic} and
extremely difficult to be answered.
Instead of solving this difficult problem, in this paper, we simply
assume the existence of the UVFP and address a hopefully easier but
still quite exciting question: 
How does the bulk gauge theory behave with such a UVFP? 

Such a theory should possess an approximate conformal invariance and
was shown to have a 
large anomalous dimension $\gamma_m = 
D/2-1$~\cite{Hashimoto:2000uk,Gusynin:2002cu}.
The situation has a strong resemblance to the four-dimensional walking
gauge theories~\cite{Holdom,YBM86,AY86,AKW86}, in which the 
gauge coupling is assumed
to be on the nontrivial UVFP and the fermion bilinear operator
$\bar\psi\psi$ acquires a large anomalous dimension $\gamma_m=1$.
It is known that the Nambu-Jona-Lasinio (NJL) type four-fermion
interaction becomes relevant in such a walking gauge dynamics. 
The walking gauge theory was then analyzed in an extended
coupling space including the NJL type four-fermion interaction (gauged 
NJL model)~\cite{BLL86,
KMY89,ASTW,MY89,Kondo:1992sq}.

This resemblance motivates us to study the gauged NJL model with extra
dimensions.
In this paper, we therefore focus on the dynamical chiral symmetry
breaking 
(\DxSB) in the gauged NJL model with extra dimensions.
The Lagrangian of the gauged NJL model with extra dimensions is given by
\begin{eqnarray}
  {\cal L} &=& \bar{\psi} i\fsl{D} \psi - m_0 \bar{\psi}\psi
  + \frac{G}{2 N_c} 
  \left[\,\left(\bar{\psi}\psi\right)^2 + \left(\bar{\psi}i\Gamma_A\tau^i\psi
  \right)^2\,\right]
  \nonumber \\ &&
  - \frac{1}{2}\tr \left(F_{MN} F^{MN}\right), \label{GNJL1}
\end{eqnarray}
where $M,N=0,1,2,3,5,\cdots,D$ and $\Gamma_A$ is the chirality matrix
in $D$ dimensions, $\tau^i$ $(i=1,2,3)$ the Pauli matrices, 
and $G$ the four-fermion coupling. 
The gauge group is $SU(N_c)$ and 
$F_{MN}$ denotes the field strength.
The gauge coupling is assumed to be on the UVFP\@.
For simplicity, we take the number of flavor as $N_f=2$, i.e.,
$\psi=(\psi_u,\psi_d)^T$ in the flavor space. 
The number of dimensions $D$ is assumed to be even, $D=6,8,10,\cdots$,
so as to introduce chiral fermions in the bulk. 
Extra $\delta (= D-4)$ spatial dimensions are compactified 
at a TeV-scale $R^{-1}$.

It is convenient to rewrite the Lagrangian Eq.~(\ref{GNJL1}) by using
auxiliary fields $\sigma$, $\pi_i$,
\begin{eqnarray}
  {\cal L} &=& \bar{\psi} i \fsl{D} \psi 
  - \bar{\psi} \left( \sigma + i \Gamma_A \tau^i \pi_i \right) \psi
  - \frac{N_c}{2 G} \left( \sigma^2 + \pi_i^2 \right)
  \nonumber \\ &&
  + \frac{N_c}{G}m_0 \sigma
  - \frac{1}{2} \tr \left(F_{MN} F^{MN}\right). \label{GNJL2} 
\end{eqnarray}
In order to see the equivalence of Eq.~(\ref{GNJL1}) and
Eq.~(\ref{GNJL2}), we just need to eliminate the auxiliary fields $\sigma$
and $\pi_i$ through their Euler equations
\begin{equation}
  \sigma = - \frac{G}{N_c}\, \bar{\psi}\psi + m_0, \quad 
  \pi_i = - \frac{G}{N_c} \, \bar{\psi}i\Gamma_A \tau^i\psi, \label{aux}
\end{equation}
in Eq.~(\ref{GNJL2}).
The vacuum expectation value (VEV) of
$\sigma$ 
\begin{equation}
  \VEV{\sigma} = -\frac{G}{N_c} \VEV{\bar\psi\psi} + m_0 
  \label{eq:vev_sigma0}
\end{equation}
is proportional to the chiral condensate in the chiral limit $m_0=0$.
Note that the model with $m_0=0$ possesses global $SU(2)_+ \times
SU(2)_-$ chiral symmetry,  
\begin{equation}
  \psi_\pm \to \psi'_\pm = e^{i\theta_\pm^i \frac{\tau^i}{2}}\psi_\pm, \quad 
  \psi_\pm \equiv \frac{1\pm\Gamma_A}{2}\psi.
\end{equation}
The bare mass term $m_0$ will be taken to be zero in the analysis
of the {\DxSB} in this paper.

\section{Phase structure}
\label{sec3}
\subsection{SD equation}

We express the bulk fermion propagator as
\begin{equation}
  i S^{-1}(p) = A(-p^2)\, [\fsl{p} - \Sigma(-p^2)].
\end{equation}
Nonvanishing $\Sigma \ne 0 $ implies {\DxSB}, which can
be investigated through the (improved) ladder Schwinger-Dyson (SD)
equation.
We need to take into account the running of the
bulk gauge coupling in the SD equation. In the 
SD equation there are three different
momenta, (Euclidean) square of which are  $x\equiv -p^2$, $y\equiv -q^2$ 
and $z \equiv -(p-q)^2$ for the external and the loop momenta of 
the fermion and the gauge boson momentum, respectively. 
Hence there exist various ways to incorporate the running effects. 
The simplest one~\cite{imp-simplest} is
to take $ g_{(4+\delta)D}^2(\mu) \to g_{(4+\delta)D}^2({\rm max}(x,y))$, 
which has been widely used for four-dimensional models
in Landau gauge where the SD equation dictates $A(-p^2)=1$, so that it is 
consistent with  the vector Ward-Takahashi (WT) identity. In the case of 
extra dimensions we also used this in Ref.~\cite{Hashimoto:2000uk}. 
However, this is not consistent with the chiral WT 
identity~\cite{Kugo:1992pr}, and hence we adopt as in 
Ref.~\cite{Gusynin:2002cu}  an ansatz of Ref.\cite{Kugo:1992pr}, namely  
the gauge boson momentum is identified as the renormalization scale of 
the gauge coupling strength, 
\begin{equation}
  g_{(4+\delta)D}^2(\mu) \to 
  g_{(4+\delta)D}^2(z) = \frac{\hat g^2(\mu = \sqrt{z})}{z^{\delta/2}},
  \label{imp_ladder}
\end{equation}
where 
$\hat g$ is the dimensionless bulk gauge coupling.
We assume here that the dimensionless bulk gauge coupling is on its UVFP,
$\hat g = g_*$.
The assumption is justified in case that the cutoff $\Lambda$ is very
large and the UVFP really exists.
The SD equation is then given by
\begin{eqnarray}
 A(x) &=& 1+\frac{\kappa_D}{x}\int_0^{\Lambda^2}\!\!\! dy 
            \frac{y^{D/2-1}A(y)}{A^2(y) y + B^2(y)}
            \frac{\min(x,y)}{[\max(x,y)]^{D/2-1}} \nonumber \\[3mm]
      &&    \qquad \times
            \left(\frac{(D-1)(D-4)}{D}+\xi\right), 
       \label{eq:SDeq_A}     \\[3mm]
 B(x) &=& \sigma + (D-1+\xi)\kappa_D \int_0^{\Lambda^2}\!\!\! dy 
          \frac{y^{D/2-1}B(y)}{A^2(y) y + B^2(y)} \nonumber \\
      &&    \qquad \times \frac{1}{[\max(x,y)]^{D/2-1}}, 
\label{eq:SDeq_B}
\end{eqnarray}
where 
we have introduced ultraviolet cutoff
$\Lambda$ and
the gauge fixing parameter is denoted as $\xi$.
$B(x)$ is defined by
\begin{equation}
   B(x) \equiv A(x) \Sigma(x).
\end{equation}
We denote the gauge coupling strength by $\kappa_D$,
\begin{equation}
 \kappa_D \equiv C_F \Omega_{\rm NDA} g_*^2,
\label{eq:def_kappaD}
\end{equation}
with $C_F$ being the quadratic Casimir of the fundamental representation,
\begin{equation}
  C_F = \frac{N_c^2-1}{2N_c}. 
  \label{Casimir}
\end{equation}
It is understood that $\sigma$ stands for the VEV of the auxiliary
field $\VEV{\sigma}$ in Eq.~(\ref{eq:SDeq_B}).
Hereafter, we will adopt this shorthand notation without discriminating
the auxiliary fields ($\sigma$, $\pi$) from their VEVs
($\VEV{\sigma}$, $\VEV{\pi}$). When $G=0$ and hence $\sigma =m_0$, the SD
equations Eqs.~(\ref{eq:SDeq_A}) 
and (\ref{eq:SDeq_B}) 
are reduced to those of the
bulk gauge theory without four-fermion interactions in a form given in 
\cite{Gusynin:2002cu}.

We choose the gauge fixing parameter $\xi$ as~\cite{Gusynin:2002cu}
\begin{equation}
 \xi = -\frac{(D-1)(D-4)}{D}. \label{NLG}
\end{equation}
With this choice of gauge fixing parameter, the fermion wave function
factor becomes trivial, i.e., $A(x) \equiv 1$  and $B(x)\equiv
\Sigma(x)$. 
The SD equation then reads
\begin{eqnarray}
 \Sigma(x)
      &=& \sigma + \dfrac{4(D-1)}{D} 
          \kappa_D \int_0^{\Lambda^2}\!\!\! dy 
          \frac{y^{D/2-1}\Sigma(y)}{y + \Sigma^2(y)} \nonumber \\
      &&    \qquad \times \frac{1}{[\max(x,y)]^{D/2-1}}.
      \label{imp_ladder_SD}
\end{eqnarray}

The VEV of $\sigma$ is given by Eq.~(\ref{eq:vev_sigma0}).
We thus obtain 
\begin{equation}
  \sigma 
         = m_0 + \frac{g}{\Lambda^{D-2}} \int_0^{\Lambda^2}
           dx x^{D/2-1} \dfrac{\Sigma(x)}{x + \Sigma^2(x)},
  \label{eq:vev_sigma}
\end{equation}
where we defined the {\it dimensionless} four-fermion coupling constant $g$ 
as
\begin{equation}
  g \equiv 2^{D/2} N_f G \Lambda^{D-2} \NDA , \quad (N_f=2),
\end{equation}
and used
\begin{align}
 \VEV{\bar\psi\psi} &= - N_f \Tr S(p) \nonumber \\
 &= -2^{D/2} N_c N_f \int \frac{d^D p}{i(2\pi)^D}
   \frac{\Sigma(-p^2)}{-p^2+\Sigma^2}. \label{chi-cond}
\end{align}

The SD equation (\ref{imp_ladder_SD}) for the mass function
is equivalent to the differential equation
\begin{eqnarray}
\lefteqn{
 x^2 \frac{d^2}{dx^2}\Sigma(x) + \frac{D}{2} x \frac{d}{dx} \Sigma(x)
} \nonumber \\[3mm] && \quad \quad 
  +\frac{2(D-1)(D-2)}{D} \kappa_D \frac{x \Sigma(x)}{x+\Sigma^2} = 0, 
   \label{sde_diff} 
\end{eqnarray}
with the infrared boundary condition (IRBC)
\begin{equation}
x^{D/2} \frac{d}{dx} \Sigma(x)\bigg|_{x = 0} = 0, \label{IR-BC}
\end{equation}
and the ultraviolet boundary condition (UVBC)
\begin{equation}
   \left[\, (D/2-1) + x \frac{d}{dx} \,\right] \Sigma(x)
   \bigg|_{x  = \Lambda^2} = (D/2-1)\sigma. \label{UV-BC}
\end{equation}
The derivative of Eq.~(\ref{imp_ladder_SD}) at the cutoff scale
$\Lambda^2$ is given by 
\begin{eqnarray}
 \lefteqn{ \hspace*{-1.5cm}
  \left. x\frac{d}{dx} \Sigma(x) \right|_{x=\Lambda^2}
  = -\dfrac{2(D-1)(D-2)}{D\Lambda^{D-2}} \kappa_D 
  } \nonumber\\
  & & \qquad \times
      \int_0^{\Lambda^2} dx 
      x^{D/2-1} \dfrac{\Sigma(x)}{x+\Sigma^2(x)}.
  \label{eq:derivative_Sigma}
\end{eqnarray}
From Eq.~(\ref{eq:vev_sigma}) and Eq.~(\ref{eq:derivative_Sigma}) we see
\begin{equation}
 \sigma = m_0 - \frac{D}{2(D-1)(D-2)} \frac{g}{\kappa_D}
          x \frac{d}{dx} \Sigma(x)\bigg|_{x = \Lambda^2} .
\label{sigma} 
\end{equation}
The UVBC then reads
\begin{eqnarray}
  \left[\, (D/2-1) + \left(1+\frac{D}{4(D-1)}\frac{g}{\kappa_D}\right)
  x\frac{d}{dx}\,\right] \Sigma(x) \bigg|_{x = \Lambda^2}
  \nonumber \\[3mm] = (D/2-1)m_0. \qquad \label{UV-BC2} 
\end{eqnarray}

For $\kappa_D=0$, the SD equation reduces to the gap equation of the
pure NJL model.
See the appendix of Ref.~\cite{Hashimoto:2000uk} for the analysis of
the $D(>4)$ dimensional pure NJL model. 

\subsection{Bifurcation technique}
\label{sec-bifurcation}
Even if we start with the
chiral limit $m_0 = 0$, we expect the dynamical chiral symmetry
breaking $\Sigma \ne 0$ takes place if the NJL coupling exceeds
certain critical value.
In order to fully analyze the behavior of the dynamical chiral phase
transition in this model, we need to solve the nonlinear SD
equation Eq.~(\ref{imp_ladder_SD}) as it stands.
Such a task turns out to be almost impossible within analytical
methods, however.
Instead of solving the nonlinear SD equation, here we first adopt the
bifurcation technique~\cite{kn:Atki87}, restricting the solution much 
smaller than the cutoff scale $\Sigma \ll \Lambda$.
The bifurcation technique is justified at least for the determination
of the critical coupling and the leading asymptotic behavior of the
mass function $\Sigma(x)$.
In this section, we employ the bifurcation technique for the sake of
simplicity to avoid unnecessarily complicated expressions in
discussing the decay constant, effective Yukawa coupling,
and the propagator of the composite scalar, which are determined only
through the dominant term of $\Sigma(x)$.
On the other hand, the sub-dominant term is relevant to the scaling
relation and the lowest interaction term of the effective potential.
The bifurcation method, however, does not always lead to the correct
sub-dominant term.
Thus we will perform more sophisticated analysis later in
Section~\ref{sec-moreabout} to determine the sub-dominant term of $\Sigma(x)$.

In the bifurcation technique, 
the $\Sigma^2$ terms in the denominators of Eq.~(\ref{imp_ladder_SD})
and Eq.~(\ref{eq:vev_sigma}) are replaced by an infrared cutoff
$\Sigma_0$
\begin{equation}
 \Sigma(x)
    = \sigma + \dfrac{4(D-1)}{D} 
               \kappa_D \int_{\Sigma_0^2}^{\Lambda^2}\!\!\! dy 
          \dfrac{y^{D/2-2}\Sigma(y)}{[\max(x,y)]^{D/2-1}},
\label{eq:integ_eq}
\end{equation}
with
\begin{equation}
  \sigma 
         = \frac{g}{\Lambda^{D-2}} \int_{\Sigma_0^2}^{\Lambda^2}
           dx x^{D/2-2} \Sigma(x), \quad m_0=0. \label{sigma_bifurcation}
\end{equation}
We also assume $\Sigma_0\ll \Lambda$ and $\Sigma_0 \sim
\Sigma(\Sigma_0^2)$.

The integral equation Eq.~(\ref{eq:integ_eq}) is equivalent with a set
of the {\it linear} differential equation
\begin{eqnarray}
\lefteqn{
 x^2 \frac{d^2}{dx^2}\Sigma(x) + \frac{D}{2} x\frac{d}{dx} \Sigma(x)
} \nonumber \\[3mm] && \qquad \quad 
  +\frac{2(D-1)(D-2)}{D} \kappa_D \Sigma(x) = 0, 
\label{eq:diff_eq}
\end{eqnarray}
and boundary conditions
\begin{equation}
 \left. \frac{d}{dx} \Sigma(x) \right|_{x = \Sigma_0^2} = 0, 
 \label{eq:ir-bc}
\end{equation}
and
\begin{equation}
  \left[\, D/2-1 + \left(1+\frac{D}{4(D-1)}\frac{g}{\kappa_D}\right)
  x\frac{d}{dx}\,\right] \Sigma(x) \bigg|_{x = \Lambda^2}
  = 0.
 \label{eq:uv-bc}
\end{equation}

The solution of Eq.~(\ref{eq:diff_eq}) behaves differently 
for $\kappa_D < \kappa_D^{\rm crit}$ and for 
$\kappa_D > \kappa_D^{\rm crit}$.
Here critical value $\kappa_D^{\rm crit}$ is obtained as~\cite{Gusynin:2002cu}
\begin{equation}
   \kappa_D^{\rm crit} \equiv \dfrac{D}{32}\dfrac{D-2}{D-1}. \label{kd-crit}
\end{equation}
For the sub-critical $\kappa_D<\kappa_D^{\rm crit}$, the solution of
Eq.~(\ref{eq:diff_eq}) is given by a power-damping form,
\begin{equation}
  \Sigma(x) = 
    c_1 \Sigma_0 \left(\dfrac{x}{\Sigma_0^2}\right)^{-\frac{\nu}{2}(1-\omega)}
  + d_1 \Sigma_0 \left(\dfrac{x}{\Sigma_0^2}\right)^{-\frac{\nu}{2}(1+\omega)},
\label{eq:bifurc1}
\end{equation}
where $c_1$ and $d_1$ are real constants.
We also define  $\nu$ and $\omega$ as\footnote{
We use notations similar to those in Ref.~\cite{Kondo:1992sq}.} 
\begin{equation}
  \nu  \equiv \dfrac{D}{2}-1, 
  \label{eq:nudef}
\end{equation}
and 
\begin{equation}
  \omega \equiv \sqrt{1-\kappa_D / \kappa_D^{\rm crit}},
  \label{eq:omegadef}
\end{equation}
respectively.
For the super-critical $\kappa_D > \kappa_D^{\rm crit}$, on the other
hand, the solution starts to oscillate,
\begin{align}
  \Sigma(x) &= 
    \tilde c_1 \Sigma_0 
    \left(\dfrac{x}{\Sigma_0^2}\right)^{-\frac{\nu}{2}(1-i\tilde\omega)}
  + \mbox{h.c.},
  \nonumber\\
  &= 2|\tilde c_1| \Sigma_0 \left(\dfrac{x}{\Sigma_0^2}\right)^{-\frac{\nu}{2}}
      \sin\left(\frac{1}{2}\nu\tilde\omega\ln\dfrac{x}{\Sigma_0^2}
                +\theta\right),
\end{align}
with $\tilde c_1$ being a complex constant.
The parameter $\tilde \omega$ and the angle $\theta$ are defined as
\begin{equation}
  \tilde \omega \equiv \sqrt{\kappa_D / \kappa_D^{\rm crit}-1},
\end{equation}
and 
\begin{equation}
  e^{2i\theta} = -\tilde c_1/\tilde c_1^*,
\end{equation}
respectively.

\begin{figure}[tbp]
  \begin{center}
    \resizebox{0.45\textwidth}{!}{
     \includegraphics{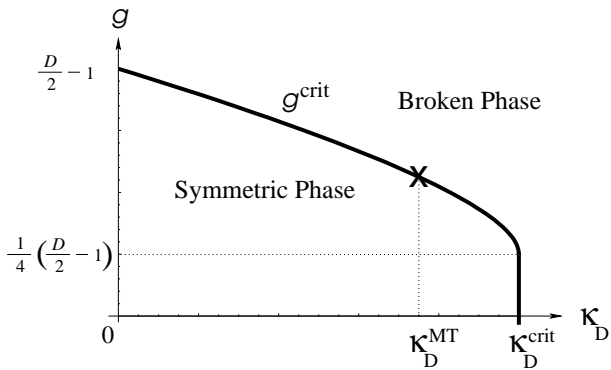}}
    \caption{The critical line in $(\kappa_D,g)$ plane.
             The critical line is given by $g^{\rm crit}=\nu(1+\omega)^2/4$
             with $\omega = \sqrt{1-\kappa_D/\kappa_D^{\rm crit}}$ 
             and $\nu=D/2-1$ for $0 \leq \kappa_D \leq \kappa_D^{\rm crit}$.
             The chiral symmetry is unbroken below the critical line.
             For $\kappa_D > \kappa_D^{\rm crit}$  the broken phase is 
             realized irrespective of the size of $g$. \label{phase}
             \label{fig:critical_line}
             }
  \end{center}
\end{figure}

Let us first discuss the case with $\kappa_D < \kappa_D^{\rm crit}$.
The IRBC Eq.~(\ref{eq:ir-bc}) leads to
\begin{equation}
  d_1 = - \frac{1-\omega}{1+\omega} c_1.
\label{eq:ir-bc2}
\end{equation}
If we normalize the mass function such as $\Sigma(\Sigma_0^2)=\Sigma_0$,
we can determine concretely $c_1$ and $d_1$. 
It is convenient to rewrite the UVBC Eq.~(\ref{eq:uv-bc}) in the
following form,
\begin{equation}
  \left. \left[
      \nu + \left( 
        1 + \dfrac{4}{1-\omega^2} \frac{g}{\nu}
      \right)x\frac{d}{dx}
  \right] \Sigma(x) \right|_{x=\Lambda^2} = 0.
  \label{eq:uv-bc2}
\end{equation}
Combining Eq.~(\ref{eq:bifurc1}) and Eq.~(\ref{eq:uv-bc2}), we obtain
\begin{equation}
  \left(
    \dfrac{\Sigma_0^2}{\Lambda^2} 
  \right)^{\nu\omega} 
  = -\dfrac{ (1-\omega) c_1 }{ (1+\omega) d_1 }
     \dfrac{4g - \nu (1+\omega)^2}{4g - \nu (1-\omega)^2}.
\label{eq:scaling0}
\end{equation}
Eq.~(\ref{eq:ir-bc2}) then leads to the scaling relation\footnote{
The scaling relation Eq.~(\ref{eq:scaling1}) is not correct near the
pure NJL limit.
Actually, it fails to reproduce the scaling behavior of the $D$($>4$)
dimensional pure NJL model~\cite{Hashimoto:2000uk} in the
$\kappa_D\rightarrow 0$ limit.
See Section~\ref{sec-moreabout} for more detailed analysis of the
scaling behavior.
}
\begin{equation}
  \left(
    \dfrac{\Sigma_0^2}{\Lambda^2} 
  \right)^{\nu\omega} 
  = \dfrac{4g - \nu (1+\omega)^2}{4g - \nu (1-\omega)^2}.
\label{eq:scaling1}
\end{equation}
The nontrivial solution $\Sigma_0 \ne 0$ exists only when the NJL
coupling exceeds the critical line, 
\begin{equation}
  g^{\rm crit} = \frac{\nu}{4} (1+\omega)^2.
\label{eq:g_crit}
\end{equation}
See Figure~\ref{fig:critical_line} for the phase diagram in the
$\kappa_D$-$g$ plane.
Note that the negative sign in Eq.~(\ref{eq:ir-bc2}) is essentially
important.
If the sign of $d_1$ were identical to $c_1$, we would not obtain
positive value in Eq.~(\ref{eq:scaling0}) for $g>g^{\rm crit}$,
indicating the instability of the vacuum.
We will be back to this problem later in Sec.~\ref{eff_pot}.

From Eq.~(\ref{eq:bifurc1}) and Eq.~(\ref{eq:ir-bc2}), we also obtain
the asymptotic form of the nontrivial solution,
\begin{equation}
  \Sigma(x) \sim c_1 \Sigma_0 
  \left(\dfrac{x}{\Sigma_0^2}\right)^{-\frac{\nu}{2}(1-\omega)},
  \qquad
  \mbox{for $x\gg \Sigma_0^2$}.
\label{eq:asymp_sol}
\end{equation}

We next consider the condensate $\sigma$, which can be calculated from 
Eq.~(\ref{UV-BC}).
We obtain 
\begin{equation}
  \sigma = \dfrac{1+\omega}{2} c_1 \Sigma_0 
  \left(\dfrac{\Lambda^2}{\Sigma_0^2}\right)^{-\frac{\nu}{2}(1-\omega)},
\end{equation}
neglecting the sub-dominant term.
The asymptotic solution Eq.~(\ref{eq:asymp_sol}) can be written in
terms of $\sigma$,
\begin{equation}
  \Sigma(x) \sim \dfrac{2}{1+\omega} \sigma
  \left(\dfrac{x}{\Lambda^2}\right)^{-\frac{\nu}{2}(1-\omega)}.
\label{eq:asymp_sol2}
\end{equation}
The mass function $\Sigma$ is thus proportional to the order parameter
$\sigma$. 
We will use Eq.~(\ref{eq:asymp_sol2}) later in Sec.~\ref{sec-yukawa} to
calculate the Yukawa vertex of the composite scalar field $\sigma$.

We next turn to the case with $\kappa_D > \kappa_D^{\rm crit}$.
From the IRBC Eq.~(\ref{eq:ir-bc}) we find
\begin{equation}
  e^{2i\theta} = \frac{1+i\tilde\omega}{1-i\tilde\omega}.
\end{equation}
On the other hand, the UVBC Eq.~(\ref{eq:uv-bc}) gives 
\begin{equation}
  \sin\left(\frac{1}{2}\nu\tilde\omega\ln\dfrac{\Lambda^2}{\Sigma_0^2}
  + \theta + \theta'\right) = 0,
\label{eq:uv-bc3}
\end{equation}
with $\theta'$ being given by
\begin{equation}
  e^{2i(\theta+\theta')} = 
                  \dfrac{4g - \nu ( 1 + i\tilde\omega )^2}
                        {4g - \nu ( 1 - i\tilde\omega )^2}.
\end{equation}
We note that nontrivial solution $\Sigma_0 \ne 0$ exists in
Eq.~(\ref{eq:uv-bc3}), irrespective of the size of the NJL coupling. 
This could be easily understood in an analytical manner by looking at
the region $\tilde\omega\ll 1$, where we can use an approximation
\begin{equation}
  \dfrac{1}{\nu\tilde \omega}(\theta + \theta') = \frac{2}{\nu - 4g}.  
\end{equation}
Eq.~(\ref{eq:uv-bc3}) then leads to a scaling relation (zero-node solution)
\begin{eqnarray}
  \dfrac{\Sigma_0}{\Lambda} 
  &=& \exp\left(-\dfrac{\pi - (\theta +\theta')}
                       {\nu\tilde\omega}
          \right), 
  \nonumber\\
  &=& \exp\left(\frac{2}{\nu - 4g} -\dfrac{\pi}{\nu\tilde\omega}\right).
\label{eq:scaling2}
\end{eqnarray}
For $g=\nu/4$ we find 
$\Sigma_0/\Lambda=\exp(\tilde\omega/2)\exp[-\pi/(2\nu\tilde\omega)]$.

We next consider the continuum limit ($\Lambda\rightarrow\infty$) in
the present model. 
As we stated before, we simply admit the existence of the UVFP in the
gauge coupling strength in the present analysis. 
One might thus think that we should be able to obtain a finite theory
even in the $\Lambda\rightarrow\infty$ limit thanks to the UVFP\@.
This is a non-trivial problem, however, in the {\DxSB} vacuum.
As we see in Eq.~(\ref{eq:scaling1}) and Eq.~(\ref{eq:scaling2}), the
dynamical mass of the fermion ($\Sigma_0$) is proportional to the cutoff
$\Lambda$ and thus diverges in the $\Lambda\rightarrow\infty$ limit if
both $g$ and $\kappa_D$ are fixed. 
In other words, in order to keep $\Sigma_0$ finite, the NJL coupling  
$g$ needs to approach its critical value Eq.~(\ref{eq:g_crit}) in the 
$\Lambda\rightarrow\infty$ limit.
This procedure, known as the Wilsonian renormalization, allows us to
define the continuum limit of the gauged NJL model.
Hereafter we use the word ``renormalization'' in this sense.

Recall that the {\DxSB} always takes place regardless the NJL
coupling $g$ for $\kappa_D > \kappa_D^{\rm crit}$.  
There does not exist critical NJL coupling with this $\kappa_D$.
What does happen in the continuum limit for  
$\kappa_D > \kappa_D^{\rm crit}$ then?
Instead of $g$, one may think about tuning $\kappa_D$ to its critical 
value $\kappa_D^{\rm crit}$.
This procedure cannot be done, however, since $\kappa_D$ is
essentially given by the UVFP of the gauge coupling strength in
Eq.~(\ref{eq:def_kappaD}). 
The value of $\kappa_D$ should be fixed once the discrete model
parameters such as $N_c$ and $N_f$ are determined.
We thus conclude that there is no finite continuum limit for 
$\kappa_D > \kappa_D^{\rm crit}$ at least within the ladder
approximation.
We probably need to introduce other higher dimensional interactions in
order to define the continuum limit with 
$\kappa_D > \kappa_D^{\rm crit}$. 

We also note that $\kappa_D = \kappa_D^{\rm crit}$ cannot be achieved
for $D=6,8$ within the approximations we used.\footnote{
For $D=10$ it happens $\kappa_D=\kappa_D^{\rm crit}$,
for instance, for $N_c=81$ and $N_f=20$.}
We thus concentrate upon the $\kappa_D < \kappa_D^{\rm crit}$ case hereafter.

\subsection{Decay constant}
\label{sec-fpi}

There appear massless Nambu-Goldstone (NG) fields in the {\DxSB}
vacuum. 
The interaction of the NG field is then described by the low
energy theorem and its decay constant.
In this subsection, we estimate the size of the decay constant by
using an approximation proposed by Pagels and Stokar (PS)~\cite{Pagels:hd}.
The decay constant can be written solely
in terms of the mass function $\Sigma$ within the PS approximation.
We know that the PS approximation works reasonably well in the
{\DxSB} of four-dimensional QCD\@. 

The $D$-dimensional generalization of the PS formula is given by
\begin{equation}
 F_\pi^2 = 2^{D/2}\NDA
 \int_0^{\Lambda^2} \!\!\!\! dx \,
 x^{D/2-1}\frac{\Sigma^2(x)-\frac{x}{D}
 \frac{d}{dx} \Sigma^2(x)}{(x+\Sigma^2)^2},
\end{equation}
where $F_\pi$ denotes the decay constant in $D$ dimensions
\begin{equation}
 \langle 0| J_M^i (0)|\pi_j(q)\rangle = -iq_M F_\pi \delta_{ij}.
\end{equation}
The four-dimensional decay constant $f_\pi$
is obtained through the matching condition
\begin{equation}
  f_\pi^2 = \frac{(2\pi R)^\delta}{2^n} F_\pi^2 .
\end{equation}

It is straightforward to estimate the decay constant by using the
asymptotic behavior of the mass function Eq.~(\ref{eq:asymp_sol}),
\begin{equation}
  F_\pi^2 \sim \Sigma_0^{2+2\nu(1-\omega)}
  \int^{\Lambda^2}_{\Sigma_0^2} dx x^{\nu\omega-2}.
\end{equation}
Note that $F_\pi$ diverges in the $\Lambda\rightarrow\infty$ limit for
\begin{equation}
  \omega \geq \frac{1}{\nu}.
\label{eq:MTcondition1}
\end{equation}
The condition Eq.~(\ref{eq:MTcondition1}) can be rewritten in terms of
$\kappa_D$,
\begin{equation}
  \kappa_D \leq \kappa_D^{\rm MT},
\label{eq:MTcondition2}
\end{equation}
with the marginal triviality (MT) point $\kappa_D^{\rm MT}$ being defined as 
\begin{equation}
  \kappa_D^{\rm MT} \equiv 
  \left(\,1-\frac{1}{\nu^2}\,\right) \kappa_D^{\rm crit}.
\end{equation}

The NG field interaction is suppressed by $1/F_\pi$ according to the
low energy theorem.
For $\kappa_D \leq \kappa_D^{\rm MT}$ the NG field is therefore
decoupled from the rest of system and becomes trivial in the continuum 
limit ($\Lambda\rightarrow\infty$).

On the other hand, surprisingly enough, the decay constant $F_\pi$
remains finite for 
\begin{equation}
  \kappa_D^{\rm MT} < \kappa_D < \kappa_D^{\rm crit},
\label{eq:MTcondition3}
\end{equation}
even in the continuum limit $\Lambda\rightarrow\infty$.
Once the condition Eq.~(\ref{eq:MTcondition3}) is satisfied, the NG
field enjoys non-trivial interactions in the ``renormalized'' theory
of the higher dimensional gauged NJL model. 

The MT point $\kappa_D^{\rm MT}$ is indicated by a
cross in the phase diagram Figure~\ref{fig:critical_line}.
At this point the decay constant $F_\pi$ diverges as
$F_\pi^2 \sim \ln \Lambda$,
while
$F_\pi^2 \sim \Lambda^{2(\nu\omega-1)}$ 
for $0 < \kappa_D < \kappa_D^{\rm MT}$.

In $D \to 4$ we find $\kappa_D^{\rm MT} \to 0$,
namely the MT point coincides with the pure NJL point, 
where it is well-known $F_\pi^2$ diverges logarithmically.
On the other hand, in the $D\rightarrow\infty$ limit 
$\kappa_D^{\rm MT} \to \kappa_D^{\rm crit}$,
and hence the region of the finite $F_\pi$ is squeezed out.

\subsection{Yukawa coupling and scalar propagator}
\label{sec-yukawa}

In addition to the NG bosons, a composite scalar boson also
appears in the spectrum of the NJL type models.
As we will show later, the auxiliary scalar field $\sigma$ of
Eq.~(\ref{GNJL2}) actually acquires non-trivial propagator at the loop
level.
In this subsection, we adopt a simple method proposed by Appelquist,
Terning, and Wijewardhana (ATW)~\cite{Appelquist:1991kn} to calculate
roughly the properties of the composite scalar.
We also describe more sophisticated method based on
Ref.~\cite{Gusynin:1997cw} in Appendix A\@.
These results qualitatively agree with each other. 

The auxiliary field $\sigma$ propagator at $|p^2|\gg \Sigma_0^2$ is
given by 
\begin{eqnarray}
\lefteqn{ iD_\sigma^{-1}(p) = } \nonumber \\
&& -N_c N_f \int\frac{d^Dq}{i(2\pi)^D}
 \tr\left[\,\Gamma_s(q+p,q) \frac{1}{\fsl{q}} 
             \frac{1}{\fsl{q}-\fsl{p}} \,\right] \nonumber \\ &&
 + \mbox{constant} ,
\label{eq:aux_prop}
\end{eqnarray}
where $\Gamma_s(q+p,q)$ is the Yukawa type vertex of the auxiliary
field, and
$q$ and $p$ are the $\psi$ and $\sigma$ momenta, respectively. 
The basic difficulty in the calculation of Eq.~(\ref{eq:aux_prop})
arises from our lack of knowledge of $\Gamma_s(q+p,q)$.
In Appendix A, we calculate analytically $\Gamma_s(q+p,q)$ 
with certain approximations.
As for the effective Yukawa vertex at $p=0$,
on the other hand, we easily find 
\begin{equation}
 \Gamma_s(q,q)=\frac{d}{d \sigma}\Sigma(-q^2) \sim \frac{2}{1+\omega}
  \left(\frac{-q^2}{\Lambda^2}\right)^{-\frac{\nu}{2}(1-\omega)},
 \label{yukawa}
\end{equation}
where we used Eq.~(\ref{eq:asymp_sol2}).
For the relation between $\Sigma$ and $\Gamma_s$, see, e.g.,
Ref.~\cite{Gusynin:1988cf}.

ATW showed that Eq.~(\ref{eq:aux_prop}) can be approximated by 
\begin{eqnarray}
\lefteqn{ iD_\sigma^{-1}(p) = } \nonumber \\
&& -N_c N_f \int\frac{d^Dq}{i(2\pi)^D}
 \tr\left[\,\Gamma_s(-q^2) \frac{1}{\fsl{q}} \Gamma_s(-q^2)
             \frac{1}{\fsl{q}-\fsl{p}} \,\right] \nonumber \\ &&
 + \mbox{constant} ,
\label{eq:aux_prop2}
\end{eqnarray}
within reasonable assumptions.
Note that Eq.~(\ref{eq:aux_prop2}) contains Yukawa vertex only at
$p=0$, 
\begin{equation}
  \Gamma_s(-q^2) = \Gamma_s(q,q),
\end{equation}
and we are thus able to use Eq.~(\ref{yukawa}) in the  evaluation of
the $\sigma$ propagator.

After the Wick rotation and performing the angular integrals, we
obtain 
\begin{eqnarray}
\lefteqn{ \hspace*{-0.4cm}
 iD_\sigma^{-1}(p)-iD_\sigma^{-1}(0)  
 } \nonumber \\
&& \hspace*{-0.5cm}
 = -2^{D/2} N_c N_f \int\frac{d^Dq}{i(2\pi)^D}
 \left[\,\Gamma_s^2(-q^2) \frac{p \cdot q-p^2}{q^2(p-q)^2} \,\right], 
   \nonumber\\
&& \hspace*{-0.5cm}
 =  2^{D/2} N_c N_f \NDA \int_0^{\Lambda^2}\!\!\!\! dy \, y^{D/2-2}
  \Gamma_s^2(y) K (x,y),  \label{sigma_prop}
\end{eqnarray}
where we defined
\begin{displaymath}
  x \equiv -p^2, \quad y \equiv -q^2.
\end{displaymath}
The kernel $K$ is given by
\begin{eqnarray}
\lefteqn{\hspace*{-5mm} K (x,y) = } \nonumber \\
&& \hspace*{-6mm}
 \frac{2}{D}\frac{\min(x,y)}{\max(x,y)}
  \hyper{2}{2-D/2}{D/2+1}{\frac{\min(x,y)}{\max(x,y)}}
  \nonumber \\ && \hspace*{-6mm}
 -\, \frac{x}{\max(x,y)}
   \hyper{1}{2-D/2}{D/2}{\frac{\min(x,y)}{\max(x,y)}}.
\end{eqnarray}
For even dimensions we obtain 
\begin{eqnarray}
K (x,y) &=& -\sum_{\ell=0}^{\delta/2+1}
     \frac{(-\delta/2-1)_{\ell}}{(\delta/2+2)_{\ell}}
      \left(\frac{y}{x}\right)^{\ell} \theta(x-y) \nonumber \\
& & \hspace*{-8mm}
    +\sum_{\ell=1}^{\delta/2+1}
     \frac{(-\delta/2-1)_{\ell}}{(\delta/2+2)_{\ell}}
      \left(\frac{x}{y}\right)^{\ell} \theta(y-x),
\end{eqnarray}
where $(\alpha)_\ell$ is defined as
\begin{equation}
  (\alpha)_\ell = (\alpha+\ell-1)(\alpha+\ell-2)\cdots (\alpha+1)\alpha
\label{eq:hyperdef}
\end{equation}
and $(\alpha)_0 = 1$.
Substituting Eq.~(\ref{yukawa}) for Eq.~(\ref{sigma_prop}), 
we obtain the composite scalar propagator
for $\omega \ne 1/\nu$ ($\kappa_D \ne \kappa_D^{\rm MT}$):
\begin{eqnarray}
\lefteqn{\hspace*{-8mm}
  iD_\sigma^{-1}(p)-iD_\sigma^{-1}(0) =
  2^{D/2} N_c N_f \NDA \left(\frac{2}{1+\omega}\right)^2
 } \nonumber \\
&& \hspace*{-7mm} 
   \times \Lambda^{D-2}\left[\,C_{\nu\omega}
    \left(\frac{x}{\Lambda^2}\right)^{\nu\omega}
   +\sum_{\ell=1}^{\delta/2+1} C_\ell
    \left(\frac{x}{\Lambda^2}\right)^{\ell}
  \,\right], 
\end{eqnarray}
where 
\begin{equation}
 C_{\nu\omega} \equiv
  - \frac{1}{\nu\omega} - \sum_{\ell=1}^{\delta/2+1}
     \frac{(-\delta/2-1)_{\ell}}{(\delta/2+2)_{\ell}}
     \frac{2\nu\omega}{(\nu\omega)^2-\ell^2}
\end{equation}
and
\begin{equation}
  C_\ell \equiv
  \frac{(-\delta/2-1)_{\ell}}{(\delta/2+2)_{\ell}}\frac{1}{\nu\omega-\ell},
  \quad (\ell \geq 1). 
\end{equation}
At the marginal triviality point $\omega=1/\nu$
($\kappa_D=\kappa_D^{\rm MT}$), we find
\begin{eqnarray}
\lefteqn{
  iD_\sigma^{-1}(p)-iD_\sigma^{-1}(0) =
  2^{D/2} N_c N_f \NDA \left(\frac{2}{1+\omega}\right)^2
 } \nonumber \\
&& \times \Lambda^{D-2}\left[\,
    C_0 \left(\frac{x}{\Lambda^2}\right)
     \ln \left(\frac{\;\Lambda^2}{x}\right)
   +C'_1 \left(\frac{x}{\Lambda^2}\right) \right. \nonumber \\ 
&&  \qquad \qquad \qquad \qquad \left.
   +\sum_{\ell=2}^{\delta/2+1} C_\ell
    \left(\frac{x}{\Lambda^2}\right)^{\ell}
  \,\right], 
\end{eqnarray}
where 
\begin{equation}
  C_0 \equiv -\frac{D-2}{D}
\end{equation}
and
\begin{equation}
 C'_1 \equiv
  - \frac{C_0}{2} - \frac{1}{\nu\omega} - \sum_{\ell=2}^{\delta/2+1}
     \frac{(-\delta/2-1)_{\ell}}{(\delta/2+2)_{\ell}}
     \frac{2\nu\omega}{(\nu\omega)^2-\ell^2} .
\end{equation}

We next consider the $\Lambda\rightarrow\infty$ limit.
It is convenient to renormalize the auxiliary field $\sigma$ 
\begin{equation}
 \sigma_R \equiv Z_\sigma^{-1/2}\sigma, 
\end{equation}
in such a limit.
We define the renormalization constant $Z_\sigma$ so as to keep the
renormalized propagator finite,
\begin{equation}
  Z_\sigma^{-1} =
  \left\{
    \begin{array}{l@{\quad}c}
      {\displaystyle
        \left(\dfrac{\Lambda}{\mu}\right)^{2\nu(1-\omega)},
      } & 
      {\displaystyle
        (\kappa_D^{\rm MT} < \kappa_D < \kappa_D^{\rm crit}),
      } \\[3ex]
      {\displaystyle
        \left(\dfrac{\Lambda}{\mu}\right)^{2(\nu-1)} 
        \ln\left( \dfrac{\Lambda^2}{\mu^2} \right),
      } & 
      {\displaystyle
        (\kappa_D = \kappa_D^{\rm MT}),
      } \\[3ex]
      {\displaystyle
        \left(\dfrac{\Lambda}{\mu}\right)^{2(\nu-1)},
      } & 
      {\displaystyle
        (0 < \kappa_D < \kappa_D^{\rm MT}).
      } 
    \end{array}
  \right.
  \label{eq:Z_wav}
\end{equation}
The renormalized propagator is then given by
\begin{eqnarray}
  \lefteqn{ \hspace*{-4mm}
    \dfrac{(iD_{\sigma(R)}^{-1}(p)-iD_{\sigma(R)}^{-1}(0))}
          {2^{D/2} N_c N_f \NDA} 
    \left(\dfrac{1+\omega}{2}\right)^{2} 
    = 
  }\nonumber \\[2mm]
  & & 
  \left\{
    \begin{array}{l@{\quad}c}
      {\displaystyle
        C_{\nu\omega} \mu^{2\nu}
        \left(\dfrac{x}{\mu^2}\right)^{\nu\omega},
      } & 
      {\displaystyle
        (\kappa_D^{\rm MT} < \kappa_D < \kappa_D^{\rm crit}),
      } \\[3ex]
      {\displaystyle
        C_{0} \mu^{2\nu}
        \left(\dfrac{x}{\mu^2}\right),
      } & 
      {\displaystyle
        (\kappa_D = \kappa_D^{\rm MT}),
      } \\[3ex]
      {\displaystyle
        C_{\ell=1} \mu^{2\nu}
        \left(\dfrac{x}{\mu^2}\right),
      } & 
      {\displaystyle
        (0 < \kappa_D < \kappa_D^{\rm MT}),
      } 
    \end{array}
  \right. \nonumber\\
\end{eqnarray}
in the $\Lambda\rightarrow\infty$ limit.

What does happen for the Yukawa type vertex, then?
From Eq.~(\ref{yukawa}) and Eq.~(\ref{eq:Z_wav}) we obtain
\begin{eqnarray}
  \lefteqn{
    \Gamma_s^{(R)}(-q^2) 
    = Z_\sigma^{1/2} \Gamma_s(-q^2)
    \sim 
  } \nonumber \\[2mm]
  & & 
  \left\{
    \begin{array}{l@{\quad}c}
      {\displaystyle
        \frac{2}{1+\omega}
        \left(\frac{-q^2}{\mu^2}\right)^{-\frac{\nu}{2}(1-\omega)}.
      } & 
      {\displaystyle
        (\kappa_D^{\rm MT} < \kappa_D < \kappa_D^{\rm crit}),\quad
      } \\[4ex]
      {\displaystyle
        0
      } & 
      {\displaystyle
        (0 < \kappa_D \le \kappa_D^{\rm MT}),
      } 
    \end{array}
  \right.  \nonumber \\ 
  \label{eq:ren_yukawa} 
\end{eqnarray}
in the $\Lambda\rightarrow\infty$ limit.
We note that the composite scalar thus decouples from the rest of the
system for $\kappa_D \le \kappa_D^{\rm MT}$ in the continuum limit.
On the other hand, for $\kappa_D^{\rm MT}< \kappa_D < \kappa_D^{\rm
  crit}$, the composite scalar interacts with fermions in a
non-trivial manner through the Yukawa vertex
Eq.~(\ref{eq:ren_yukawa}). 
These behaviors are consistent with our analysis of the NG field
(decay constant) in the previous subsection.

\section{More about the mass function}
\label{sec-moreabout}

For the determination of the precise scaling behavior around the
critical line, it is not enough to know the dominant asymptotic
behavior of mass function Eq.~(\ref{eq:asymp_sol}).
Actually, lacking the sub-dominant term,
the bifurcation method 
does not reproduce  
exactly the same scaling relation (gap equation) as that of 
the pure NJL model in the vanishing gauge coupling limit
even in four dimensions~\cite{Nonoyama}.
We also need information of sub-dominant terms.
The sub-dominant terms are also required in the calculation of the
effective potential.
Unfortunately, the bifurcation technique is not enough for such a
purpose. 
In this section, we evaluate the sub-dominant terms first in
the well-known linearized SD equation, and next in the power expansion
method.

It is known that the linearized approximation works well in four
dimensions. By using the linearized ansatz, we can solve analytically
the ladder SD equation. We obtain the coefficient of the sub-dominant
term  as well as its exponent.
The power expansion method reflects the correct asymptotic behavior of
the nonlinear SD equation. However, we can determine only the exponent of
the sub-dominant term in the method. The coefficient of the sub-dominant
term is left as an unknown parameter.
In a certain region of the gauge coupling, it turns out that
the exponent of the sub-dominant term is different depending on the
methods. We thus describe both results.

\subsection{Linearized approximation}

We first consider the linearized SD equation~\cite{Fomin,Miransky:vk}, in
which the $\Sigma^2$ term in the denominator is replaced by a constant
$\Sigma_0^2 \equiv \Sigma^2(x=0)$.
This approximation works well in the
determination of the dominant asymptotic solution at $x\gg \Sigma_0^2$.
It is also known that the scaling behavior determined from the
linearized SD equation closely approximates to the numerical results of
the non-linear SD equation in the case of the four-dimensional gauged
NJL model.

The linearized SD equation is given by
\begin{eqnarray}
 \Sigma(x)
      &=& \sigma + \dfrac{4(D-1)}{D} 
          \kappa_D \int_0^{\Lambda^2}\!\!\! dy 
          \frac{y^{D/2-1}\Sigma(y)}{y + \Sigma_0^2} \nonumber \\
      & &    \qquad \times \frac{1}{[\max(x,y)]^{D/2-1}},
  \label{eq:linearized_SD}
      \\
  \sigma 
      &=& m_0 + \frac{g}{\Lambda^{D-2}} \int_0^{\Lambda^2}
           dx x^{D/2-1} \dfrac{\Sigma(x)}{x + \Sigma_0^2},
\end{eqnarray}
and a subsidiary condition,
\begin{equation}
  \Sigma_0 = \Sigma(x=0).
  \label{eq:subsidiary}
\end{equation}

The integral equation Eq.~(\ref{eq:linearized_SD}) is equivalently
rewritten in terms of a set of differential equation and boundary
conditions,
\begin{equation}
 x^2 \frac{d^2}{dx^2}\Sigma + (\nu+1) x \frac{d}{dx} \Sigma
  + \frac{\nu^2(1-\omega^2)}{4}
    \frac{x \Sigma}{x+\Sigma_0^2} = 0, 
  \label{eq:SDdifflinear}
\end{equation}
\begin{eqnarray}
  x^{\nu+1} \frac{d}{dx} \Sigma(x)\bigg|_{x = 0} &=& 0, 
  \label{eq:IR-BC2}
  \\
   \left[\, \nu + x \frac{d}{dx} \,\right] \Sigma(x)
   \bigg|_{x  = \Lambda^2} &=& \nu \sigma, 
  \quad
  \label{eq:UV-BC2}
\end{eqnarray}
where
$\nu$ and $\omega$ are defined in Eq.~(\ref{eq:nudef}) and
Eq.~(\ref{eq:omegadef}). 
The UVBC Eq.~(\ref{eq:UV-BC2}) can also be expressed as
\begin{equation}
  \left[\, \nu + \left(1+\dfrac{4}{1-\omega^2}\frac{g}{\nu}\right)
  x\frac{d}{dx}\,\right] \Sigma(x) \bigg|_{x = \Lambda^2}
  = \nu m_0. \quad \label{eq:UV-BC3} 
\end{equation}

Note that the differential equation (\ref{eq:SDdifflinear}) possesses
three regular singular points at $x=0, -\Sigma_0^2, \infty$.
The solution is then expressed by using the Gauss hypergeometric
functions. 
It is now easy to find the solution of Eq.~(\ref{eq:SDdifflinear})
satisfying the subsidiary condition Eq.~(\ref{eq:subsidiary}) and the
IRBC Eq.~(\ref{eq:IR-BC2}),
\begin{equation}
  \Sigma(x) = \Sigma_0 
  \hyper{\frac{\nu}{2}(1+\omega)}{\frac{\nu}{2}(1-\omega)}{\nu+1}
        {-\frac{x}{\Sigma_0^2}},
\label{eq:solution_in_hypergeo}
\end{equation}
where $F$ is the Gauss hypergeometric function
\begin{equation}
  F(\alpha,\beta,\gamma ; z) 
  \equiv \sum_{n=0}^{\infty} 
  \dfrac{(\alpha)_n (\beta)_n}{(\gamma)_n} \dfrac{z^n}{n!}
\label{eq:hypergeo_def}
\end{equation}
with $(\alpha)_n$, $(\beta)_n$, $(\gamma)_n$ being defined in
Eq.~(\ref{eq:hyperdef}).  
We assumed here $\kappa_D < \kappa_D^{\rm crit}$ and hence a real and
positive $\omega$.

The UVBC Eq.~(\ref{eq:UV-BC3}) determines $\Sigma_0$ and 
thus the scaling relation.
We need to know the behavior of the mass function
Eq.~(\ref{eq:solution_in_hypergeo}) in the ultraviolet region 
$x \gg \Sigma_0^2$.  
For such a purpose, it is convenient to use a well-known
formula~\cite{Bateman}, 
\begin{eqnarray}
\lefteqn{
  F(\alpha,\beta,\gamma ; z) =
} \nonumber\\
  & &
     \dfrac{\Gamma(\gamma) \Gamma(\beta-\alpha)}
            {\Gamma(\beta) \Gamma(\gamma-\alpha)} 
      (-z)^{-\alpha}
      F(\alpha,\alpha-\gamma+1,\alpha-\beta+1 ; \frac{1}{z})
  \nonumber\\
  & & +
      \dfrac{\Gamma(\gamma) \Gamma(\alpha-\beta)}
            {\Gamma(\alpha) \Gamma(\gamma-\beta)} 
      (-z)^{-\beta}
      F(\beta,\beta-\gamma+1,\beta-\alpha+1 ; \frac{1}{z}).
  \nonumber\\
  & &
\label{eq:hypergeo_formula}
\end{eqnarray}
Combining Eq.~(\ref{eq:hypergeo_def}) and
Eq.~(\ref{eq:hypergeo_formula}), we expand the solution
Eq.~(\ref{eq:solution_in_hypergeo}) for $x\gg \Sigma_0^2$,
\begin{eqnarray}
  \Sigma(x) &=& \Sigma_0
  \left(\dfrac{x}{\Sigma_0^2}\right)^{-\frac{\nu}{2}(1-\omega)} 
  \sum_{n=1}^\infty c_n \left(\dfrac{x}{\Sigma_0^2}\right)^{-n+1}
  \nonumber\\
  & & \hspace*{-1cm}
  +\Sigma_0
  \left(\dfrac{x}{\Sigma_0^2}\right)^{-\frac{\nu}{2}(1+\omega)} 
  \sum_{n=1}^\infty d_n \left(\dfrac{x}{\Sigma_0^2}\right)^{-n+1},
\label{eq:linearized_SD_asymptotic}
\end{eqnarray}
with $c_n$ and $d_n$ being given by
\begin{eqnarray}
  c_n &=& 
      \dfrac{(-1)^{n-1}}{(n-1)!}
      \dfrac{\Gamma(\nu+1)\Gamma(\nu\omega)}
                {\Gamma(\frac{\nu}{2}(1+\omega))
                 \Gamma(\frac{\nu}{2}(1+\omega)+1)}
      \nonumber\\
      & & \times \dfrac{
           \left(\frac{\nu}{2}(1-\omega)\right)_{n-1}
           \left(-\frac{\nu}{2}(1+\omega)\right)_{n-1}
          }{\left(-\nu\omega + 1 \right)_{n-1}},
\label{eq:c_n}
       \\
  d_n &=& 
      \dfrac{(-1)^{n-1}}{(n-1)!}
      \dfrac{\Gamma(\nu+1)\Gamma(-\nu\omega)}
                {\Gamma(\frac{\nu}{2}(1-\omega))
                 \Gamma(\frac{\nu}{2}(1-\omega)+1)}
      \nonumber\\
      & & \times \dfrac{
           \left(\frac{\nu}{2}(1+\omega)\right)_{n-1}
           \left(-\frac{\nu}{2}(1-\omega)\right)_{n-1}
          }{\left(\nu\omega + 1 \right)_{n-1}}.
\label{eq:d_n}
\end{eqnarray}
The $c_1$ term in
Eq.~(\ref{eq:linearized_SD_asymptotic}) gives the leading
contribution to the mass function $\Sigma(x)$ in the asymptotic
region $x \gg \Sigma_0^2$ for $\kappa_D < \kappa_D^{\rm crit}$.
The next-to-leading contribution, on the other hand, comes from $c_2$
or $d_1$, depending on the value of $\kappa_D$.
For $\kappa_D^{\rm MT} < \kappa_D < \kappa_D^{\rm crit}$ $d_1$ term
gives the next-to-leading contribution, while $c_2$ term becomes the
next-to-leading for $\kappa_D<\kappa_D^{\rm MT}$.
At $\kappa_D = \kappa_D^{\rm MT}$ the power of $c_2$ and $d_1$ terms
goes same in the expression of Eq.~(\ref{eq:linearized_SD_asymptotic}).
Although the coefficients $c_2 \propto 1/(-\nu\omega+1)$ and 
$d_1 \propto \Gamma(-\nu\omega)$ diverge at 
$\kappa_D = \kappa_D^{\rm MT}$, i.e., $\omega=1/\nu$,
they cancel out each other, so that the logarithmic term appears,
\begin{eqnarray}
  \Sigma(x) &=& c_1 \Sigma_0
   \left(\dfrac{x}{\Sigma_0^2}\right)^{-\frac{1}{2}(\nu-1)}
  \nonumber \\ && 
  -\frac{\nu^2-1}{4} \, c_1 \Sigma_0
   \left(\dfrac{x}{\Sigma_0^2}\right)^{-\frac{1}{2}(\nu+1)}
   \ln \frac{x}{\Sigma_0^2} \nonumber \\ && 
  + \cdots, \hspace*{2.0cm} (\kappa_D = \kappa_D^{\rm MT}).
\end{eqnarray}

In order to determine the scaling relation near the critical line, 
we compare the leading ($c_1$) and the next-to-leading ($d_1$
or $c_1$ depending on $\kappa_D$) terms in the UVBC
Eq.~(\ref{eq:UV-BC3}). 
For $m_0=0$, we find
\begin{equation}
  \left(\dfrac{\Sigma_0}{\Lambda}\right)^{2\nu\omega}
  =  -\dfrac{(1-\omega^2)}{4\omega}\dfrac{c_1}{d_1}
        \left(
          1 - \dfrac{g^{\rm crit}}{g}
        \right),
\label{eq:lin_scaling1}
\end{equation}
for $\kappa_D^{\rm MT} < \kappa_D < \kappa_D^{\rm crit}$, and
\begin{equation}
  \left(\dfrac{\Sigma_0}{\Lambda}\right)^{2}
  =  -\dfrac{\nu (1-\omega^2)}{4}\dfrac{c_1}{c_2}
        \left(
          1 - \dfrac{g^{\rm crit}}{g}
        \right),
\end{equation}
for $\kappa_D < \kappa_D^{\rm MT}$, where we used
$g \simeq g^{\rm crit}$. 
Note here that the signs of $c_1$ and $d_1$ ($c_2$) are opposite to each
other for $\kappa_D^{\rm MT}<\kappa_D$ ($\kappa_D<\kappa_D^{\rm MT}$).
We thus find
\begin{equation}
  \dfrac{\Sigma_0}{\Lambda} \sim \left\{
    \begin{array}{l@{\quad}c}
      {\displaystyle
        \left(
          1 - \dfrac{g^{\rm crit}}{g}
        \right)^{\frac{1}{2\nu\omega}}
      }, & 
      {\displaystyle
        (\kappa_D^{\rm MT} < \kappa_D < \kappa_D^{\rm crit}),
      } \\[4ex]
      {\displaystyle
        \sqrt{
          1 - \dfrac{g^{\rm crit}}{g}
        }
      },
      & 
      {\displaystyle
        ( 0 \leq \kappa_D < \kappa_D^{\rm MT} ).
      }
    \end{array}
  \right. \label{gapeq-linear}
\end{equation}
Note here that Eq.~(\ref{gapeq-linear}) in $\kappa_D\rightarrow 0$
limit agrees with the scaling relation obtained in the $D$($>4$)
dimensional pure NJL model~\cite{Hashimoto:2000uk}.

The chiral condensate is calculated from Eq.~(\ref{UV-BC}),
\begin{eqnarray}
  \sigma &=& 
  \frac{(1+\omega)}{2} c_1  \Sigma_0
   \left(\dfrac{\Sigma_0}{\Lambda}\right)^{\nu (1-\omega)} 
  \nonumber\\ 
  & &
  +
  \frac{(1-\omega)}{2} d_1 \Sigma_0
   \left(\frac{\Sigma_0}{\Lambda}\right)^{\nu (1+\omega)} +\cdots,
\label{eq:linearized_sigma1}
\end{eqnarray}
for $\kappa_D^{\rm MT} < \kappa_D < \kappa_D^{\rm crit}$, and
\begin{eqnarray}
  \sigma &=&
  \frac{(1+\omega)}{2} c_1  \Sigma_0
   \left(\dfrac{\Sigma_0}{\Lambda}\right)^{\nu (1-\omega)}
  \nonumber\\ 
  &&
  +\left(\frac{1+\omega}{2}-\frac{1}{\nu}\right) c_2 \Sigma_0
   \left(\dfrac{\Sigma_0}{\Lambda}\right)^{\nu (1-\omega)+2} + \cdots,
  \label{eq:linearized_sigma2}
  \nonumber\\ 
\end{eqnarray}
for $ 0 < \kappa_D < \kappa_D^{\rm MT}$.

Further details of our analysis on the linearized SD equation are given
in Appendix~\ref{sec-appendix2}. 
 
\subsection{Power expansion method}

We next try to investigate the sub-dominant asymptotic solution
directly from the non-linear SD equation Eq.~(\ref{imp_ladder_SD})
without using the linearizing ansatz of the previous 
subsection.~\cite{Bardeen:sv}
The non-linear differential equation Eq.~(\ref{sde_diff}) and the
boundary conditions Eqs.~(\ref{IR-BC}) and (\ref{UV-BC}) can be
rewritten as
\begin{eqnarray}
  & & \hspace*{-1.2cm}
  x^2 \dfrac{d^2\Sigma}{dx^2} + (\nu +1 ) x\dfrac{d\Sigma}{dx} +
  \dfrac{\nu^2(1-\omega^2)}{4} \dfrac{x\Sigma}{x+\Sigma^2}=0, 
\label{eq:PE_diff}
  \\
  & & \hspace*{-1.2cm}
  \left. x^{\nu+1} \dfrac{d}{dx} \Sigma(x) \right|_{x=0} = 0, 
\label{eq:PE_IRBC}
  \\
  & & \hspace*{-1.2cm}
  \left. \left[
    \nu +
  \left(1+\dfrac{4}{1-\omega^2}\frac{g}{\nu}\right)x\frac{d}{dx}
  \right] \Sigma(x) \right|_{x=\Lambda^2} = \nu m_0. 
\label{eq:PE_UVBC}
\end{eqnarray}
For $x\gg \Sigma^2(x)$ the last term in the non-linear differential
equation Eq.~(\ref{eq:PE_diff}) can be expanded as
\begin{equation}
  \dfrac{x\Sigma}{x+\Sigma^2} = \Sigma - \dfrac{\Sigma^3}{x} +
  \dfrac{\Sigma^5}{x^2} + \cdots.
\label{eq:PE_expansion1}
\end{equation}
Eq.~(\ref{eq:PE_expansion1}) allows us formally to expand the
solution $\Sigma$ of Eq.~(\ref{eq:PE_diff}),
\begin{equation}
  \Sigma(x) = \Sigma_1(x) + \Sigma_2(x) + \cdots,
\label{eq:PE_expansion2}
\end{equation}
where $\Sigma_1(x)$ is the solution of a homogeneous linear differential
equation
\begin{equation}
  \left[
    x^2 \dfrac{d^2}{dx^2} + (\nu+1) x\dfrac{d}{dx} 
    + \dfrac{\nu^2(1-\omega^2)}{4}
  \right] \Sigma_1(x) = 0,
\label{eq:PE_diff1}
\end{equation}
while $\Sigma_2(x)$ is the inhomogeneous part of the solution of 
\begin{eqnarray}
& & 
  \left[
    x^2 \dfrac{d^2}{dx^2} + (\nu+1) x\dfrac{d}{dx} 
    + \dfrac{\nu^2(1-\omega^2)}{4}
  \right] \Sigma_2(x) 
\nonumber\\
 & & \qquad\qquad
    =\dfrac{\nu^2(1-\omega^2)}{4}\dfrac{\Sigma_1^3(x)}{x}. 
\label{eq:PE_diff2}
\end{eqnarray}
The higher terms in the expansion Eq.~(\ref{eq:PE_expansion2}) are
given in a similar manner to Eq.~(\ref{eq:PE_diff2}) just by replacing
the R.H.S. to $\Sigma_1^2\Sigma_2$, $\Sigma_1\Sigma_2^2$, $\Sigma_2^3$,
$\Sigma_1^5$, $\cdots$.

What does the expansion Eq.~(\ref{eq:PE_expansion2}) mean then?
The solution of Eq.~(\ref{eq:PE_diff1}) is given by
\begin{equation}
  \Sigma_1(x) = 
  c_1 \Sigma_0 
  \left( \dfrac{x}{\Sigma_0^2} \right)^{-\frac{1}{2}\nu(1-\omega)}
  + d_1 \Sigma_0
  \left( \dfrac{x}{\Sigma_0^2} \right)^{-\frac{1}{2}\nu(1+\omega)},
\label{eq:PE_sol1}
\end{equation}
with $\Sigma_0$ being a constant with a mass dimension.
We thus obtain 
\begin{eqnarray}
\lefteqn{
  \Sigma_2(x) =} \nonumber \\ && 
  c_2 \Sigma_0 
  \left( \dfrac{x}{\Sigma_0^2} \right)^{-\frac{3}{2}\nu(1-\omega)-1}
  +c_2' \Sigma_0 
  \left( \dfrac{x}{\Sigma_0^2} \right)^{-\frac{3}{2}\nu +
  \frac{1}{2}\nu\omega -1}
  \nonumber\\
  & &
  +d_2 \Sigma_0 
  \left( \dfrac{x}{\Sigma_0^2} \right)^{-\frac{3}{2}\nu(1+\omega)-1}
  +d_2' \Sigma_0 
  \left( \dfrac{x}{\Sigma_0^2} \right)^{-\frac{3}{2}\nu -
  \frac{1}{2}\nu\omega -1} \hspace*{-1.5cm} ,
\end{eqnarray}
with $c_2 \propto c_1^3$, $c_2'\propto c_1^2 d_1$, 
$d_2 \propto d_1^3$, $d_2'\propto c_1 d_1^2$. 
Note that the exponent of the power damping of $\Sigma_2(x)$ is
steeper than that of $\Sigma_1(x)$ for $x \gg \Sigma_0^2$.
It is also obvious that the higher terms in the expansion
Eq.~(\ref{eq:PE_expansion2}) falls quicker even than $\Sigma_2(x)$ for
$x\gg \Sigma_0^2$.
The expansion Eq.~(\ref{eq:PE_expansion2}) can therefore be understood
as an ``power expansion'' in this sense.

Since we are interested only in the dominant term ($c_1$) and in the
sub-dominant term ($d_1$ or $c_2$ depending on $\kappa_D$), we
restrict our analysis to these three terms. 
The power damping behavior of $c_1$ and $d_1$ terms agree with those
in the linearized SD equation.
We note, however, the exponent of the power damping of the $c_2$ term
differs from the the result of the linearized SD equation.
Actually, the $c_2$ term becomes sub-dominant for 
$\kappa_D < \kappa_D^{\rm PE}$,
\begin{equation}
  \kappa_D^{\rm PE} \equiv
  \left[
    1 - \frac{1}{4}\left(1+\frac{1}{\nu}\right)^2
  \right] \kappa_D^{\rm crit}.
\end{equation}
We also note
\begin{equation}
  \kappa_D^{\rm PE} < \kappa_D^{\rm MT},
\end{equation}
for $D>4$.

We first consider the case $\kappa_D > \kappa_D^{\rm PE}$, where the
$d_1$ term gives the sub-dominant contribution.
Since the power expansion cannot be adopted in the infrared region, we
are not able to use the IRBC Eq.~(\ref{eq:PE_IRBC}) to determine the
coefficient $c_1$  and $d_1$.
If we assume that $d_1$ possesses opposite sign to $c_1$, we find a
scaling relation similar to Eq.~(\ref{eq:lin_scaling1}). 
For $d_1/c_1>0$, on the other hand, the R.H.S. of
Eq.~(\ref{eq:lin_scaling1}) becomes negative.
Eq.~(\ref{eq:lin_scaling1}) thus cannot be understood as a scaling
relation.
This problem may be related with the vacuum instability which will be
discussed in Sec.\ref{eff_pot}.
Hereafter we assume the result in the linearized SD equation,
$d_1/c_1<0$, remains to be valid even in the power expansion method
for $\kappa_D > \kappa_D^{\rm MT}$.

We next turn to the $\kappa_D < \kappa_D^{\rm PE}$ case, where the
$c_2$ term gives the sub-dominant contribution.
From Eq.~(\ref{eq:PE_diff2}), we find
\begin{equation}
  c_2 = -\frac{\nu^2 (1-\omega^2)}
              {4(\nu+1-\nu\omega)(2\nu\omega-\nu-1)} c_1^3,
\label{eq:c2c1}
\end{equation}
and thus $c_2/c_1<0$.
The scaling relation is then given by
\begin{equation}
  \left(\dfrac{\Sigma_0}{\Lambda}
  \right)^{2\nu(1-\omega)+2}
  = - \dfrac{\nu(1-\omega^2)}{4(\nu+1-\nu\omega)}
      \dfrac{c_1}{c_2} \left(
        1 - \dfrac{g^{\rm crit}}{g}
      \right), \label{eq:pow_scaling2}
\end{equation}
with this $\kappa_D$, where we used $g \simeq g^{\rm crit}$.
We thus find
\begin{equation}
  \dfrac{\Sigma_0}{\Lambda} \sim \left\{
    \begin{array}{l@{\quad}c}
      {\displaystyle
        \left(
          1 - \dfrac{g^{\rm crit}}{g}
        \right)^{\frac{1}{2\nu\omega}}
      }, & 
      {\displaystyle
        (\kappa_D^{\rm PE} < \kappa_D < \kappa_D^{\rm crit}),
      } \\[4ex]
      {\displaystyle
        \left(
          1 - \dfrac{g^{\rm crit}}{g}
        \right)^{\frac{1}{2\nu(1-\omega)+2}}
      },
      & 
      {\displaystyle
        ( 0 \leq \kappa_D < \kappa_D^{\rm PE} ).
      }
    \end{array}
  \right. \label{gapeq-pow}
\end{equation}

As in the linearized approximation,
the coefficient $c_2$ in Eq.~(\ref{eq:c2c1}) diverges at
$\kappa_D = \kappa_D^{\rm PE}$, i.e, $\omega = 1/2+1/(2\nu)$.
This implies that the logarithmic term appears at the point.
We can easily confirm that $\Sigma_2(x)$ given by
\begin{equation}
  \Sigma_2(x) = -\frac{(\nu-1)(3\nu+1)}{8(\nu+1)}\,c_1^3 \Sigma_0
  \left(\dfrac{x}{\Sigma_0^2}\right)^{-\frac{3\nu+1}{4}}
  \ln \frac{x}{\Sigma_0^2},
\end{equation}
satisfies Eq.~(\ref{eq:PE_diff2}).

The chiral condensation is calculated from Eq.~(\ref{UV-BC}),
\begin{eqnarray}
  \sigma &=& \frac{1}{2}(1+\omega)c_1 \Sigma_0
  \left(\frac{\Lambda^2}{\Sigma_0^2}\right)^{-\frac{\nu}{2}(1-\omega)}
  \nonumber \\ && +
  \frac{1}{2}(1-\omega)d_1  \Sigma_0
  \left(\frac{\Lambda^2}{\Sigma_0^2}\right)^{-\frac{\nu}{2}(1+\omega)}
  +\cdots,
  \qquad
  \label{sig_pw1}
\end{eqnarray}
for $\kappa_D^{\rm PE} < \kappa_D$, while
\begin{eqnarray}
  \sigma &=& \frac{1}{2}(1+\omega)c_1 \Sigma_0
  \left(\frac{\Lambda^2}{\Sigma_0^2}\right)^{-\frac{\nu}{2}(1-\omega)}
  \nonumber \\ && -
  \frac{1}{2}\left(1-3\omega+\frac{2}{\nu}\right)c_2 \Sigma_0
  \left(\frac{\Lambda^2}{\Sigma_0^2}\right)^{-\frac{3\nu}{2}(1-\omega)-1}
  +\cdots,
  \nonumber \\
  \label{sig_pw2}
\end{eqnarray}
for $0 < \kappa_D < \kappa_D^{\rm PE}$.

\section{Effective potential}
\label{eff_pot}

We here outline our method to calculate the effective potential
$V(\sigma)$ in the gauged NJL model. 
See Refs.~\cite{eff_pot, MY97, Gusynin:2002cu} for details.
For simplicity we assume $\pi_i\equiv 0$ in this analysis.
It is easy to restore the $\pi_i$ degree of freedom by using the
chiral symmetry, i.e., $\sigma^2 \to \sigma^2+\pi_i^2$. 

We start with the partition function $W[J]$,
\begin{equation}
  W[J] \equiv \frac{1}{i} \ln \int [d\psi d\bar\psi] [{\rm gauge}]
    \exp \left( i\int d^D x ({\cal L} + J \sigma) \right),
\label{eq:defWJ}
\end{equation}
with ${\cal L}$ being the Lagrangian of the auxiliary field
description of the gauged NJL model Eq.~(\ref{GNJL2}).
For a constant external field $J$, the partition function can be
written as
\begin{equation}
  W[J] = \int d^D x w(J).
\end{equation}
We note
\begin{equation}
  \dfrac{d}{d J} w(J) = \sigma,
\label{eq:wJ}
\end{equation}
with $\sigma$ in the R.H.S. being understood as the VEV of $\sigma$.
Solving $J$ as a function of $\sigma$ in Eq.~(\ref{eq:wJ}), we define
the effective potential $V(\sigma)$,
\begin{equation}
  V(\sigma) = J \sigma - w(J),
\end{equation}
and thus obtain
\begin{equation}
  \dfrac{d}{d \sigma} V(\sigma) = J, \quad
  \mbox{or}
  \quad
  V(\sigma) = \int^\sigma d\sigma J.
\label{eq:localpot1}
\end{equation}

We next describe how we calculate $J$ in terms of $\sigma$.
Looking at Eq.~(\ref{GNJL2}) and Eq.~(\ref{eq:defWJ}), we find the
effect of constant $J$ can be taken into account by replacing the bare
mass $m_0$,
\begin{equation}
  \frac{N_c}{G} m_0 \rightarrow \frac{N_c}{G} m_0 + J.
\label{eq:replace}
\end{equation}
in the SD equation.
From Eq.~(\ref{sigma}) we obtain
\begin{equation}
 \dfrac{G}{N_c} J = -m_0 +\sigma + \dfrac{4}{1-\omega^2}
 \dfrac{g}{\nu^2} \left. x \dfrac{d}{dx}\Sigma(x)
 \right|_{x=\Lambda^2}.
\end{equation}
The derivative at the cutoff 
$\left. \frac{d}{dx}\Sigma(x)\right|_{x=\Lambda^2}$ can be calculated
from the UVBC Eq.~(\ref{UV-BC}),
\begin{equation}
  \left. x \dfrac{d}{dx}\Sigma(x)
 \right|_{x=\Lambda^2}
 = \nu \left[ \sigma - \Sigma(\Lambda^2) \right],
\end{equation}
which leads to a compact expression for $J$,
\begin{equation}
 \dfrac{G}{N_c} J = -m_0 +\sigma + \dfrac{4}{1-\omega^2}
 \dfrac{g}{\nu} \left[ \sigma - \Sigma(\Lambda^2) \right].
\label{eq:expressionJ}
\end{equation}
We thus obtain the effective potential $V(\sigma)$ 
in terms of $\sigma$ and $\Sigma(\Lambda^2)$:
\begin{eqnarray}
\lefteqn{\hspace*{-1cm}
 \left(2^{D/2}N_cN_f\Lambda^D\NDA\right)^{-1}V(\sigma) = 
} \nonumber \\[3mm]
&&
 -\frac{1}{g}\frac{m_0 \sigma}{\Lambda^2}+\frac{1}{2g}
  \dfrac{\sigma^2}{\Lambda^2} \nonumber \\[3mm]
&&
+\frac{4}{\nu(1-\omega^2)}\frac{1}{\Lambda^2}
 \int d\sigma \, [\sigma-\Sigma(\Lambda^2)] . \label{pot_gen}
\end{eqnarray}
In the pure NJL limit $\kappa_D=0$, i.e., $\omega=1$, we should return to
Eq.~(\ref{imp_ladder_SD}).
Plugging with $\Sigma(x)=\sigma$ and 
Eqs.~(\ref{eq:vev_sigma}), (\ref{eq:replace}), we easily find
\begin{eqnarray}
\lefteqn{ \hspace*{-5mm}
 \left(2^{D/2}N_cN_f\Lambda^D\NDA\right)^{-1}V(\sigma)  
} \nonumber \\[3mm]
&& \hspace*{-6mm}
= -\frac{1}{g}\frac{m_0 \sigma}{\Lambda^2}+\frac{1}{2g}
  \dfrac{\sigma^2}{\Lambda^2} 
-\frac{1}{\Lambda^2}
 \int d\sigma \, \int_0^1 dz
  \frac{z^{\nu}\,\sigma}{z+\frac{\sigma^2}{\Lambda^2}} , \nonumber \\[3mm]
&& \hspace*{-6mm}
= -\frac{1}{g}\frac{m_0 \sigma}{\Lambda^2}
  +\frac{1}{2}\left(\frac{1}{g}-\frac{1}{\nu}\right)
  \dfrac{\sigma^2}{\Lambda^2} 
  + \frac{1}{4(\nu-1)} \dfrac{\sigma^4}{\Lambda^4} \nonumber \\[3mm]
&& \hspace*{-6mm} \qquad 
+ {\cal O}\left(\frac{\sigma^6}{\Lambda^6}\right), \qquad (\kappa_D=0) .
 \label{pot_NJL}
\end{eqnarray}
Since the sub-dominant term of $\Sigma(\Lambda^2)$ with nonvanishing 
gauge coupling depends on the approximations, 
we calculate $V(\sigma)$ separately.

\subsection{Linearized approximation}

In the calculation of  the effective potential $V(\sigma)$, we need to 
express $\Sigma(\Lambda^2)$ in term of $\sigma$.
Let us start with the solution of the linearized SD equation for
$\kappa_D > \kappa_D^{\rm MT}$.
Combining Eq.~(\ref{eq:linearized_sigma1}) with
\begin{equation}
  \Sigma(\Lambda^2) 
  = c_1 \Sigma_0
    \left(\dfrac{\Sigma_0}{\Lambda}\right)^{\nu(1-\omega)}
  + d_1 \Sigma_0
    \left(\dfrac{\Sigma_0}{\Lambda}\right)^{\nu(1+\omega)} + \cdots,
\end{equation}
we obtain
\begin{equation}
  \Sigma(\Lambda^2) = \frac{2}{1+\omega}\sigma
  + \frac{2\omega}{1+\omega} d_1 \Lambda \left(
       \dfrac{2c_1^{-1}}{1+\omega}
       \frac{\sigma}{\Lambda}
    \right)^{\frac{\nu(1+\omega)+1}{\nu(1-\omega)+1}} + \cdots.
\label{eq:Sigma_at_Lambda1}
\end{equation}
We are now able to calculate the effective potential from
Eq.~(\ref{eq:localpot1}), Eq.~(\ref{eq:expressionJ}) and
Eq.~(\ref{eq:Sigma_at_Lambda1}),
\begin{eqnarray}
\lefteqn{ \hspace*{-3mm}
 \left(2^{D/2}N_cN_f\Lambda^D\NDA\right)^{-1}V(\sigma) = 
} \nonumber \\[3mm]
&&
 -\frac{1}{g}\frac{m_0 \sigma}{\Lambda^2}+\frac{1}{2}
  \left(\frac{1}{g}-\frac{1}{g^{\rm crit}}\right)
  \dfrac{\sigma^2}{\Lambda^2}
\nonumber\\
& & \qquad\qquad\qquad
 +A_1 \left(\frac{\sigma}{\Lambda}
       \right)^{2+\frac{2\nu\omega}{\nu(1-\omega)+1}} + \cdots,
\label{eq:linearized_pot1}
\end{eqnarray}
for $\kappa_D^{\rm MT} < \kappa_D  < \kappa_D^{\rm crit}$.
Here the coefficient $A_1$ is defined as
\begin{equation}
A_1 \equiv  -\frac{4}{1-\omega^2}
             \frac{\omega}{1+\omega}\frac{\nu(1-\omega)+1}{\nu (\nu+1)}\,
         d_1 \left(\frac{2c_1^{-1}}{1+\omega}
              \right)^{1+\frac{2\nu\omega}{\nu(1-\omega)+1}}. 
\end{equation}

It is straightforward to perform similar analysis for 
$0<\kappa_D <\kappa_D^{\rm MT}$.
We obtain
\begin{eqnarray}
\lefteqn{\hspace*{-5mm}
 \left(2^{D/2}N_cN_f\Lambda^D\NDA\right)^{-1}V(\sigma) = 
} \nonumber \\[3mm]
&&
 -\frac{1}{g}\frac{m_0 \sigma}{\Lambda^2}+\frac{1}{2}
  \left(\frac{1}{g}-\frac{1}{g^{\rm crit}}\right)
  \dfrac{\sigma^2}{\Lambda^2}
\nonumber\\
&&\qquad\qquad\quad
 +A_2
 \left(\frac{\sigma}{\Lambda}\right)^{2+\frac{2}{\nu(1-\omega)+1}}
 +\cdots,
\label{eq:linearized_pot2}
\end{eqnarray}
for $0<\kappa_D<\kappa_D^{\rm MT}$.
The coefficient $A_2$ is defined as
\begin{equation}
  A_2 \equiv \frac{1}{2(\nu\omega-1)} 
              \frac{\nu(1-\omega)+1}{\nu(1-\omega)+2} \, c_1^2  
              \left(\frac{2c_1^{-1}}{1+\omega}
               \right)^{2+\frac{2}{\nu(1-\omega)+1}}.
\end{equation}

The calculation at $\kappa_D=\kappa_D^{\rm MT}$ is a little bit
involved, since we need to take account of both $d_1$ and $c_2$.
By using formulas given in Appendix~\ref{sec-appendix2}, we obtain
\begin{eqnarray}
\lefteqn{\hspace*{-5mm}
 \left(2^{D/2}N_cN_f\Lambda^D\NDA\right)^{-1}V(\sigma) = 
} \nonumber \\[3mm]
&&
 -\frac{1}{g}\frac{m_0 \sigma}{\Lambda^2}+\frac{1}{2}
  \left(\frac{1}{g}-\frac{1}{g^{\rm crit}}\right)
  \dfrac{\sigma^2}{\Lambda^2}
\nonumber\\
&&\qquad\qquad\quad
 +A'_2 \left(\frac{\sigma}{\Lambda}
       \right)^{2+\frac{2}{\nu}} \ln
 \left(\frac{\Lambda}{\sigma}\right) 
 +\cdots,
\label{eq:linearized_pot3}
\end{eqnarray}
for $\kappa_D=\kappa_D^{\rm MT}$ and
the coefficient $A'_2$ is given by
\begin{equation}
A'_2 \equiv  \frac{1}{\nu+1} \,c_1^2 
              \left(\frac{2\nu c_1^{-1}}{\nu+1}
              \right)^{2+\frac{2}{\nu}} .
\end{equation}

Using Eqs.~(\ref{eq:c_n}) and (\ref{eq:d_n}), we find the coefficients 
$A_1$, $A_2$ and $A'_2$ are all positive definite in the linearized SD
equation.
The potential is therefore stabilized for large value of $\sigma$ even 
if we truncate the potential and neglect $+\cdots$ terms in
Eqs.~(\ref{eq:linearized_pot1}),  (\ref{eq:linearized_pot2}) and
(\ref{eq:linearized_pot3}).

\subsection{Power expansion method}

The effective potential $V(\sigma)$ can also be calculated in the
power expansion method.
We should keep it in mind, however, that the coefficients $c_1$ and
$d_1$ cannot be determined with the power expansion method, unlike the
calculation in the linearized SD equation. 

We here summarize our results of the effective potential in the power
expansion method:
\begin{eqnarray}
\lefteqn{
 \left(2^{D/2}N_cN_f\Lambda^D\NDA\right)^{-1}V(\sigma) = 
} \nonumber \\[3mm]
&&
 -\frac{1}{g}\frac{m_0 \sigma}{\Lambda^2}+\frac{1}{2}
  \left(\frac{1}{g}-\frac{1}{g^{\rm crit}}\right)
  \dfrac{\sigma^2}{\Lambda^2}
\nonumber\\
& & \qquad\qquad\qquad
 +\tilde A_1 \left(\frac{\sigma}{\Lambda}
       \right)^{2+\frac{2\nu\omega}{\nu(1-\omega)+1}} + \cdots, 
\label{eff-pot-pow1}
\end{eqnarray}
for $\kappa_D^{\rm PE} < \kappa_D < \kappa_D^{\rm crit}$, and
\begin{eqnarray}
\lefteqn{
 \left(2^{D/2}N_cN_f\Lambda^D\NDA\right)^{-1}V(\sigma) = 
} \nonumber \\[3mm]
&&
 -\frac{1}{g}\frac{m_0 \sigma}{\Lambda^2}+\frac{1}{2}
  \left(\frac{1}{g}-\frac{1}{g^{\rm crit}}\right)
  \dfrac{\sigma^2}{\Lambda^2}
\nonumber\\
& & \qquad\qquad\qquad
 +\tilde A_2 \, \frac{\sigma^4}{\Lambda^4}+\cdots, \label{eff_pot_pw2}
\end{eqnarray}
for 
$0< \kappa_D < \kappa_D^{\rm PE}$, and
\begin{eqnarray}
\lefteqn{
 \left(2^{D/2}N_cN_f\Lambda^D\NDA\right)^{-1}V(\sigma) = 
} \nonumber \\[3mm]
&&
 -\frac{1}{g}\frac{m_0 \sigma}{\Lambda^2}+\frac{1}{2}
  \left(\frac{1}{g}-\frac{1}{g^{\rm crit}}\right)
  \dfrac{\sigma^2}{\Lambda^2}
\nonumber\\
& & \qquad\qquad\qquad
 +\tilde A'_2 \, \frac{\sigma^4}{\Lambda^4}
   \ln \left(\frac{\Lambda}{\sigma}\right)+\cdots, \label{eff_pot_pw3}
\end{eqnarray}
for $\kappa_D = \kappa_D^{\rm PE}$.
The coefficients $\tilde{A}_1$, $\tilde{A}_2$ and 
$\tilde{A}'_2$ are defined by
\begin{eqnarray}
\tilde 
A_1 &\equiv&  -\frac{4}{1-\omega^2}
             \frac{\omega}{1+\omega}\frac{\nu(1-\omega)+1}{\nu(\nu+1)}\,
         d_1 \left(\frac{2c_1^{-1}}{1+\omega}
              \right)^{\frac{\nu(1+\omega)+1}{\nu(1-\omega)+1}} 
\!\!\! ,
\nonumber\\
& &\\  
\tilde
A_2 &\equiv& \frac{1}{4\,(2\nu\omega-\nu-1)}\,
              \left(\frac{2}{1+\omega}\right)^4 , \\
A'_2 &\equiv& \frac{1}{\nu+1}\,
              \left(\frac{4\nu}{3\nu+1}\right)^4 ,
\end{eqnarray}
respectively.

We emphasize here, due to the lack of our knowledge of $c_1$ and
$d_1$, the coefficient $\tilde A_1$ cannot be calculated in the power
expansion method.
On the other hand, we find $\tilde A_2 > 0$ for
$0<\kappa_D<\kappa_D^{\rm PE}$, independently of $c_1$. 

We comment that 
the exponent of the lowest interaction term ($\tilde A_1$ term)
in Eq.~(\ref{eq:linearized_pot1}) 
is the same as that of $A_1$ in Eq.~(\ref{eff-pot-pow1}).
In passing, the bifurcation solution also leads to the same exponent of 
the lowest interaction term.
Thus the lowest interaction term in the effective potential 
is insensitive to the approximations of the SD equation
for $\kappa_D^{\rm MT} < \kappa_D < \kappa_D^{\rm crit}$.
Note that $\kappa_D^{\rm PE} < \kappa_D^{\rm MT}$.
On the other hand, for $0  < \kappa_D \leq \kappa_D^{\rm MT}$,
the exponent of the lowest interaction term is different depending on
the approximations.

\section{Nontrivial window}
\label{sec-RG}

In Sec.~\ref{sec3} we found ``renormalizability'' or ``nontriviality'' 
in the sense that 
the decay constant $F_\pi$ 
and the renormalized Yukawa coupling $\Gamma_s^{(R)}$
become finite in the continuum 
limit for $\kappa_D^{\rm MT} < \kappa_D < \kappa_D^{\rm crit}$.

Here we further study the renormalization and 
discuss the renormalization group (RG) flow, based on 
the effective potential derived in Sec.~\ref{eff_pot}. 
The renormalization of the effective potential is performed
in a way similar to the four-dimensional gauged NJL model~\cite{Kondo:1992sq}.
This method is applicable not only in the broken phase, but also in
the symmetric phase.
Note that we already renormalized the $\sigma$ field in Eq.~(\ref{eq:Z_wav}) 
so that its kinetic term is kept finite.

For $\kappa_D^{\rm MT} < \kappa_D < \kappa_D^{\rm crit}$
we require the bare parameters ($m_0$ and $g$) in the effective
potential to depend on the cutoff ($m_0(\Lambda)$ and $g(\Lambda)$)
such that
\begin{equation}
  \dfrac{m_0(\Lambda)}{g(\Lambda)} \Lambda^{\nu(1+\omega)} = \mbox{const.} ,
\end{equation}
and 
\begin{equation}
   \left(\,\dfrac{1}{g(\Lambda)}-\dfrac{1}{g^{\rm crit}}\,\right) 
  \,\Lambda^{2\nu\omega}
   = \mbox{const.} . 
\end{equation}
The effective potential is thus kept finite even in 
the continuum limit $\Lambda \to \infty$ 
(Wilsonian renormalization).
We define the renormalized parameters $m_R, g_R$ as
\begin{equation}
  \dfrac{m_R(\mu)}{g_R(\mu)} \mu^{\nu(1+\omega)}=
  \dfrac{m_0(\Lambda)}{g(\Lambda)} \Lambda^{\nu(1+\omega)},
 \label{ren_m1} 
\end{equation}
and
\begin{equation}
   \left(\,\dfrac{1}{g_R(\mu)}-\dfrac{1}{g^{\rm crit}}\,\right)
  \,\mu^{2\nu\omega} 
 =
   \left(\,\dfrac{1}{g(\Lambda)}-\dfrac{1}{g^{\rm crit}}\,\right) 
  \,\Lambda^{2\nu\omega} ,
   \label{ren_g1}
\end{equation}
with $\mu$ being the renormalization scale.
The effective potential Eq.~(\ref{eq:linearized_pot1}) 
(linearizing approximation) is thus renormalized as
\begin{eqnarray}
\lefteqn{
 \left(2^{D/2}N_cN_f\mu^D\NDA\right)^{-1}V_R(\sigma_R) = 
} \nonumber \\[3mm]
&&
 -\frac{1}{g_R}\frac{m_R \sigma_R}{\mu^2}+\frac{1}{2}
  \left(\frac{1}{g_R}-\frac{1}{g^{\rm crit}}\right)
  \dfrac{\sigma_R^2}{\mu^2}
\nonumber\\
& & \qquad\qquad\qquad\quad
 +A_1 \left(\frac{\sigma_R}{\mu}
       \right)^{2+\frac{2\nu\omega}{\nu(1-\omega)+1}},
\label{eq:renormalized_pot1}
\end{eqnarray}
for $\kappa_D^{\rm MT} < \kappa_D  < \kappa_D^{\rm crit}$,
and similarly is Eq.~(\ref{eff-pot-pow1}) (power expansion method).
Note that the ``$+\cdots$'' terms in Eqs.~(\ref{eq:linearized_pot1}) 
and (\ref{eff-pot-pow1}) are
decoupled in the $\Lambda\rightarrow\infty$ limit.
The $\sigma$ field has non-trivial self-interaction in 
the renormalized effective potential. 
Recall that $A_1 > 0$.
The potential is therefore stabilized
and the VEV of $\sigma_R$ remains finite 
for $g_R > g^{\rm crit}$.
We emphasize that the form of the renormalized effective potential 
$V_R(\sigma_R)$ does not depend on the approximations,
although the coefficient of the interaction term does.

From Eqs.~(\ref{ren_m1}) and (\ref{ren_g1})
we obtain the beta function of the dimensionless four-fermion coupling $g_R$,
\begin{equation}
  \beta (g_R) = 
   2\nu\omega g_R \left( 1 - \dfrac{g_R}{g^{\rm crit}} \right),
  \label{beta-g1}
\end{equation}
and the anomalous dimension of the fermion mass $\gamma_m$,
\begin{eqnarray}
  \gamma_m (g_R) &=& 
  -\dfrac{\beta(g_R)}{g_R} + \nu ( 1 + \omega) , \nonumber \\
 &=&
  \nu \left[\, 1 - \omega + 2\omega \dfrac{g_R}{g^{\rm crit}}\,\right] ,
 \label{beta-m1}
\end{eqnarray}
for $\kappa_D^{\rm MT} < \kappa_D < \kappa_D^{\rm crit}$. 
They 
take the same form as those in the four-dimensional gauged NJL 
model with fixed gauge coupling~\cite{Kondo:1992sq} 
up to the factor $\nu=D/2-1\to 1$ ($D\to 4$).
We note the beta function of the NJL coupling possesses a UVFP 
$g_R=g^{\rm crit}\,(=\nu(1+\omega)^2/4)$. 
At the fixed point of $g_R$ the anomalous dimension $\gamma_m$
reads 
\begin{equation}
\gamma_m = \nu(1+\omega). \label{ano-FP1}
\end{equation}   

Let us turn to the region $0 < \kappa_D < \kappa_D^{\rm MT}$.
In this region the decay constant $F_\pi$ diverges as
we mentioned.
It is thus expected that the same renormalization procedure
as in the region $\kappa_D^{\rm MT} < \kappa_D < \kappa_D^{\rm crit}$
will break down.
In fact, if we formally performed such a renormalization,
we would get~\footnote{
At $\kappa_D = \kappa_D^{\rm MT}$ the renormalized effective potential
would take the same form as Eq.~(\ref{eq:renormalized_pot2}). Although
the definition of $g_R$ and $m_R$ in Eqs.~(\ref{ren_m2}) and (\ref{ren_g2})
should be modified by logarithmic factors,
the beta function and the anomalous dimension are unchanged from
Eqs.~(\ref{beta-g2}) and (\ref{beta-m2}).
If one took the limit $\kappa_D \searrow \kappa_D^{\rm MT}$ of the
expression in Eq.~(\ref{eq:renormalized_pot1}),
one would find the pole in the coefficient $A_1$. 
This pole actually is logarithmic divergence $\ln \Lambda$ when
the sub-sub-leading term is properly incorporated in the limit
$\Lambda \to \infty$ as in the pure NJL limit of 
the four-dimensional gauged NJL model~\cite{Kondo:1992sq}.
After the renormalization of the $\sigma$ field Eq.~(\ref{eq:Z_wav}),
we find the interaction term vanishes in $V_R(\sigma_R)$, 
consistently with Eq.~(\ref{eq:renormalized_pot2}).} 
\begin{eqnarray}
\lefteqn{
 \left(2^{D/2}N_cN_f\mu^D\NDA\right)^{-1}V_R(\sigma_R) = 
} \nonumber \\[3mm]
&&
 -\frac{1}{g_R}\frac{m_R \sigma_R}{\mu^2}+\frac{1}{2}
  \left(\frac{1}{g_R}-\frac{1}{g^{\rm crit}}\right)
  \dfrac{\sigma_R^2}{\mu^2},
\label{eq:renormalized_pot2}
\end{eqnarray}
where we defined $m_R$ and $g_R$ as
\begin{equation}
  \dfrac{m_R (\mu)}{g_R (\mu)}\,\mu^{\nu+1} =
  \dfrac{m_0 (\Lambda)}{g (\Lambda)}\,\Lambda^{\nu+1}, 
  \label{ren_m2} 
\end{equation}
and
\begin{equation}
  \left(\,\dfrac{1}{g_R(\mu)}-\dfrac{1}{g^{\rm crit}}\,\right) \mu^2
 = \left(\,\dfrac{1}{g(\Lambda)}-\dfrac{1}{g^{\rm crit}}\,\right)\Lambda^2.
  \label{ren_g2}
\end{equation}
This renormalization would imply the beta function 
and the anomalous dimension:
\begin{equation}
  \beta (g_R) =
  2 g_R \left( 1 - \dfrac{g_R}{g^{\rm crit}} \right),
  \label{beta-g2}
\end{equation}
and
\begin{eqnarray}
  \gamma_m (g_R) &=&
 -\dfrac{\beta(g_R)}{g_R} + \nu + 1 \nonumber \\
&=& \nu - 1 + 2\dfrac{g_R}{g^{\rm crit}}.   \label{beta-m2}
\end{eqnarray}
The anomalous dimension $\gamma_m$ would read 
\begin{equation}
\gamma_m = \nu+1 \label{ano-FP2}
\end{equation}   
at the UVFP of $g_R$.

However, the self-interaction term in $V_R(\sigma_R)$ 
Eq.~(\ref{eq:renormalized_pot2}) disappears at $\Lambda \to \infty$
in our renormalization procedure both in the linearized approximation
and in the power expansion method.
Thus the renormalized potential is not stabilized
for $g_R > g^{\rm crit}$ and hence the renormalization breaks down.
The Yukawa interaction does also vanish and hence
interactions of $\sigma$ and $\pi$ are all trivial.

Let us next consider the dynamical dimension of composite operators.
Within the ladder approximation adopted in this paper, 
various composite operators have the dynamical dimensions,
\begin{eqnarray}
\mbox{dim}((\bar\psi\psi)^2) &=& 2\,\mbox{dim}(\bar\psi\psi), \\
\mbox{dim}((\bar\psi\psi)^4) &=& 4\,\mbox{dim}(\bar\psi\psi), \\
\mbox{dim}(\partial_M (\bar\psi\psi)\partial^M (\bar\psi\psi)) &=& 
2\,\mbox{dim}(\bar\psi\psi)+2,
\end{eqnarray}
with 
\begin{equation}
  \mbox{dim}(\bar\psi\psi) = D-1-\gamma_m . 
\end{equation}
Using Eqs.~(\ref{ano-FP1}) and (\ref{ano-FP2}), we find 
\begin{equation}
\mbox{dim}((\bar\psi\psi)^2) =
\left\{\begin{array}{l@{\qquad}c}
D-2\nu\omega, & (\kappa_D^{\rm MT} < \kappa_D < \kappa_D^{\rm crit}), \\[3mm]
D-2, & (0 \leq \kappa_D \leq \kappa_D^{\rm MT})  ,
\end{array}\right.  
\end{equation}   
\begin{equation}
\mbox{dim}((\bar\psi\psi)^4) =
\left\{\begin{array}{l@{\qquad}c}
2D-4\nu\omega, & (\kappa_D^{\rm MT} < \kappa_D < \kappa_D^{\rm crit}), \\[3mm]
2D-4, & (0 \leq \kappa_D \leq \kappa_D^{\rm MT})  ,
\end{array}\right.  
\end{equation}   
\begin{eqnarray}
\lefteqn{ \hspace*{-0.5cm}
\mbox{dim}(\partial_M (\bar\psi\psi)\partial^M (\bar\psi\psi)) = 
} \nonumber \\[2mm] &&
\left\{\begin{array}{l@{\qquad}c}
D+2-2\nu\omega, & (\kappa_D^{\rm MT} < \kappa_D < \kappa_D^{\rm crit}), \\[3mm]
D, & (0 \leq \kappa_D \leq \kappa_D^{\rm MT})  .
\end{array}\right. \nonumber \\
\end{eqnarray}
The four-fermion operator $(\bar{\psi}\psi)^2$ is relevant in 
the whole region $0 \leq \kappa_D < \kappa_D^{\rm crit}$, which justifies
the inclusion of the four-fermion interaction in the study of 
the phase structure of walking gauge theories.

For the region $ \kappa_D^{\rm MT} < \kappa_D < \kappa_D^{\rm crit}$
both operators $(\bar\psi\psi)^4$ and 
$\partial_M (\bar\psi\psi)\partial^M (\bar\psi\psi)$ are irrelevant,
i.e.,
dim($(\bar\psi\psi)^4$) $> D$ and 
dim($(\partial_M (\bar\psi\psi)\partial^M (\bar\psi\psi)$) $> D$.
This is consistent with our renormalization procedure 
without introducing such operators.

On the other hand, for $0 \leq \kappa_D \leq \kappa_D^{\rm MT}$,
the kinetic operator $\partial_M (\bar\psi\psi)\partial^M (\bar\psi\psi)$
becomes marginal, 
dim($(\partial_M (\bar\psi\psi)\partial^M (\bar\psi\psi)$) $= D$,
while the eight-fermion operator $(\bar\psi\psi)^4$ is irrelevant,
dim($(\bar\psi\psi)^4$) $> D$. 
This is another symptom of breakdown of our renormalization in this region.
We may need to incorporate the marginal operator 
$\partial_M (\bar\psi\psi)\partial^M (\bar\psi\psi)$
as a counter term, which is similar to the situation in the four-dimensional
pure NJL model where both  $\partial_\mu (\bar\psi\psi)\partial^\mu (\bar\psi\psi)$
and $(\bar\psi\psi)^4$ are marginal~\cite{Kondo:1992sq}.

\begin{figure}[tbp]
  \begin{center}
    \resizebox{0.45\textwidth}{!}{
     \includegraphics{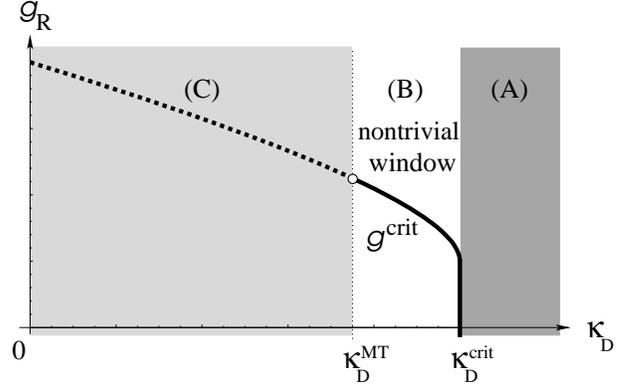}}
    \caption{Phase diagram in $(\kappa_D, g_R)$ plane. 
             The critical line $g^{\rm crit}$ is the same as that in
             Fig.~\ref{fig:critical_line}. Each point on this line 
             describes a different theory having different $N_c, N_f$.
             In the region (A) $\kappa_D > \kappa_D^{\rm crit}$ 
             the dynamical mass diverges in the continuum limit, 
             i.e., no finite theory exists.
             In the region (B)
             $\kappa_D^{\rm MT} < \kappa_D < \kappa_D^{\rm crit}$ 
             (nontrivial window),
             the dynamical mass can be made finite and 
             the self-interaction of $\sigma_R$ and
             the Yukawa interaction remain nontrivial in the continuum limit.
             In the region (C) $0 \leq \kappa_D \leq \kappa_D^{\rm MT}$ 
             the renormalized effective potential is not stabilized
             for the broken phase $g_R > g^{\rm crit}$ and hence 
             the renormalization breaks down.
             In this region the self-interaction of $\sigma_R$ and
             the Yukawa interaction vanish and are trivial. 
             \label{phase-R}}
  \end{center}
\end{figure}

We comment on further symptom of breakdown of our renormalization procedure
for $0 \leq \kappa_D \leq \kappa_D^{\rm MT}$.
If we renormalized consistently the effective action in this region,
we would expect the renormalization of $\bar\psi\psi$ as
\begin{equation}
  m_R \VEV{(\bar\psi\psi)_R} = m_0 \VEV{(\bar\psi\psi)_0}.
  \label{psi-R}
\end{equation}
Note that $m_0 \sim \Lambda^{-\gamma_m}=\Lambda^{-\nu-1}$,
while Eq.~(\ref{chi-cond}) yields
\begin{equation}
  \VEV{(\bar{\psi}\psi)_0} \propto
  \int_0^{\Lambda^2} dx x^\nu \frac{\Sigma(x)}{x+\Sigma^2}
  \sim \Lambda^{\nu (1+\omega)},
\label{chi-cond2}
\end{equation}
where we used the asymptotic behavior of the dominant solution 
in Eq.~(\ref{eq:asymp_sol}).
Thus Eq.~(\ref{psi-R}) diverges as 
$m_0 \VEV{(\bar{\psi}\psi)_0} \sim \Lambda^{\nu\omega-1}$.
This can be understood by noting $(\bar\psi\psi)_R \propto \sigma_R$,
the VEV of which is divergent due to the lack of the interaction term
as stabilizer in $V_R(\sigma_R)$ Eq.~(\ref{eq:renormalized_pot2}).
This reflects necessity of other operators such as 
the kinetic operator $\partial_M (\bar\psi\psi)\partial^M (\bar\psi\psi)$
as counter terms. 
See Appendix~\ref{app-D} for further details.
%
%
\begin{table}[tb]
  \begin{center}
    \resizebox{0.45\textwidth}{!}{
     \includegraphics{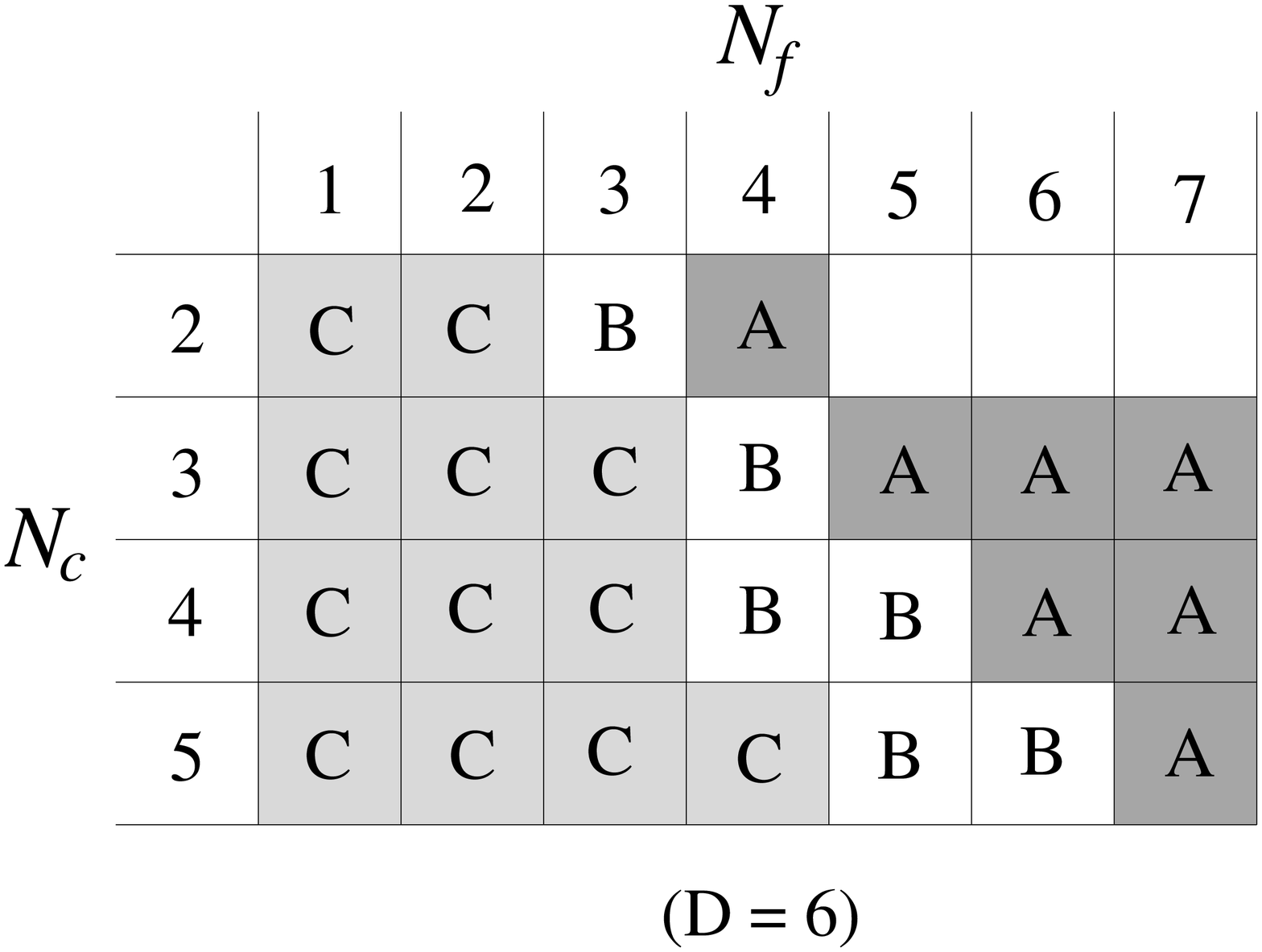}}
    \caption{Classification of models ($D=6$).
          The entries labeled by A, B and C stand for the models in
          the regions (A) $\kappa_D > \kappa_D^{\rm crit}$, 
          (B) $\kappa_D^{\rm MT} < \kappa_D < \kappa_D^{\rm crit}$
          and (C) $0 \leq \kappa_D \leq \kappa_D^{\rm MT}$, respectively. 
          No entry denotes absence of the UVFP of the gauge coupling $g_*$.
          \label{tab1}}
  \end{center}
\end{table}
%
%
\begin{table}[tb]
  \begin{center}
    \resizebox{0.45\textwidth}{!}{
     \includegraphics{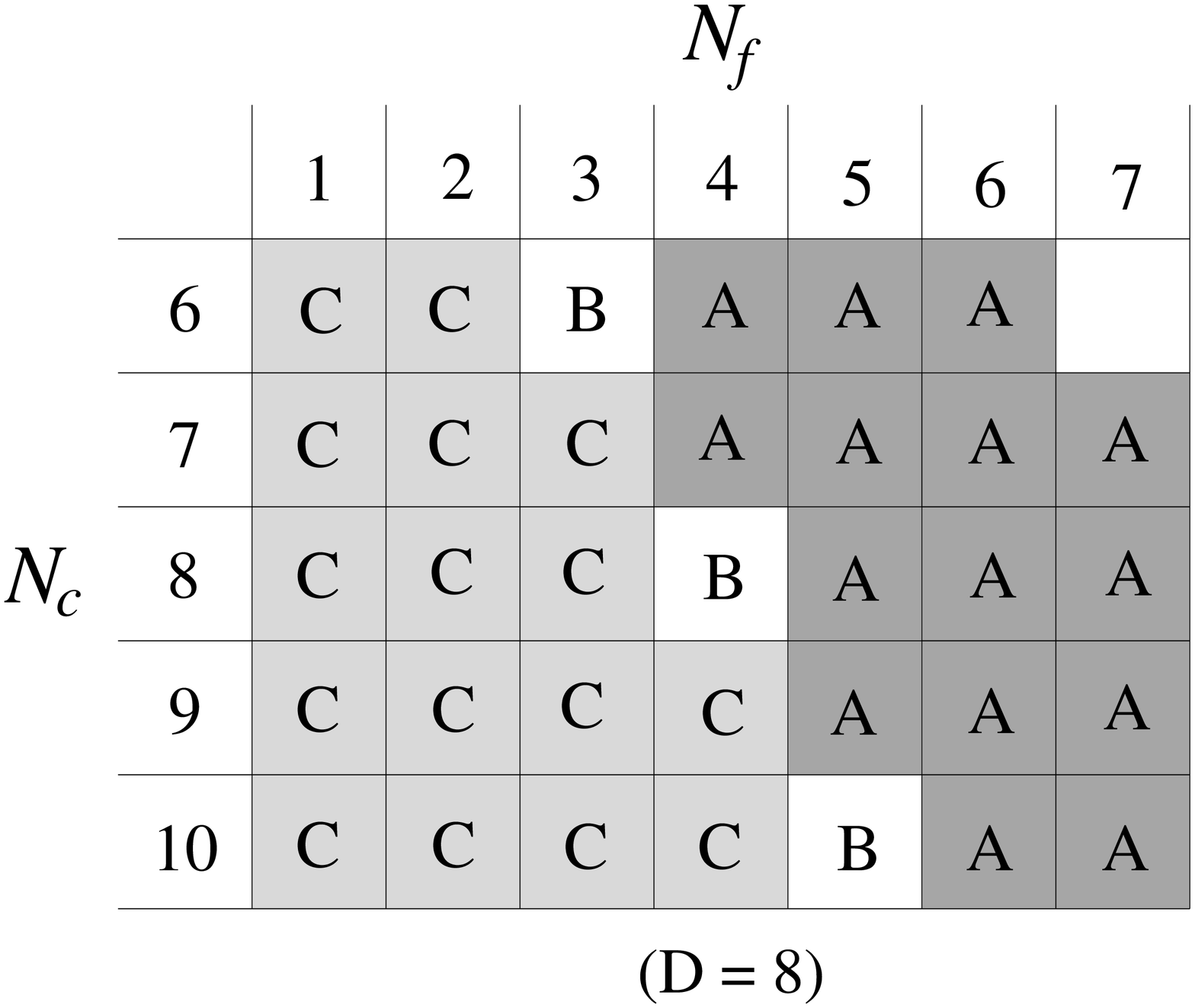}}
    \caption{Classification of models ($D=8$).
          The entries labeled by A, B and C stand for the models in
          the regions (A) $\kappa_D > \kappa_D^{\rm crit}$, 
          (B) $\kappa_D^{\rm MT} < \kappa_D < \kappa_D^{\rm crit}$
          and (C) $0 \leq \kappa_D \leq \kappa_D^{\rm MT}$, respectively. 
          No entry denotes absence of the UVFP of the gauge coupling $g_*$.
          For $2 \leq N_c \leq 5$ the nontrivial window is not found.
          \label{tab2}}
  \end{center}
\end{table}
%
%
\begin{table}[tb]
  \begin{center}
    \resizebox{0.45\textwidth}{!}{
     \includegraphics{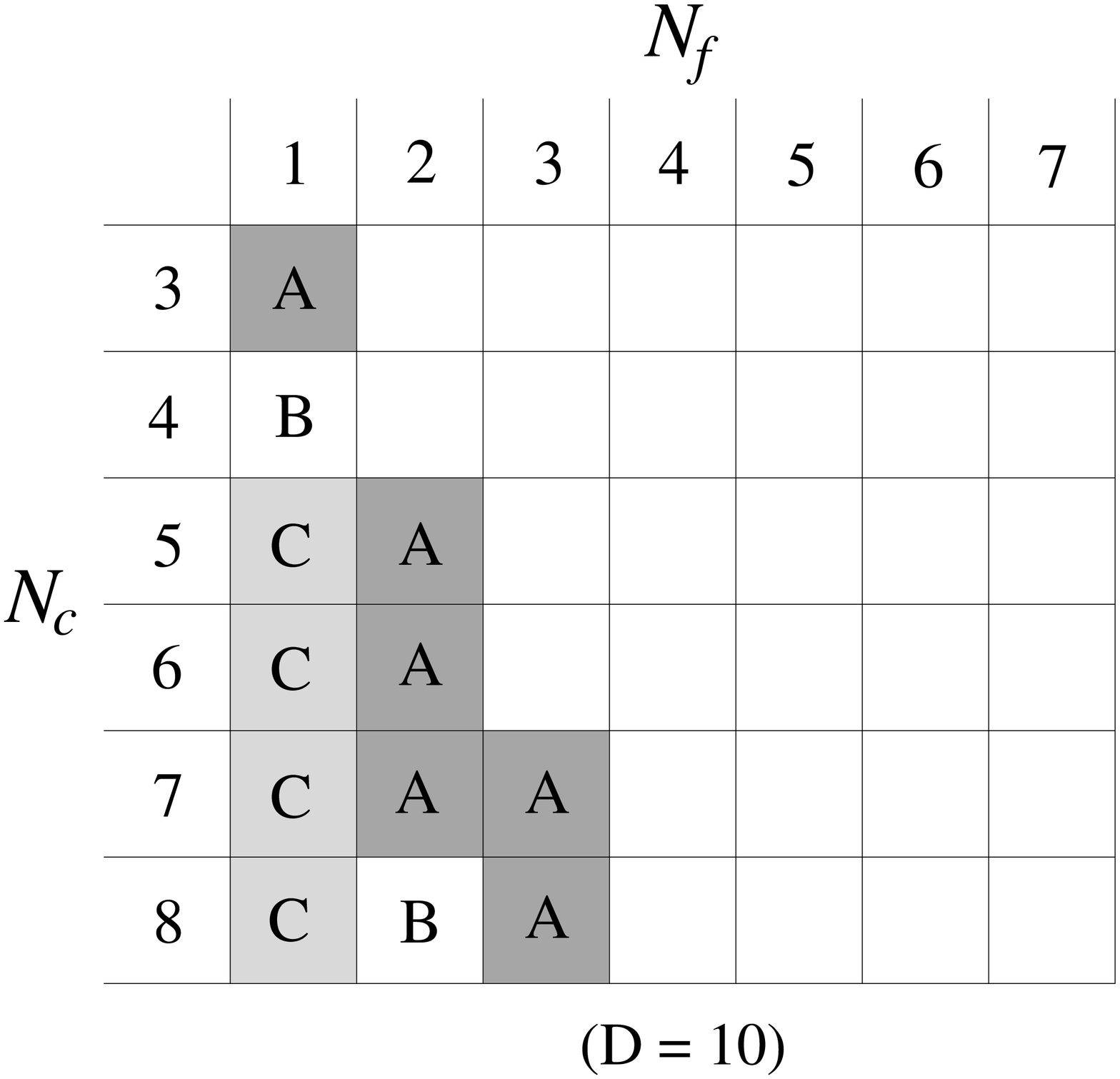}}
    \caption{Classification of models ($D=10$).
          The entries labeled by A, B and C stand for the models in
          the regions (A) $\kappa_D > \kappa_D^{\rm crit}$, 
          (B) $\kappa_D^{\rm MT} < \kappa_D < \kappa_D^{\rm crit}$
          and (C) $0 \leq \kappa_D \leq \kappa_D^{\rm MT}$, respectively. 
          No entry denotes absence of the UVFP of the gauge coupling $g_*$.
          For $N_c=2$ there is no UVFP.
          \label{tab3}}
  \end{center}
\end{table}

We depict our results mentioned above in Fig.~\ref{phase-R} and 
Tables~\ref{tab1}--\ref{tab3}~\footnote{
There could be a possible ambiguity of the estimate of the
concrete number of $N_f,N_c$ 
of the nontrivial window: It could arise from the ambiguity of the estimate of 
$\kappa_D^{\rm crit}$ in the improved ladder SD equation with
different momentum identification for the running 
gauge coupling~\cite{Hashimoto:2000uk,Gusynin:2002cu}.
Although, as we mentioned in Sect.\ref{sec3}, our momentum 
identification Eq.~(\ref{imp_ladder}) with nonlocal gauge fixing
has much advantage over 
the simplest one~\cite{imp-simplest}, 
$ g_{(4+\delta)D}^2(\mu) \to 
  g_{(4+\delta)D}^2({\rm max}(x,y))$, with Landau
gauge fixing, one might use the latter. Then one could get
somewhat smaller values:\cite{Hashimoto:2000uk}
$\kappa_6^{\rm crit}=0.122$, $\kappa_8^{\rm crit}=0.146$, and
$\kappa_{10}^{\rm crit}=0.163$,
compared with those in the
present paper, $\kappa_D^{\rm crit} = D(D-2)/[32(D-1)]$.
This would shift the nontrivial window for a fixed $N_c$ 
to a somewhat smaller $N_f$
region. For example, the model with $D=6,N_c=3,N_f=3$ would enter 
the nontrivial window.
}:
\begin{itemize}
\item[(A)] The region $\kappa_D > \kappa_D^{\rm crit}$ -- 
{\it no finite continuum theory.} 

As we discussed in Sec.~\ref{sec3}, the gauge coupling strength 
on the UVFP $\kappa_D$ is not a continuous parameter,
so that $\kappa_D$ cannot be fine-tuned arbitrarily close to 
$\kappa_D^{\rm crit}$ to make the dynamical mass finite in the
continuum limit ($\Lambda \to \infty$).
For example, models with $D=6,N_c=3,N_f=5,6,7$ fall into 
this category. 

\item[(B)] The region
$\kappa_D^{\rm MT} < \kappa_D < \kappa_D^{\rm crit}$ -- 
{\it nontrivial window.} 

The dynamical mass can be made finite by fine-tuning of
the four-fermion coupling. 
Once the dynamical mass is made finite,
the decay constant of the NG boson is also finite
and so is the Yukawa coupling of the NG boson.
The self-interaction of $\sigma_R$ and
the Yukawa interaction remain nontrivial.
We also note that the self-interaction of $\sigma_R$ is conformal symmetric.
The renormalized four-fermion coupling $g_R$ has a UVFP
$g_R(\infty)=g^{\rm crit}$ and the anomalous dimension $\gamma_m$ is 
very large, $\nu < \gamma_m=\nu(1+\omega) < \nu+1$.
For example, models with $D=6,N_c=3,N_f=4$ fall into 
this category. 
The nontrivial window  
gets closed, $\kappa_D^{\rm MT} \to \kappa_D^{\rm crit}$, when $D\to \infty$.

\item[(C)] The region
$0 \leq \kappa_D \leq \kappa_D^{\rm MT}$ -- 
{\it trivial or nonrenormalizable.}

Even if we made the dynamical mass finite by fine-tuning
the four-fermion coupling,
the decay constant does diverge in the continuum limit. 
The self-interaction of $\sigma_R$ and
the Yukawa interaction vanish and are trivial.
The renormalized effective potential is not stabilized
for $g_R > g^{\rm crit}$.
Without introducing additional operators other than the four-fermion
operator, the theory cannot be renormalized.
For example, models with $D=6,N_c=3,N_f=1,2,3$ fall into 
this category.
Note that 
$\kappa_D^{\rm MT}$ is reduced to the pure NJL point in the
four-dimensional NJL model,
$\kappa_D^{\rm MT} \to 0$, in the limit $D \to 4$.
\end{itemize}

So far we focused our discussions on the theory with the gauge coupling 
exactly on the UVFP $\kappa_D=C_F g_*^2 \NDA$.
Each point on the critical line $g^{\rm crit}$ describes a different 
theory (different $N_c, N_f$).
Now that we have shown existence of the nontrivial window,
we shall study the RG flows of $(\hat g^2(\mu), \; g_R(\mu))$
of a particular theory specified by one point on the critical line.
In order to obtain the RG flows in the vicinity of the UVFP 
$(\hat g^2(\infty)=g_*^2, \; g_R(\infty)=g^{\rm crit})$, 
we define variations $\delta \hat g^2$ and $\delta g_R$ as
\begin{eqnarray}
  \delta \hat g^2(\mu) & \equiv & \hat g^2(\mu) - g_*^2, \\
  \delta g_R(\mu) & \equiv & g_R(\mu) - g^{\rm crit}.
\end{eqnarray}
It is enough to study linearized RGEs~\cite{Wilson:1973jj}
\begin{equation}
 \dfrac{\partial}{\partial \ln \mu}
\left(
 \begin{array}{@{}c@{}}
  \delta \hat g^2 \\[2mm]
  \delta g_R
 \end{array}
 \right)
 = {\cal M}
 \left(
 \begin{array}{@{}c@{}}
  \delta \hat g^2 \\[2mm] \delta g_R
 \end{array}
 \right),
\end{equation}
with
\begin{equation}
{\cal M} = \left( \begin{array}{cc}
 - 2(\nu-1) & 0 \\ \Delta & -2\nu\omega
 \end{array}
 \right), 
\end{equation}
neglecting higher order terms of $\delta \hat g^2$ and 
$\delta g_R$.
The diagonal components are obtained
from Eqs.~(\ref{rge_ED3}) and (\ref{beta-g1}).
The off-diagonal one $\Delta$ is a constant to be determined later.
The eigenvalues $\lambda_i$ $(i=1,2)$ and eigenvectors 
${\mathbf v}_i, {\mathbf w}_i$ are defined as
\begin{equation}
  {\mathbf v}_i^T {\cal M} = \lambda_i\, {\mathbf v}_i^T, \quad
  {\cal M} {\mathbf w}_i = \lambda_i\, {\mathbf w}_i. 
\end{equation}
We obtain
\begin{equation}
 \lambda_1 = -2(\nu-1), \quad \lambda_2 = -2\nu\omega
\end{equation}
and
\begin{equation}
 {\mathbf v}_1 =  \left(
 \begin{array}{c}
 1 \\ 0
 \end{array}
 \right), \quad
 {\mathbf v}_2 =  \left(
 \begin{array}{c}
 \Delta \\ 2(\nu-1-\nu\omega)
 \end{array}
 \right),
\end{equation}
\begin{equation}
 {\mathbf w}_1 =  \left(
 \begin{array}{@{}c@{}}
 -2(\nu-1-\nu\omega) \\ \Delta
 \end{array}
 \right), \quad
 {\mathbf w}_2 =  \left(
 \begin{array}{c}
 0 \\ 1
 \end{array}
 \right).
\end{equation}
We here note 
\begin{equation}
 |\lambda_1| > |\lambda_2|, \label{lam1-lam2}
\end{equation}
since  $|\lambda_1| - |\lambda_2| = 2 (\nu-1-\nu\omega) > 0$ 
for $D \geq 6$ ($\nu \geq 2$)
in the nontrivial window ($\nu\omega < 1$).
In order to estimate $\Delta$, we need to determine the effective
potential off the UVFP, $\hat g^2 \ne g_*^2$, which can be obtained by
solving the SD equation.
Instead of doing this, here we simply replace the UVFP value $g_*^2$ in 
Eq.~(\ref{ren_g1}) with the running bulk gauge coupling $\hat g^2$:
\begin{equation}
   \left(\,\dfrac{1}{g_R(\mu)}-\dfrac{1}{\frac{\nu}{4}(1+\hat\omega)^2}
 \,\right) \,\mu^{2\nu\hat\omega} 
 = \mbox{const.},
 \label{ren_g3}
\end{equation}
with
\begin{equation}
 \hat\omega \equiv \sqrt{1-\frac{C_F \hat g^2(\mu) \NDA}{\kappa_D^{\rm crit}}}.
\end{equation}
Differentiating Eq.~(\ref{ren_g3}) with respect to $\ln \mu$,
we obtain a rough estimate of $\Delta$ as 
\begin{equation}
  \Delta = 2\,(\nu-1-\nu\omega)\,\frac{1-\omega}{\omega}
  \frac{g^{\rm crit}}{g_*^2} .
\end{equation}
The eigenvector ${\bf w}_1$ is then parallel to
the tangent of the critical line $g^{\rm crit}$.

\begin{figure}[tbp]
  \begin{center}
    \resizebox{0.45\textwidth}{!}{
     \includegraphics{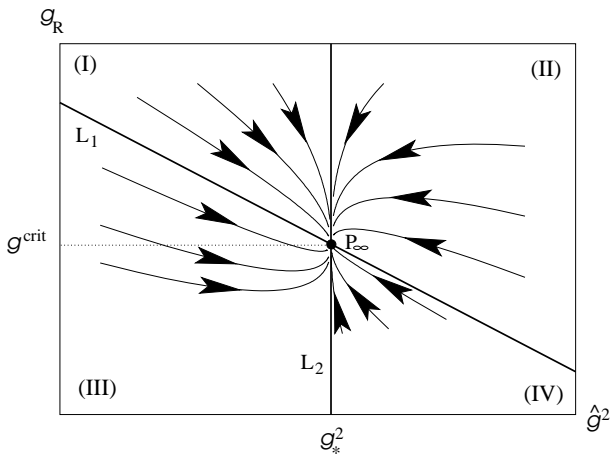}}
    \caption{RG flows of a theory with $D=6,N_c=3,N_f=4$.             
      The point $P_\infty$ denotes the UVFP, $(\hat g^2(\infty)=g_*^2, 
      g_R(\infty)=g^{\rm crit})$, of this theory.
      The arrows of the flows indicate the direction toward 
      the UV limit. The lines $L_1$ and $L_2$ correspond to
      the eigenvectors ${\mathbf w}_1$ and ${\mathbf w}_2$, 
      respectively. $L_1$ coincides with the tangent of the critical line
      $g^{\rm crit}$.
      \label{RGflow}}
  \end{center}
\end{figure}

Once we choose the model, ($N_c, N_f$), the UVFP value $g_*$ is fixed 
and so is the critical value of the four-fermion coupling $g^{\rm crit}$. 
RG flows for a typical $(N_c, N_f)$ are depicted in Fig.~\ref{RGflow}
in the vicinity of the point 
$P_\infty \equiv (\hat g^2=g_*^2,g_R=g^{\rm crit})$.
From Eq.~(\ref{lam1-lam2}) we can see a gross feature of RG flows
independently of the value of $\Delta$:
The RG flows approach this point 
as increasing $\mu$ more rapidly in the direction of ${\mathbf w}_1$ 
than that of ${\mathbf w}_2$. 
There are four regions (I,II,III and IV)
which are separated by two lines $L_1$
(parallel to ${\mathbf w}_1$) and $L_2$ (parallel to ${\mathbf w}_2$).
The \DxSB{} takes place in the regions I and II.
The line $L_1$ coincides with
the tangent of the critical line. While the chiral
symmetry is unbroken in the region III\@, the region IV is not precisely
the symmetric phase:  
Although the NJL coupling in the region IV is not strong enough for
the \DxSB{}, the gauge coupling strength grows strong in the infrared
there. 
We thus expect \DxSB{} also in the region IV, 
with a fermion dynamical mass being of order  $\Lambda_{\overline{\rm MS}}^{(D)}$
$(\ll \Lambda)$~\cite{Gusynin:2002cu} which is a scale parameter of this gauge theory analogously 
to $\Lambda_{\rm QCD}$.\footnote{
Thus $L_1$ as the border between region II and IV is not a phase boundary in the
exact sense but has a similar feature in the sense that the dynamical mass 
grows rapidly from $\Lambda_{\overline{\rm MS}}^{(D)}$ (instead of $0$)
to $\Lambda$ as we
cross the line from region IV to II. A similar phenomenon also takes place in 
the four-dimensional gauged NJL model with the QCD-like gauge coupling.
}
Note that the theory is controlled by the UVFP $P_\infty$, where the
composite fields $\sigma$ and $\pi$ enjoy nontrivial interactions.
Thus the theory in the nontrivial window 
is consistently renormalized even off the UVFP $\hat g^2 \ne g_*^2$.

\section{Summary and discussion}
\label{summary}
We have discussed phase structure of the $D$-dimensional gauged NJL model
with compactified extra $\delta (=D-4)$ 
dimensions on the TeV scale, based on the improved ladder SD equation
with the running gauge coupling given by the truncated KK effective 
theory. Such a running (dimensionless) 
gauge coupling has a nontrivial UV fixed point (UVFP) and the
theory behaves as a four-fermion theory coupled to 
a walking gauge theory with the gauge coupling almost
constant near the value of the UVFP in a wide energy region above
the compactification scale. The central assumption of this paper
was that the existence of the UVFP is not an artifact of 
the KK effective theory and may have some reality in a more elaborate
nonperturbative approach.

Solving the SD equation in the bulk
by setting the gauge coupling at the value of
the UVFP as a reasonable approximation, 
we found the critical line similar to that
of the four-dimensional gauged NJL model:
\begin{equation}
g^{\rm crit} = \frac{\frac{D}{2} -1}{4}
 \left( 1+ \sqrt{1-\kappa_D/\kappa_D^{\rm crit} } \right)^2\, ,
\end{equation}
with $\kappa_D^{\rm crit}=\frac{D}{32}\frac{D-2}{D-1}$, which 
takes the same form as 
that of the four-dimensional one with a fixed gauge coupling
for $D\rightarrow 4$ in the prefactor.  
In spite of the formal resemblance to the four-dimensional case,
however,
the UVFP value $\kappa_D$, given by Eq.~(\ref{eq:def_kappaD}) with 
Eqs.~(\ref{bprime}) and (\ref{UV-FP}),
is determined as a function of $N_c$ and $N_f$ and hence
is not a free parameter but is a fixed quantity once we specify the model.

Here we should emphasize the following:
In the pure gauge theories with extra dimensions
the ``nontrivial'' theory defined at the critical point is only formal, 
since $\kappa_D$ cannot be fine-tuned arbitrarily close 
to the critical value~\cite{Hashimoto:2000uk,Gusynin:2002cu}. 
The dynamical mass cannot be made finite in the continuum limit 
$(\Lambda \to \infty)$.
Then, even if we assume the existence of the UVFP beyond the truncated
KK effective theory, the renormalizability of higher dimensional gauge 
theories with $(N_c,N_f)$ such that $\kappa_D > \kappa_D^{\rm crit}$ 
(region (A) in Fig.~\ref{phase-R}) is only formal~\footnote{ 
Nevertheless, such a theory
can still be useful
for model building as an effective theory with finite cutoff 
$\Lambda$. See, for example, the tMAC analysis of the $D=8$ Top Mode
Standard Model with extra dimension~\cite{Hashimoto:2003ve}.}.
On the other hand, 
we do have a continuous parameter, the four-fermion
coupling $g$, in the gauged NJL model and hence the nontrivial theory
can be defined by fine-tuning $g$ arbitrarily close to the critical line.
Thus the inclusion of the four-fermion interactions may provide 
an interesting new possibility for higher dimensional gauge theories 
which were long considered nonrenormalizable and trivial theories 
based on the perturbation.

Remarkably enough, we in fact found 
the nontrivial window (region (B) in Fig.~\ref{phase-R})
for $(N_c,N_f)$ such that 
\begin{equation}
 \kappa_D^{\rm MT} <\kappa_D <\kappa_D^{\rm crit} , 
\end{equation}
\begin{equation}
\kappa_D^{\rm MT}= \left( 1-\frac{1}{(D/2-1)^2}\right) \kappa_D^{\rm crit}\, ,
\end{equation}
where the four-fermion theory in the presence of walking gauge interactions
becomes
``nontrivial'' and ``renormalizable'', similarly to the
four-dimensional gauged NJL model, in the sense that 
the decay constant $F_\pi$ and 
couplings of composite $\pi$ and $\sigma$ as well as the
induced Yukawa coupling of the fermion (with mass $M$) 
to those composites remain finite in the
continuum limit $\Lambda/M \rightarrow \infty$.
We explicitly performed renormalization of the kinetic term of 
the composite bosons and the effective potential. 
The renormalized four-fermion coupling has a UVFP at
$g_R(\infty)=g^{\rm crit}$,
where
the theory has a large anomalous dimension 
\begin{equation}
 \gamma_m = \left(\frac{D}{2} -1\right)
 \left(1+ \sqrt{1-\kappa_D/\kappa_D^{\rm crit}}\right)\, . 
\end{equation} 
The renormalized effective potential indeed has  nontrivial interactions
which are {\it conformal invariant}, thanks to the large anomalous dimension.

It is rather surprising that even in higher dimensions 
the four-fermion operators become relevant thanks to 
the large anomalous dimension, 
$D/2 > \gamma_m > D/2-1$.
In fact, the dynamical dimension of $(\bar\psi\psi)^2$ operator
is relevant, dim($(\bar\psi\psi)^2$) $< D$.
On the other hand, the $(\bar\psi\psi)^4$  operator and the kinetic term 
$\partial_M (\bar\psi\psi)\partial^M (\bar\psi\psi)$ are 
irrelevant, dim($(\bar\psi\psi)^4$) 
$=2D-2(D-2)\sqrt{1-\kappa_D/\kappa_D^{\rm crit}} > D$, and
dim($\partial_M (\bar\psi\psi)\partial^M (\bar\psi\psi))
= D+2-(D-2)\sqrt{1-\kappa_D/\kappa_D^{\rm crit}} 
> D$.
Thus the theory can be renormalized without operators other than
$(\bar\psi\psi)^2$.
We further gave the RG flow off the UVFP in Fig.~\ref{RGflow}
which is consistent with our renormalization performed on the UVFP.

On the other hand, 
for $(N_c,N_f)$ such that $0 \leq \kappa_D \leq \kappa_D^{\rm MT}$ 
(region (C) in Fig.~\ref{phase-R}),
the decay constant $F_\pi$ diverges even if we fine-tune 
the four-fermion coupling close to the critical value $g^{\rm crit}$
to make finite the dynamical mass of the fermion.
At the MT point $\kappa_D=\kappa_D^{\rm MT}$ the divergence of 
$F_\pi^2$ is logarithmic, $F_\pi^2 \sim \ln \Lambda$, which 
is similar to the situation at the pure NJL point of 
the four-dimensional case.
The MT point is reduced to the pure NJL point $\kappa_D^{\rm MT} 
\rightarrow 0$ for $D\rightarrow 4$.
Without 
introducing other operators such as
$(\bar\psi\psi)^4$, $\partial_M (\bar\psi\psi)\partial^M(\bar\psi\psi)$, 
etc., the renormalization breaks down in the region (C) 

\begin{figure}[tbp]
  \begin{center}
    \resizebox{0.3\textwidth}{!}{
     \includegraphics{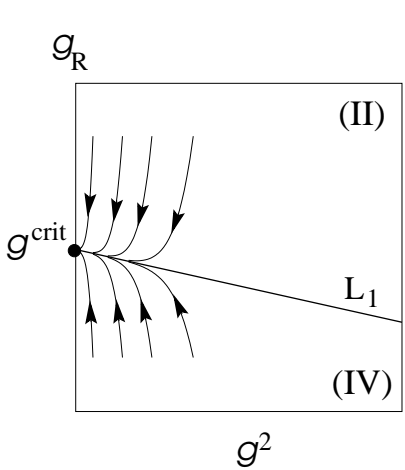}}
    \caption{RG flows of a theory for $D=4$.
      The arrows of the flows indicate the direction toward 
      the UV limit. The line $L_1$ corresponds to
      the eigenvector ${\mathbf w}_1$ (tangent of the critical line of the
      four-dimensional gauged NJL model with fixed gauge 
      coupling~\cite{KMY89,ASTW}).
      The black circle is the pure NJL point.\label{RGflow-D4}}
  \end{center}
\end{figure}
%
%
\begin{table}[tb]
  \begin{center}
    \resizebox{0.45\textwidth}{!}{
     \includegraphics{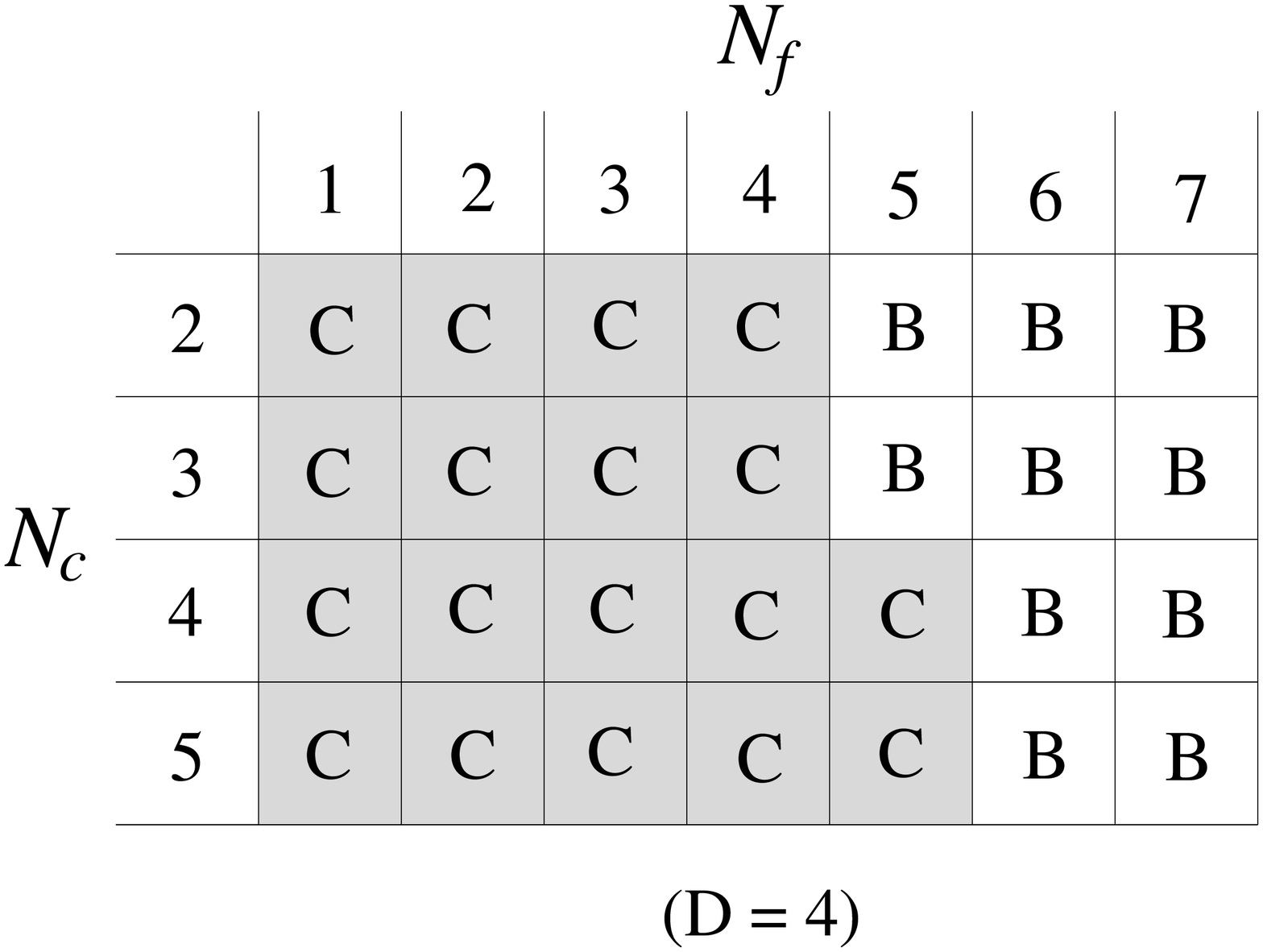}}
    \caption{Classification of models ($D=4$).
          The entries labeled by B and C stand for the models in
          the regions (B) $A=18C_F/(11N_c-2N_f)>1$
          and (C) $A \leq 1$, respectively. 
          \label{tab4}}
  \end{center}
\end{table}
At this point it is worthwhile mentioning the formal limit $D \to 4$ of the
nontrivial window.
It was shown~\cite{Kondo:1991yk,Krasnikov:1992zn,
Kondo:1992sq,Harada:1994wy,Kubota:1999jf}  that 
the four-dimensional
gauged NJL model becomes renormalizable and nontrivial, when
the one-loop gauge coupling runs 
as specified by:
\begin{eqnarray}
&&A \equiv \frac{6C_F}{-b} \, > \,1, 
\label{moderate walking}\\
  &&\left( C_F=\frac{N_c^2-1}{2N_c}, \quad
  -b =\frac{11N_c - 2N_f}{3} \right), \nonumber
\end{eqnarray}
where $A$ measures the running speed of the 
coupling~\cite{BMSY87,Yamawaki:1996vr}:$A \gg 1$ is the walking theory and 
$A\to \infty \, (b \to 0) $ 
corresponds to the non-running (standing limit of walking
coupling) case with vanishing beta function.
The condition Eq.~(\ref{moderate walking}) 
was first obtained~\cite{Kondo:1991yk} 
by evaluating the $F_\pi^2$, 
taking account of the {\it logarithmic corrections} to the asymptotic
behavior of the  mass function~\cite{MY89}:
\begin{equation}
  \Sigma(x) \sim  c_1 \Sigma_0 
  \left(\ln \left(\dfrac{x}{\Sigma_0^2}\right)\right)^{-\frac{A}{2}},
\end{equation}
instead of Eq.~(\ref{eq:asymp_sol}),
and then 
in the RG group analysis in the language of the equivalent gauged
Yukawa model~\cite{Krasnikov:1992zn,
Kondo:1992sq,Harada:1994wy} and also 
in the nonperturbative RG equation\cite{Kubota:1999jf}.  

Here we show that the formal limit of $D\to 4$ of the nontrivial
window in this paper coincides with that characterized by  
precisely the same condition as Eq.~(\ref{moderate walking}), now
{\it without logarithmic corrections}
which were rather delicate factors in the previous arguments in
the four-dimensional model.
Instead we consider $D=4 + \delta$ with $0<\delta \ll 1$
in the spirit of $\epsilon$ expansion~\cite{Wilson,Wilson:1973jj}. 
The nontrivial window is defined by
\begin{equation}
\nu \omega <1,
\end{equation}
which reads
\begin{equation}
\frac{\kappa_D}{\kappa_D^{\rm crit}} >\delta.
\end{equation}
On the other hand, from Eqs.~(\ref{UV-FP}) and (\ref{eq:def_kappaD}) we have
\begin{equation}
\frac{\kappa_D}{\kappa_D^{\rm crit}}= \frac{6C_F}{-b}\,\delta= A \,\delta
\end{equation}
Then we obtain Eq.~(\ref{moderate walking}), $A >1$.
The nontrivial window in the formal limit $D \to 4 \;(\delta \to 0)$ 
thus coincides with the condition of the renormalizability/nontriviality of
the four-dimensional gauged NJL models, namely those 
coupled to the (moderately) walking 
gauge theory with $A>1$. 
Combining the condition of asymptotic freedom $N_f<11N_c/2$ and  $A>1$, 
we thus have theories within the nontrivial window in $D\to 4$:
\begin{equation}
\frac{9}{2}\frac{1}{N_c}<N_f-N_c <\frac{9}{2}N_c.
\label{nt4}
\end{equation}
See Table~\ref{tab4} for the theories 
satisfying Eq.~(\ref{nt4}) (or Eq.~(\ref{moderate walking})). 
On the other hand, the trivial region (C) ($0\le\kappa_D\le\kappa_D^{\rm MT}$)
coincides with the condition of the triviality/nonrenormalizability, 
$A \leq 1$, in the $D\to 4$ limit.
Thus in the $D\to4$ limit  
where $\kappa_D^{\rm MT}\to 0$ and $\kappa_D\to 0$ 
(for $A\,<\,\infty$)\footnote{If one formally changes the theory 
$(N_c, N_f)$
depending upon
the dimension $\delta (\to 0) $ in such a way that 
$A\sim 4\lambda/\delta\to \infty$, one could have a limit arriving at the four-dimensional
gauged NJL model with fixed (standing) gauge coupling $\lambda$: 
$\kappa_D/\kappa_D^{\rm crit} \to 4 \lambda \,(0<\lambda<\lambda^{\rm crit}=1/4)$. },
the UVFP's for both regions (C) ($0\le\kappa_D\le\kappa_D^{\rm MT}$)  and 
(B) ($\kappa_D^{\rm MT}<\kappa_D< \kappa_D^{\rm crit}$)
in the phase diagram of Fig. \ref{phase-R} 
shrink to a single point of the pure NJL point and hence the distinction
among them is not obviously visible compared with the case with extra 
dimensions. 
In other words, the renormalizability/nontriviality of the
four-dimensional gauged NJL model with $A>1$ gauge theories~\cite{Kondo:1991yk,Krasnikov:1992zn,
Kondo:1992sq,Harada:1994wy,Kubota:1999jf}   
is a four-dimensional manifestation of the nontrivial window, in sharp
contrast to the theory with $A \leq 1$.
Thus {\it the nontrivial window is not a peculiarity nor an artifact of
the extra dimensions but is rather a universal
feature of the gauged NJL model}.

The RG flows of the nontrivial window in this limit are only 
in regions II and IV of Fig.~\ref{RGflow} and are similar to
Fig.~\ref{RGflow} but with a crucial difference: Since $|\lambda_1| <|\lambda_2|$ in this
limit, the flows first approach $L_1$ instead of $L_2$ and then converge
toward the pure NJL point (Fig.~\ref{RGflow-D4}). This is actually consistent with the gross feature obtained in
the four-dimensional model.~\cite{Kondo:1991yk,Aoki:1999dv,Kubota:1999jf}

In conclusion, we have shown that the four-fermion theories in $D>4$ dimensions
are renormalizable and nontrivial, {\it when coupled to $D$-dimensional
walking gauge theories}, with the parameters $(N_c,N_f)$ being 
in the nontrivial window,
which we assumed have a nontrivial ultraviolet 
fixed point as given by the truncated KK effective theory.
The fact that the four-fermion theory in dimensions $D<4$, {\it lower than 4}, 
is renormalizable and nontrivial
has been known for long time since 
Wilson~\cite{Wilson} (For a proof on the $D=3$ theory, see
Ref.\cite{RWP}) and  was also known for the arbitrary 
$D(<4)$ dimensions~\cite{Kikukawa:1989fw,Hands:1992be,Kondo:1992sq}. Further 
in $D=4$ it was known~\cite{Kondo:1991yk,Krasnikov:1992zn,Kondo:1992sq,Harada:1994wy,Kubota:1999jf} that 
the four-fermion theory when coupled with certain
walking gauge theories is renormalizable and nontrivial.
Here we have shown for the first time that 
the four-fermion theory in $D>4$, {\it higher than 4}, 
also shares the same
feature when coupled to walking gauge theories characterized by the 
nontrivial window.

This renormalizability/nontriviality is not just a formal matter,
but implies {\it cutoff insensitivity} of the physics prediction,
which may be useful for model buildings even with a finite cutoff. 
This is rather surprising, since the theory
is dominated by the dynamics in the ultraviolet region where both
the gauge and the four-fermion interactions become strongly coupled. 
Explicit model buildings based on this observation were
in fact attempted for the four-dimensional case (``Top Mode Walking
GUTs'')~\cite{Inukai:1996dt,Yamawaki:1996vr}. 
  
The phase structure of the gauged NJL model in the bulk with 
compactified extra dimensions may be useful for model buildings
such as the Top Mode Standard Model, bulk technicolor, etc..
For instance, the $D=6$ Standard Model gauge interactions are not enough
to trigger the top quark condensate~\cite{Hashimoto:2000uk,Gusynin:2002cu,Hashimoto:2003ve}, while
the gauged NJL model can do work due to the additional attractive
interactions as in the case of the original Top Mode Standard Model.
Although introduction of such an ad hoc four-fermion interaction 
may be less attractive than the scenario without it, 
the KK modes of the top quark still naturally reduce the top quark 
mass prediction of the original model 
as was emphasized in Ref.\cite{Cheng:1999bg}. 
Since the electroweak symmetry breaking is still 
a central mystery of the modern 
particle theory, it would be  useful to consider all possible 
dynamical scenarios before the LHC will take off. The gauged 
NJL model with extra dimensions may become one of the dormant
volcano to give an explosion into rich TeV physics.   

\section*{Acknowledgments}
The work is supported in part by the JSPS Grant-in-Aid for the
Scientific Research (B)(2) 14340072 (K.Y. and M.T.),
(C)(2) 16540226 (M.T.) and  
by KRA PBRG 2002-070-C00022 (M.H.).

\appendix 

\section{Propagator of the composite scalar}
\label{app1}

We have calculated the auxiliary field propagator in 
Sec.~\ref{sec-yukawa} by using 
the resummation technique~\cite{Appelquist:1991kn}. 
We may take yet another choice proposed in Ref.~\cite{Gusynin:1997cw}
in which we obtain the scalar propagator as the lowest order of 
the Chebyshev expansion. 
While the resummation technique is operative in a weak gauge coupling 
region, the Chebyshev expansion method is valid in the whole region.
In four dimensions the results of the Chebyshev expansion method are 
similar to those of the resummation technique.~\cite{Gusynin:1997cw}
We here demonstrate that the wave function renormalization constant
for the auxiliary field agrees with the result of the resummation
technique shown in Eq.~(\ref{eq:Z_wav}) even in extra dimensions.

The composite scalar propagator is written in terms of 
the vacuum polarization function $\Pi_{\rm S}$ for the composite scalar:
\begin{equation}
i D_{\sigma}^{-1}(q)=-\frac{N_c}{G}+\Pi_{\rm S}(q),
\label{scal_prop:eq}
\end{equation}
where the vacuum polarization function is given by
\begin{equation}
\Pi_{\rm S}(q) = N_c N_f
\int \frac{d^D k}{i(2\pi)^D}\,\tr
\left[S(k+q){\chi_{\rm S}}(k+q,k)S(k) \right],
\label{sde_scalvac}
\end{equation}
with $\chi_{\rm S}$ being the Bethe-Salpeter (BS) amplitude for the scalar. 
Neglecting the explicit breaking of the $D$ dimensional Lorentz
symmetry owing to the compactification,
we can decompose generally the BS amplitude $\chi_{\rm S}$ 
into four Lorentz scalar functions $F_i$ $(i=1,2,3,4)$,
\begin{eqnarray}
\lefteqn{ \hspace*{-5mm}
 \chi_{\rm S}(p+q,p) = } \nonumber \\ && 
 F_1(p+q,p)+\left(\fsl q\fsl p-\fsl p\fsl q\right)
 F_2(p+q,p) \nonumber \\[1mm]
 && + (\fsl p+\fsl q) F_3(p+q,p)+\fsl p F_4(p+q,p), 
\label{bs}
\end{eqnarray}
where $p$ and $q$ denote momenta of the fermion and the scalar bound state, 
respectively. 
Although one might suspect that more structure functions 
are needed in $D(>4)$ dimensions, 
other terms are zero or reduced into the above four,
because the BS amplitude includes only two momenta. 
We analyze the scalar propagator and the BS equation 
in the symmetric phase, i.e., 
\begin{equation}
 S(p) = S_0(p) = \frac{i}{\fsl{p}}.   
\end{equation}
We then obtain the vacuum polarization function as
\begin{widetext}
\begin{equation}
 2^{-\frac{D}{2}} N_c^{-1} N_f^{-1} \Pi_{\rm S}(q) =
 \int \frac{d^D k}{i(2\pi)^D} \frac{-k \cdot (k+q)}{k^2 (k+q)^2} F_1(k+q,k) 
 + 2 \int \frac{d^D k}{i(2\pi)^D}
  \frac{(k \cdot q)^2-k^2 q^2}{k^2(k+q)^2} F_2(k+q,k) . \label{PiS}
\end{equation}
On the other hand, the BS equation under the ladder approximation is 
\begin{eqnarray}
{\chi_{\rm S}}(p+q,p)&=&{\bf 1}+\int\frac{{\rm d}^Dk}{(2\pi)^D}
[-iT^a\Gamma^M] S(k+q) 
{\chi_{\rm S}}(k+q,k) S(k)[-iT^a\Gamma^N] g^2_{(4+\delta)D}D_{MN}(p-k),
\label{bs_eq}
\end{eqnarray}
where ${\bf 1}$ is the identity matrix and $D_{MN}$ is the propagator
of the gauge boson. 
In the symmetric phase the BS equation is obviously decomposed into
two simultaneous equations for $F_{1,2}$ and for $F_{3,4}$.
Since the structure functions $F_3$ and $F_4$ do not contribute to 
$\Pi_{\rm S}$ as shown in Eq.~(\ref{PiS}), 
we neglect the two hereafter.
The BS equations for $F_1$ and $F_2$ are given by
\begin{eqnarray}
 F_1 (p+q,p) &=& 1 + (D-1+\xi) C_F g_*^2 \int \frac{d^D k}{i(2\pi)^D} 
 \frac{-k \cdot (k+q)}{k^2\,(k+q)^2\,[-(p-k)^2]^{D/2-1}} F_1(k+q,k)
 \nonumber \\ && +
 2 (D-1+\xi) C_F g_*^2 \int \frac{d^D k}{i(2\pi)^D} 
 \frac{(k \cdot q)^2-k^2 q^2}{k^2\,(k+q)^2\,[-(p-k)^2]^{D/2-1}} F_2(k+q,k),
 \label{bs_eq1} \\
 F_2 (p+q,p) &=& \frac{C_F g_*^2}{2[(p \cdot q)^2-p^2 q^2]}
 \int \frac{d^D k}{i(2\pi)^D}
 \frac{\Xi (p,q,k)}{k^2\,(k+q)^2\,[-(p-k)^2]^{D/2-1}} F_1(k+q,k)
 \nonumber \\ && +
\frac{C_F g_*^2}{(p \cdot q)^2-p^2 q^2}
 \int \frac{d^D k}{i(2\pi)^D}
 \frac{-k \cdot (k+q) \, \Xi(p,q,k)}{k^2\,(k+q)^2\,[-(p-k)^2]^{D/2-1}}
 F_2(k+q,k)
 \label{bs_eq2}
\end{eqnarray}
with
\begin{eqnarray}
  \Xi(p,q,k) &\equiv& (D-3+\xi)[q^2(p \cdot k)-(p \cdot q)(q \cdot k)]
  \nonumber \\ &&
  +\frac{2(1-\xi)}{-(p-k)^2}\left\{ -q^2[k \cdot (p-k)]\,[p \cdot (p-k)]
   -[q \cdot (k-p)]\,[p^2(q \cdot k)-k^2(p \cdot q)]\right\},
\end{eqnarray}
\end{widetext}
where the bulk gauge coupling was replaced as in Eq.~(\ref{imp_ladder}) 
with $\hat g^2=g_*^2$. The gauge fixing parameter $\xi$ should be taken as
\begin{equation}
  \xi = -\frac{(D-1)(D-4)}{D}
\end{equation}
for consistency with the ladder approximation.

Let us expand $F_1$, $F_2$, and the integral kernels of $\Pi_{\rm S}$ 
and the BS equations by series of Gegenbauer polynomials, 
which is generalization of the Chebyshev polynomials.
At this stage the BS equations (\ref{bs_eq1}) and (\ref{bs_eq2})
depend on three angles $\alpha, \beta, \gamma$ defined by
\begin{equation}
\cos \alpha=\frac{p_E\cdot q_E}{|p_E||q_E|},\;
\cos \beta=\frac{p_E\cdot k_E}{|p_E| |k_E|},\;
\cos \gamma=\frac{q_E\cdot k_E}{|q_E| |k_E|},
\end{equation}
as well as an infinite chain of harmonics $f_n$ and $g_n$ for
expansion of the Gegenbauer polynomials $C_n^{\nu}(\cos\alpha)$, where
\begin{equation}
  F_1(p+q,p)=\sum_{n=0}^\infty f_n(p^2_E,q^2_E) C_n^{\nu}(\cos\alpha)
\end{equation}
and
\begin{equation}
  F_2(p+q,p)=\sum_{n=0}^\infty g_n(p^2_E,q^2_E) C_n^{\nu}(\cos\alpha),
\end{equation}
so that it is quite difficult to solve analytically the equations.
We here note that only the BS equation for the harmonics $f_0$ 
contains an inhomogeneous term (the constant unity),
because the Gegenbauer polynomials satisfy 
\begin{equation}
  \int_0^\pi d\theta \sin^{2\nu}\theta\, 
  C_n^{\nu}(\cos\theta) \, C_m^{\nu}(\cos\theta) = w_n \, \delta_{mn}
\end{equation}
with
\begin{equation}
w_n \equiv 
\frac{\pi \, \Gamma(n+2\nu)}{2^{2\nu-1}(n+\nu)\,n!\,[\Gamma(\nu)]^2},  
\end{equation}
and 
\begin{equation}
  C_0^{\nu}(\cos\theta) = 1.
\end{equation}
The inhomogeneous term certainly gives main contributions to 
the BS equations.
Hence we assume that the BS equation for $f_0$ is 
well approximated by the closed form of $f_0$, neglecting the effects
of $f_n (n \geq 1)$ and $g_n (n \geq 0)$.
The harmonics $f_n (n \geq 1)$ and $g_n (n \geq 0)$ 
are iteratively determined after $f_0$ is computed.
We then obtain the BS equation for $f_0$ as
\begin{equation}
f_0(s,t)=1 + \frac{4(D-1)}{D} \kappa_D
\int_0^{\Lambda^2} \!\!\!\! du u^{\frac{D}{2}-2}\,K_0(s,t,u)
 f_{0}(u,t),
\label{zeroth_bs_eq}
\end{equation}
where $s \equiv p_E^2$, $t \equiv q_E^2$, $u \equiv k_E^2$, and
\begin{equation}
 K_0(s,t,u) \equiv a_0(u,t) b_0(s,u) \label{kernel} 
\end{equation}
with
\begin{eqnarray}
\lefteqn{ \hspace*{-7mm}
 a_0(u,t) \equiv} \nonumber \\
&& \hspace*{-3mm}
 \frac{1}{w_0}\int_0^\pi d\gamma \sin^{2\nu}\gamma
 \, C_0^\nu(\cos\gamma) \, \frac{k_E \cdot (k_E+q_E)}{(k_E+q_E)^2} , \\[3mm]
\lefteqn{ \hspace*{-7mm}
 b_0(s,u) \equiv} \nonumber \\
&& \hspace*{-3mm}
 \frac{1}{w_0}\int_0^\pi d\beta \sin^{2\nu}\beta
 \, \frac{C_0^\nu(\cos\beta)}{[(p_E-k_E)^2]^{\frac{D}{2}-1}} . 
\end{eqnarray}
We can easily perform the angular integrations and find
\begin{eqnarray}
a_0(u,t) &=& \frac{1}{2}\,
  \left[\frac{u-t}{\max(u,t)} 
   F\left(1,2-\frac{D}{2};\frac{D}{2};\frac{\min(u,t)}{\max(u,t)}\right)
 \right. \nonumber \\[2mm]
 && \qquad \left. + 1 \phantom{\Big|}\right], \\[2mm]
b_0(s,u) &=& \frac{1}{[\max(s,u)]^{\frac{D}{2}-1}}, 
\end{eqnarray}
The vacuum polarization function is written in terms of $f_0$ as
\begin{equation}
\Pi_{\rm S}(t)=
2^{\frac{D}{2}}\Omega_{\rm NDA} N_c N_f \int_0^{\Lambda^2}\!\!\!\!
du \, u^{\frac{D}{2}-2}\, a_0(u,t) f_0(u,t).
\label{vacpol3}
\end{equation}
Noting that the integral of R.H.S. in the zeroth order 
BS equation (\ref{zeroth_bs_eq}) at $s=\Lambda^2$ is just same 
as in Eq.~(\ref{vacpol3}), 
we can rewrite the composite scalar propagator without the integral
as follows:
\begin{eqnarray}
\lefteqn{ \hspace*{-1cm}
\left(2^{\frac{D}{2}} N_c N_f \Lambda^{D-2} \NDA \right)^{-1} 
i D_{\sigma}^{-1}(t) =} \nonumber \\ &&  \qquad 
 -\frac{1}{g} + \frac{1}{\frac{4(D-1)}{D}\kappa_D}
 \left[F_{\rm UV}(\Lambda^2,t)-1\right],
\end{eqnarray}
where we divided the zeroth order harmonics $f_0$ into two parts of 
the IR and UV regions,
\begin{equation}
f_0(s,t) \equiv F_{\rm IR}(s,t)\theta(t-s) + F_{\rm UV}(s,t)\theta(s-t).
\end{equation}
The BS amplitude in zero momentum transfer,
i.e., the effective yukawa coupling $\Gamma_s$, is also obtained through
$F_{\rm UV}$,
\begin{equation}
\Gamma_s(p_E^2) = {\chi_{\rm S}}(p,p) \simeq F_{\rm UV}(p_E^2,q_E^2=0).
\end{equation}

Let us solve the zeroth order BS equation (\ref{zeroth_bs_eq})
which is equivalent to a set of the second order differential equation, 
\begin{equation}
\frac{\partial^2}{\partial s^2}f_0(s,t)
 +\frac{D}{2s}\frac{\partial }{\partial s}f_0(s,t)
 +\frac{\tilde{\kappa}_D}{s^2} a_0(s,t) f_0(s,t)=0, \label{BS_f0}
\end{equation}
and the IRBC
\begin{equation}
\left.s^{D/2}\frac{\partial}{\partial s}f_0(s,t)\right|_{s \to 0}=0,  
\label{BS-IRBC}
\end{equation}
and the UVBC
\begin{equation}
\left.s\frac{\partial }{\partial s}f_0(s,t)
      +(D/2-1)\left(f_0(s,t)-1\right)\right|_{s \to \Lambda^2}=0,  
\label{BS-UVBC}
\end{equation}
with
\begin{equation}
  \tilde{\kappa}_D \equiv \frac{2(D-1)(D-2)}{D} \kappa_D 
  =\frac{(D-2)^2(1-\omega^2)}{16} .
\end{equation}
We now replace the hypergeometric function 
in $a_0(u,t)$ by unity, i.e., 
\begin{equation}
  a_0(u,t) \to \frac{1}{2} \, \left[\frac{u-t}{\max(u,t)} + 1 \right].
  \label{app_a0}
\end{equation}
We confirm later on that the approximation works well in $D=6$.
Within the approximation the differential equations for 
the UV and IR parts are 
\begin{equation}
\frac{\partial }{\partial s^2}F_{\rm UV}+
 \frac{\nu+1}{s}\frac{\partial }{\partial s}F_{\rm UV}+
  \frac{\tilde{\kappa}_D}{s^2} \left(1-\frac{t}{2s}\right)
   F_{\rm UV}=0 , \label{bs_eq_UV}
\end{equation}
and
\begin{equation}
\frac{\partial }{\partial s^2}F_{\rm IR}+
 \frac{\nu+1}{s}\frac{\partial }{\partial s}F_{\rm IR}+
  \frac{\tilde{\kappa}_D}{2st} F_{\rm IR}=0 , \label{bs_eq_IR} 
\end{equation}
respectively, where we used 
\begin{equation}
  \nu = D/2-1 .
\end{equation}
We can solve analytically Eqs.~(\ref{bs_eq_UV}) and (\ref{bs_eq_IR})
and find 
\begin{eqnarray}
F_{\rm UV}(s,t)&=&\left(\frac{t}{s}\right)^{\frac{\nu}{2}}\left[c_1  
 I_{-\nu\omega}\left(\sqrt{2\tilde{\kappa}_D \frac{t}{s}}\right) \right.
 \nonumber \\ && \qquad \quad \left. +
 c_2 I_{\nu\omega}\left(\sqrt{2\tilde{\kappa}_D \frac{t}{s}}\right)\right],
 \label{fuv}
\end{eqnarray}
for the UV part, and 
\begin{eqnarray}
F_{\rm IR}(s,t)&=&\left(\frac{t}{s}\right)^{\frac{\nu}{2}}\left[c_3 
 J_{\nu}\left(\sqrt{2\tilde{\kappa}_D \frac{s}{t} }\right) \right.
 \nonumber \\ && \qquad \quad \left. +
 c_4Y_{\nu}\left(\sqrt{2\tilde{\kappa}_D \frac{s}{t}}\right)\right],
 \label{fir}
\end{eqnarray}
for the IR part,
where $J_{\nu}$, $Y_{\nu}$ and $I_{\pm\nu\omega}$ represent 
the Bessel functions of first and second kind, and modified one, 
respectively.
For $\omega=1/\nu$ we should use the modified Bessel function of 
the third kind, $K_1$, instead of $I_{-1}$.
The IRBC (\ref{BS-IRBC}) leads to
\begin{equation}
  c_4 = 0 .
\end{equation}
Other coefficients $c_1,c_2$, and $c_3$ are obtained from
the UVBC (\ref{BS-UVBC}) 
and smoothness conditions for $f_0(s,t)$ at $s=t$,
\begin{equation}
F_{\rm IR}(s \to t,t)=F_{\rm UV}(s \to t,t), \label{cont}
\end{equation}
and
\begin{equation}
\frac{\partial }{\partial s}F_{\rm IR}(s,t)\Biggr|_{s \to t}=
\frac{\partial }{\partial s}F_{\rm UV}(s,t)\Biggr|_{s \to t},\label{contdiff}
\end{equation}
as follows:
\begin{eqnarray}
c_1 &\equiv& 
c_1\left(\frac{t}{\Lambda^2},\omega\right), \nonumber \\
&=&
 \frac{\pi\gamma\left(\omega\right)}{2\sin(\pi\nu\omega)}
\eta^{-1}\left(\frac{t}{\Lambda^2},\omega\right),\\[2mm]
c_2 &=&
 c_1\left(\frac{t}{\Lambda^2},-\omega\right),\\
c_3 &=&
 \eta^{-1}\left(\frac{t}{\Lambda^2},\omega\right), 
\end{eqnarray}
where we defined
\begin{eqnarray}
\eta\left(\frac{t}{\Lambda^2},\omega\right)&\equiv& 
 \frac{\pi}{2\sin(\pi\nu\omega)}
\left[\gamma\left(\omega\right)
\zeta\left(\frac{t}{\Lambda^2},-\omega\right) \right. \nonumber \\
&& \qquad \;\; \left.
-\gamma\left(-\omega\right)
  \zeta\left(\frac{t}{\Lambda^2},\omega\right) \right],
 \label{Zdef} 
\end{eqnarray}
\begin{eqnarray}
\gamma\left(\omega\right)&\equiv&
 \sqrt{2\tilde{\kappa}_D}\left[
J_{\nu}\left(\sqrt{2\tilde{\kappa}_D}\right)I_{\nu\omega}^\prime
         \left(\sqrt{2\tilde{\kappa}_D}\right)\right. \nonumber \\
&& \quad \left.
+J_{\nu}^\prime\left(\sqrt{2\tilde{\kappa}_D}\right)
 I_{\nu\omega}\left(\sqrt{2\tilde{\kappa}_D}\right)\right],
\label{gammaeq}
\end{eqnarray}
\begin{eqnarray}
\zeta\left(\frac{t}{\Lambda^2},\omega\right)&\equiv&
 \frac{1}{2\nu}\left(\frac{t}{\Lambda^2}\right)^{\frac{\nu}{2}}
 \left[ \nu
  I_{\nu\omega}\left(\sqrt{\frac{2\tilde{\kappa}_D t}{\Lambda^2}}\right)
\right. \nonumber \\
&& \;\; \left.
 -\sqrt{\frac{2\tilde{\kappa}_D t}{\Lambda^2}}I_{\nu\omega}^\prime
  \left(\sqrt{\frac{2\tilde{\kappa}_D t}{\Lambda^2}}\right)\right].
\label{Geq}
\end{eqnarray}
The prime $X'(z)$ denotes $dX/dz$.
We also note the relation
\begin{equation}
 I'_\rho (z) I_{-\rho} (z) - I_\rho (z) I'_{-\rho} (z) 
 =\frac{2\sin(\pi \rho)}{\pi z}. 
\end{equation}
The BS amplitude is thus given by
\begin{eqnarray}
&& \hspace*{-12mm}
F_{\rm IR}(s,t) = \eta^{-1}\left(\frac{t}{\Lambda^2},\omega\right) 
\left(\frac{t}{s}\right)^{\frac{\nu}{2}}
J_{\nu}\left(\sqrt{\frac{2\tilde{\kappa}_D s}{t} }\right),
\label{fir2} \\
&& \hspace*{-12mm}
F_{\rm UV}(s,t)=
\frac{\pi}{2\sin(\pi\nu\omega)}
 \eta^{-1}\left(\frac{t}{\Lambda^2},\omega\right)
\left(\frac{t}{s}\right)^{\frac{\nu}{2}}\nonumber\\
&&\times\left[ 
\gamma\left(\omega\right) 
 I_{-\nu\omega}\left(\sqrt{\frac{2\tilde{\kappa}_D t}{s}}\right)
 \right. \nonumber \\
&& \left. \qquad \qquad 
-\gamma\left(-\omega\right)
 I_{\nu\omega}\left(\sqrt{\frac{2\tilde{\kappa}_D t}{s}}\right)\right].
\label{fuv2}
\end{eqnarray}
The effective yukawa coupling $\Gamma_s$ is also obtained as
\begin{equation}
\Gamma_s(p_E^2) = F_{\rm UV}(p_E^2,q_E^2=0)
=\frac{2}{1+\omega}\left(\frac{p_E^2}
{\Lambda^2}\right)^{-\frac{\nu}{2}(1-\omega)} \hspace*{-1cm},
\label{GSqzero}
\end{equation}
which agrees with Eq.~(\ref{yukawa}) calculated through 
the resummation technique.

\begin{figure}[tb]
  \begin{center}
   \resizebox{0.47\textwidth}{!}{\includegraphics{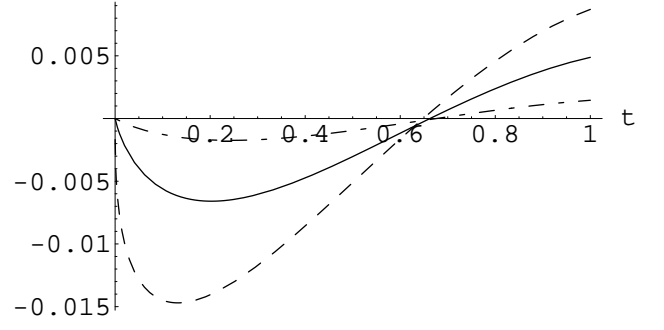}}
    \caption{Error of the zeroth order BS amplitude $f_0$ 
             arising from the approximation (\ref{app_a0}).
             The curves show $(f_0^{\rm app}-f_0)/f_0$ at $s=\Lambda^2$ 
             in the unit of $\Lambda=1$, 
             where $f_0$ and $f_0^{\rm app}$ denote 
             the solutions in Eqs.~(\ref{F_UV_d6})--(\ref{F_IR_d6}) 
             without the approximation (\ref{app_a0}) and
             the approximate expressions in Eqs.~(\ref{fir2})--(\ref{fuv2}),
             respectively. 
             The dot-dashed, solid, and dashed lines represent graphs for 
             $\tilde{\kappa}_6=\kappa_6/\kappa_6^{\rm crit}=0.02,20/33,0.98$,
             respectively. (The ACDH model in $D=6$, $N_c=3,N_f=2$, 
             corresponds to $\tilde{\kappa}_6=20/33$.)
             \label{fig2}}
  \end{center}
\end{figure}

\begin{figure}[tb]
  \begin{center}
   \resizebox{0.47\textwidth}{!}{\includegraphics{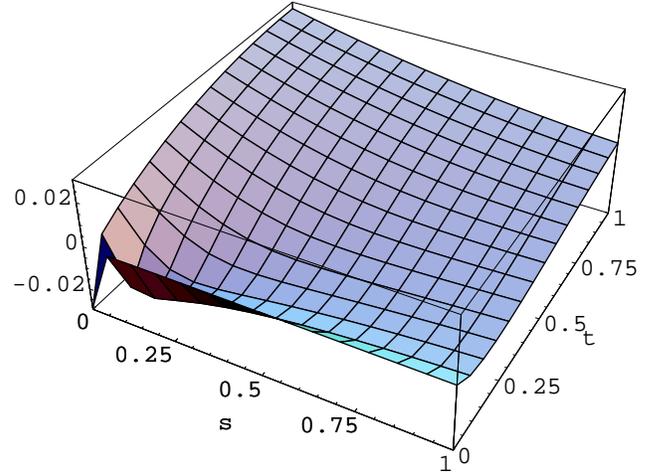}}
    \caption{Error of the zeroth order BS amplitude $f_0$ 
             arising from the approximation (\ref{app_a0}).
             The curved surface shows $(f_0^{\rm app}-f_0)/f_0$ for
             $\tilde{\kappa}_6=20/33$ ($N_c=3,N_f=2$) in the unit of 
             $\Lambda=1$, where $f_0$ and $f_0^{\rm app}$ denote 
             the solutions in Eqs.~(\ref{F_UV_d6})--(\ref{F_IR_d6}) 
             without the approximation (\ref{app_a0}) and
             the approximate expressions in Eqs.~(\ref{fir2})--(\ref{fuv2}),
             respectively.   \label{fig3}}
  \end{center}
\end{figure}

In the case of $D=6$, we can solve the BS equation without using
the approximation for $a_0$ represented in Eq.~(\ref{app_a0}).
The BS equations for $F_{\rm IR}$ and $F_{\rm UV}$ are 
\begin{eqnarray}
\lefteqn{ \hspace*{-1cm}
\frac{\partial }{\partial s^2}F_{\rm IR}(s,t)+
 \frac{3}{s}\frac{\partial }{\partial s}F_{\rm IR}(s,t) } \nonumber \\[2mm] &&
+ \frac{\tilde{\kappa}_6}{6} \left(\frac{4}{st}-\frac{1}{t^2}\right)
   F_{\rm IR}(s,t)=0 , \label{bs_eq2_IR} 
\end{eqnarray}
\begin{eqnarray}
\lefteqn{ \hspace*{-1cm}
\frac{\partial }{\partial s^2}F_{\rm UV}(s,t)+
 \frac{3}{s}\frac{\partial }{\partial s}F_{\rm UV}(s,t) } \nonumber \\[2mm] &&
+\frac{\tilde{\kappa}_6}{2s^2} \left(2-\frac{4t}{3s}+\frac{t^2}{3s^2}\right)
   F_{\rm UV}(s,t)=0 , \label{bs_eq2_UV}
\end{eqnarray}
respectively.
Noting that Eq.~(\ref{bs_eq2_UV}) is rewritten as
\begin{equation}
\frac{\partial }{\partial \bar{s}^2}F_{\rm UV}-
 \frac{1}{\bar{s}}\frac{\partial }{\partial \bar{s}}F_{\rm UV}
+ \tilde{\kappa}_6 \left(\frac{1}{\bar{s}^2}-\frac{2t}{3\bar{s}}+\frac{t^2}{6}
   \right) F_{\rm UV}=0  \label{bs_eq3_UV}  
\end{equation}
with $\bar{s} \equiv 1/s$, we obtain the solutions
\begin{eqnarray}
\lefteqn{ \hspace*{-1mm}
 F_{\rm UV}(s,t) = } \nonumber \\
&& \hspace*{-3mm}
 C_1 \left(\frac{s}{t}\right)^{\rho_+}
     \exp\left(-i\sqrt{\frac{\tilde{\kappa}_6}{6}}\frac{t}{s}\right)
     \conf{\alpha_+}{\gamma_+}{2i\sqrt{\frac{\tilde{\kappa}_6}{6}}\frac{t}{s}}
        \nonumber \\ && \hspace*{-3mm}
+C_2 \left(\frac{s}{t}\right)^{\rho_-}
     \exp\left(-i\sqrt{\frac{\tilde{\kappa}_6}{6}}\frac{t}{s}\right)
     \conf{\alpha_-}{\gamma_-}{2i\sqrt{\frac{\tilde{\kappa}_6}{6}}\frac{t}{s}},
 \nonumber \\ && \label{F_UV_d6} \\
\lefteqn{ \hspace*{-1mm}
 F_{\rm IR}(s,t) = } \nonumber \\
&& 
   C_3 \exp\left(-\sqrt{\frac{\tilde{\kappa}_6}{6}}\frac{s}{t}\right)
       \conf{\alpha_{\rm IR}}{3}{2\sqrt{\frac{\tilde{\kappa}_6}{6}}\frac{s}{t}}
        \nonumber \\ &&
 + C_4 \exp\left(-\sqrt{\frac{\tilde{\kappa}_6}{6}}\frac{s}{t}\right)
       \Psi\left(\alpha_{\rm IR},3;2\sqrt{\frac{\tilde{\kappa}_6}{6}}\frac{s}{t}
           \right), 
 \label{F_IR_d6} 
\end{eqnarray}
where $\conf{\alpha}{\gamma}{z}$ and $\Psi(\alpha,\gamma;z)$ denote 
the confluent hypergeometric function and its $\Psi$ function, respectively,
We here defined 
\begin{align}
 \rho_\pm &\equiv -1\pm\sqrt{1-\tilde{\kappa}_6}, \quad
 \gamma_\pm  \equiv  1\mp2\sqrt{1-\tilde{\kappa}_6}, \\
 \alpha_\pm &\equiv \frac{\gamma_\pm}{2}-2i\sqrt{\frac{\tilde{\kappa}_6}{6}},
 \quad \alpha_{\rm IR}  \equiv  \frac{3}{2}-2\sqrt{\frac{\tilde{\kappa}_6}{6}}.
\end{align}
We can confirm that $F_{\rm UV}$ is real, i.e.,
$F^*_{\rm UV}=F_{\rm UV}$, by using the Kummer's transformation 
$\conf{\alpha}{\gamma}{z}=e^z\conf{\gamma-\alpha}{\gamma}{-z}$. 
The IRBC (\ref{BS-IRBC}) leads to $C_4=0$ owing to
the behavior of the confluent hypergeometric $\Psi$ function around zero
\begin{eqnarray}
\lefteqn{ \hspace*{-8mm}
\Psi(\alpha,1+n;z) \simeq \frac{(n-1)!}{\Gamma(\alpha)}z^{-n} }\nonumber \\
&& \hspace*{-4mm}
+ \frac{(-1)^{n-1}}{n!\,\Gamma(\alpha-n)} \conf{\alpha}{n+1}{z}\ln z ,
 \; (z \sim 0). 
\end{eqnarray}
Other coefficients $C_1, C_2$ and $C_3$ are obtained by using 
the UVBC (\ref{BS-UVBC}) and continuous conditions 
(\ref{cont})--(\ref{contdiff}). 
We show the difference between the solutions with or without 
the approximation (\ref{app_a0}). (See Figs.~\ref{fig2} and \ref{fig3}.)
The approximation works well within a few \%.
We expect that the approximation is also valid in $D=8,10,\cdots$.

We now determine the wave function renormalization constant for
the auxiliary field $\sigma$.
We expand the function $F_{\rm UV}(\Lambda^2,q_E^2)$ in
Eq.~(\ref{fuv2}) over $q_E^2/\Lambda^2$.
Depending on the values of $\nu\omega$,
three expressions are obtained: 
\begin{eqnarray}
\lefteqn{ \hspace*{-1cm}
iD_{\sigma}^{-1}(q_E^2) = - 2^{\frac{D}{2}} N_c N_f \NDA 
} \nonumber \\
&& \times \Lambda^{D-2}
 \left[\frac{1}{g}-\frac{1}{g^{\rm crit}} + A_1
 \left(\frac{q_E^2}{\Lambda^2}\right)^{\nu\omega}\right],
\end{eqnarray}
for $0 < \omega < 1/\nu$, and
\begin{eqnarray}
\lefteqn{ \hspace*{-1cm}
iD_{\sigma}^{-1}(q_E^2) = - 2^{\frac{D}{2}} N_c N_f \NDA 
} \nonumber \\
&& \times \Lambda^{D-2}
 \left[\frac{1}{g}-\frac{1}{g^{\rm crit}} + A_2
 \frac{q_E^2}{\Lambda^2}\ln \left(\frac{\Lambda^2}{q_E^2}\right) \right], 
\end{eqnarray}
for $\omega = 1/\nu$, and
\begin{eqnarray}
\lefteqn{ \hspace*{-1cm}
iD_{\sigma}^{-1}(q_E^2) = - 2^{\frac{D}{2}} N_c N_f \NDA 
} \nonumber \\
&& \times \Lambda^{D-2}
 \left[\frac{1}{g}-\frac{1}{g^{\rm crit}} + A_3
 \frac{q_E^2}{\Lambda^2}\right], 
\end{eqnarray}
for $1/\nu < \omega \leq 1$, where we defined
\begin{equation}
  A_1 \equiv \frac{\gamma(-\omega)}{\gamma(\omega)}
             \frac{\Gamma(1-\nu\omega)}{\Gamma(1+\nu\omega)}
             \frac{16\omega}{\nu(1+\omega)^3(1-\omega)} 
             \left(\frac{\tilde{\kappa}_D}{2}\right)^{\nu\omega},
\end{equation}
\begin{equation}
  A_2 \equiv \frac{2\nu^2}{(\nu+1)^2},
\end{equation}
and
\begin{equation}
  A_3 \equiv \frac{2}{(1+\omega)^2(1-\nu\omega)}.
\end{equation}
For $\omega=1/\nu$ we also used the behavior of the Bessel function 
$K_1(z)$ around $z \sim 0$,
\begin{equation}
  K_1 (z) \sim \frac{1}{z} + \frac{z}{2}\ln z, \quad z \sim 0.  
\end{equation}
Although we have calculated in the symmetric phase so far,
we expect that the momentum dependence of $D_\sigma^{-1}(q)$ is unchanged 
even in the broken phase,
as long as neglecting effects of the mass term.
We thus obtain the wave function renormalization constant for $\sigma$ as
\begin{equation}
  \left\{
  \begin{array}{lcc}
   {\displaystyle
   Z_\sigma^{-1}=\left(\frac{\Lambda}{\mu}\right)^{2\nu(1-\omega)}
   }
   & \mbox{for} & 
   {\displaystyle 0 < \omega < \frac{1}{\nu}} \\[5mm] 
   {\displaystyle Z_\sigma^{-1}=\left(\frac{\Lambda}{\mu}\right)^{2(\nu-1)}
   \ln \left(\frac{\Lambda^2}{\mu^2}\right) }
   & \mbox{for} & 
   {\displaystyle \omega = \frac{1}{\nu}} \\[5mm]
   {\displaystyle
   Z_\sigma^{-1}=\left(\frac{\Lambda}{\mu}\right)^{2(\nu-1)}
   }
   & \mbox{for} & 
   {\displaystyle \frac{1}{\nu} < \omega \leq 1 }
  \end{array}
  \right.
\end{equation}
with
\begin{equation}
  \sigma_R = Z_\sigma^{-1/2}\sigma .
\end{equation}
The results are consistent with the analysis of the resummation technique
in Sec.~\ref{sec-yukawa}.

\begin{widetext}

\section{Linearized SD equation}
\label{sec-appendix2}

In this appendix, we summarize our results on the linearized SD
equation including the cases with $\kappa_D \ge \kappa_D^{\rm crit}$
and $\kappa_D=\kappa_D^{\rm MT}$.
We obtain the asymptotic behavior of the mass function as follows:
\begin{equation}
  \frac{\Sigma(x)}{\Sigma_0} = \left\{
  \begin{array}{l@{\qquad}c@{\qquad}c}
 {\displaystyle
  2|\tilde{c}_1|\left(\frac{x}{\Sigma_0^2}\right)^{-\frac{\nu}{2}}
  \sin\left(\frac{1}{2}\nu\tilde{\omega}\ln \frac{x}{\Sigma_0^2}+\theta\right),
  } & \mbox{for} & \kappa_D > \kappa_D^{\rm crit} \\[5mm]
 {\displaystyle
  \frac{\Gamma(D/2)}{\Gamma(\nu/2)\Gamma(1+\nu/2)}
  \left(\frac{x}{\Sigma_0^2}\right)^{-\frac{\nu}{2}}
  \left(\ln \frac{x}{\Sigma_0^2}+h_0\right),
  } & \mbox{for} & \kappa_D = \kappa_D^{\rm crit}, \\[5mm]
  {\displaystyle
  c_1 \left(\frac{x}{\Sigma_0^2}\right)^{-\frac{\nu}{2}(1-\omega)} +
  d_1 \left(\frac{x}{\Sigma_0^2}\right)^{-\frac{\nu}{2}(1+\omega)} ,
  } & \mbox{for} & 
 {\displaystyle
  \kappa_D^{\rm MT} < 
  \kappa_D < \kappa_D^{\rm crit}, 
  } \\[5mm]
 {\displaystyle 
  c_1
  \left(\frac{x}{\Sigma_0^2}\right)^{-\frac{1}{2}(\nu-1)}
  \left(1-\frac{1}{4}(\nu^2-1)\frac{\Sigma_0^2}{x}
           \ln \frac{x}{\Sigma_0^2}\right),
  } & \mbox{for} & 
 {\displaystyle
  \kappa_D = \kappa_D^{\rm MT},
  } \\[5mm]
 {\displaystyle 
  c_1 \left(\frac{x}{\Sigma_0^2}\right)^{-\frac{\nu}{2}(1-\omega)} +
  c_2 \left(\frac{x}{\Sigma_0^2}\right)^{-\frac{\nu}{2}(1-\omega)-1}, 
  } & \mbox{for} & 
 {\displaystyle
  0 < \kappa_D < \kappa_D^{\rm MT},
  } \\[5mm]
  1, & \mbox{for} & \kappa_D = 0,
  \end{array}\right.
 \label{sol_linear}
\end{equation}
where 
\begin{equation}
  \tilde{c}_1 \equiv 
  \frac{\Gamma(D/2)\Gamma(i\nu\tilde{\omega})}
       {\Gamma(\nu(1+i\tilde{\omega})/2)\Gamma(1+\nu(1+i\tilde{\omega})/2)}, 
 \quad \tilde{d}_1 = \tilde{c}_1^*, \quad 
 e^{2i\theta} = - \tilde{c}_1/\tilde{d}_1
 \quad 
 \tilde{\omega} \equiv \sqrt{\frac{\kappa_D}{\kappa_D^{\rm crit}}-1},
\end{equation}
\begin{equation}
  c_1 = \frac{\Gamma(D/2)\Gamma(\nu\omega)}
               {\Gamma(\nu(1+\omega)/2)\Gamma(1+\nu(1+\omega)/2)}, \quad
  d_1 = \frac{\Gamma(D/2)\Gamma(-\nu\omega)}
               {\Gamma(\nu(1-\omega)/2)\Gamma(1+\nu(1-\omega)/2)}, \quad
  c_2 = - \frac{\nu^2 (1-\omega^2)}{4(\nu\omega-1)} c_1 , 
\end{equation}
and
\begin{equation}
  h_0 \equiv -2\gamma - 2\psi(\nu/2) -\frac{2}{\nu}, \quad
  \psi(z) \equiv \frac{d}{dz} \ln \Gamma(z) . \label{h0}
\end{equation}
In Eq.~(\ref{h0}), $\gamma$ is the Euler's constant, $\gamma \simeq 0.5772$.

Substituting Eq.~(\ref{sol_linear}) for the UVBC (\ref{UV-BC2}) with $m_0=0$,
we obtain the gap equation 
\begin{equation}
 \left\{
  \begin{array}{l@{\qquad}c@{\qquad}c}
 {\displaystyle
  \sin \left(\nu\tilde{\omega}\ln \frac{\Lambda}{\Sigma_0}
  + \theta + \theta'\right) = 0,
  } & \mbox{for} & \kappa_D > \kappa_D^{\rm crit} \\[7mm]
 {\displaystyle
  \ln \frac{\Sigma_0^2}{\Lambda^2}=h_0+\frac{2}{\nu}-
  \frac{\frac{16}{\nu^2}}{\frac{4}{\nu}-\frac{1}{g}},
  } & \mbox{for} & \kappa_D = \kappa_D^{\rm crit}, \\[7mm]
  {\displaystyle
   \frac{g^{\rm crit}}{g}=1+\frac{4\omega d_1}{1-\omega^2}
   \frac{\left(\dfrac{\Sigma_0}{\Lambda}\right)^{2\nu\omega}}
        {c_1+\frac{1-\omega}{1+\omega}d_1
         \left(\dfrac{\Sigma_0}{\Lambda}\right)^{2\nu\omega}},
  } & \mbox{for} & 
 {\displaystyle
  \kappa_D^{\rm MT} < 
  \kappa_D < \kappa_D^{\rm crit}, 
  } \\[12mm]
 {\displaystyle 
   \frac{g^{\rm crit}}{g}=1-\nu g^{\rm crit}\left(
   1-\frac{(\nu-1)^2}{4\nu}\frac{1}{g}\right)
   \frac{\Sigma_0^2}{\Lambda^2} \ln \frac{\Lambda^2}{\Sigma_0^2},
  } & \mbox{for} & 
 {\displaystyle
  \kappa_D = \kappa_D^{\rm MT},
  } \\[7mm]
 {\displaystyle 
   \frac{g^{\rm crit}}{g}=1+\frac{4c_2}{\nu(1-\omega^2)}
   \frac{\frac{\Sigma_0^2}{\Lambda^2}}
        {c_1 + \left(1-\frac{2}{\nu(1+\omega)}\right)c_2
         \frac{\Sigma_0^2}{\Lambda^2}},
  } & \mbox{for} & 
 {\displaystyle
  0 < \kappa_D < \kappa_D^{\rm MT},
  } \\[7mm]
 {\displaystyle
   \frac{g^{\rm crit}}{g} = 1-\frac{\nu}{\nu-1}
   \frac{\Sigma_0^2}{\Lambda^2},
  } & \mbox{for} & \kappa_D = 0.
  \end{array}\right.
\end{equation}
with
\begin{equation}
  \tan \theta' = 
  \frac{\nu(1+\tilde{\omega}^2)+4g}{\nu(1+\tilde{\omega}^2)-4g} \,
  \tilde{\omega}.
\end{equation}
The scaling relation is then found as
\begin{equation}
  \frac{\Sigma_0}{\Lambda} = \left\{
  \begin{array}{l@{\qquad}c@{\qquad}c}
 {\displaystyle
  \exp\left(\frac{\theta+\theta'}{\nu\tilde{\omega}}\right)
   \exp\left(-\frac{\pi}{\nu\tilde{\omega}}\right),
  } & \mbox{for} & \kappa_D > \kappa_D^{\rm crit} \\[5mm]
 {\displaystyle
  \exp\left(\frac{h_0}{2}+\frac{1}{\nu}\right)
  \exp\left(-\frac{\frac{8}{\nu^2}}{\frac{4}{\nu}-\frac{1}{g}}\right)
  } & \mbox{for} & \kappa_D = \kappa_D^{\rm crit}, \\[5mm]
  {\displaystyle
  \left[\,\frac{c_1}{-d_1}\frac{1-\omega^2}{4\omega}
   \left(1-\frac{g^{\rm crit}}{g}\right)\,\right]^{\frac{1}{2\nu\omega}} 
  } & \mbox{for} & 
 {\displaystyle
  \kappa_D^{\rm MT} < 
  \kappa_D < \kappa_D^{\rm crit}, 
  } \\[5mm]
 {\displaystyle 
  \sqrt{\left(\omega-\frac{1}{\nu}\right)
         \left(1-\frac{g^{\rm crit}}{g}\right)}
  } & \mbox{for} & 
 {\displaystyle
  0 \leq \kappa_D < \kappa_D^{\rm MT},
  }
  \end{array}\right.
\end{equation}
where we used $g \simeq g^{\rm crit}$.
For $\kappa_D=\kappa_D^{\rm MT}$
the scaling relation cannot be written by elementary functions:
\begin{equation}
\frac{\Sigma_0^2}{\Lambda^2}\ln \frac{\Lambda^2}{\Sigma_0^2}=
\frac{4}{(1+\nu)^2}\frac{g-\frac{(\nu+1)^2}{4\nu}}{g-\frac{(\nu-1)^2}{4\nu}}.
\end{equation}

The chiral condensation is calculated from Eq.~(\ref{UV-BC}): 
\begin{equation}
  \frac{\sigma}{\Sigma_0} = \left\{
  \begin{array}{l@{\qquad}c@{\qquad}c}
 {\displaystyle
  |\tilde{c}_1| \sqrt{1+\tilde{\omega}^2}
  \left(\dfrac{\Sigma_0}{\Lambda}\right)^{\nu}
  \sin\left(\nu\tilde{\omega}\ln \frac{\Lambda\;}{\Sigma_0} + 
  \theta+\arctan \tilde{\omega} \right),
  } & \mbox{for} & \kappa_D > \kappa_D^{\rm crit} \\[5mm]
 {\displaystyle
  \frac{\Gamma(D/2)}{\Gamma(\nu/2)\Gamma(1+\nu/2)}
  \left(\dfrac{\Sigma_0}{\Lambda}\right)^{\nu}
  \left(\ln \dfrac{\Lambda\;}{\Sigma_0}+\frac{h_0}{2}+\frac{1}{\nu}\right),
  } & \mbox{for} & \kappa_D = \kappa_D^{\rm crit}, \\[5mm]
  {\displaystyle
  \frac{(1+\omega)}{2} c_1 
   \left(\dfrac{\Sigma_0}{\Lambda}\right)^{\nu (1-\omega)} +
  \frac{(1-\omega)}{2} d_1 
   \left(\frac{\Sigma_0}{\Lambda}\right)^{\nu (1+\omega)} ,
  } & \mbox{for} & 
 {\displaystyle
  \kappa_D^{\rm MT} < 
  \kappa_D < \kappa_D^{\rm crit}, 
  } \\[5mm]
 {\displaystyle 
  \frac{(1+\nu)}{2\nu} c_1 
  \left(\dfrac{\Sigma_0}{\Lambda}\right)^{\nu-1}
  \left[1-\frac{1}{4}(\nu^2-1)\frac{\Sigma_0^2}{\Lambda^2}
          \ln \frac{\Lambda^2}{\Sigma_0^2}\right],
  } & \mbox{for} & 
 {\displaystyle
  \kappa_D = \kappa_D^{\rm MT},
  } \\[5mm]
 {\displaystyle 
  \frac{(1+\omega)}{2} c_1 
   \left(\dfrac{\Sigma_0}{\Lambda}\right)^{\nu (1-\omega)}
  +\left(\frac{1+\omega}{2}-\frac{1}{\nu}\right) c_2
   \left(\dfrac{\Sigma_0}{\Lambda}\right)^{\nu (1-\omega)+2},
  } & \mbox{for} & 
 {\displaystyle
  0 < \kappa_D < \kappa_D^{\rm MT},
  } \\[5mm]
  1, & \mbox{for} & \kappa_D = 0.
  \end{array}\right.
 \label{sig_linear}
\end{equation}
Eq.~(\ref{sig_linear}) reads
\begin{equation}
  \Sigma(\Lambda^2)
 = \left\{
 \begin{array}{l@{\qquad}c@{\qquad}c}
{\displaystyle
 2\sigma - \frac{2(\nu+1)}{\nu}
  \frac{\sigma}{\ln \left(\dfrac{\Lambda}{\sigma}\right)},
 } & \mbox{for} & \kappa_D = \kappa_D^{\rm crit}, \\[5mm]
 {\displaystyle
\frac{2}{1+\omega} \sigma 
+\frac{2\omega}{1+\omega}d_1 \Lambda
 \left(\frac{2c_1^{-1}}{1+\omega}
       \frac{\sigma}{\Lambda}\right)^{1+
       \frac{2\nu\omega}{\nu(1-\omega)+1}},
 } & \mbox{for} & 
{\displaystyle \kappa_D^{\rm MT} < \kappa_D < \kappa_D^{\rm crit}, 
 } \\[5mm]
{\displaystyle 
 \frac{2\nu}{1+\nu} \sigma
-\frac{\nu-1}{\nu} c_1 \Lambda
 \left(\frac{2\nu c_1^{-1}}{1+\nu}
  \frac{\sigma}{\Lambda}\right)^{1+\frac{2}{\nu}}
 \ln \frac{\Lambda}{\sigma},
 } & \mbox{for} & 
{\displaystyle
 \kappa_D = \kappa_D^{\rm MT},
 } \\[5mm]
{\displaystyle 
\frac{2}{1+\omega}\sigma
 +\frac{2}{\nu(1+\omega)} c_2 \Lambda
 \left(\frac{2c_1^{-1}}{1+\omega}
       \frac{\sigma}{\Lambda}\right)^{1+
       \frac{2}{\nu(1-\omega)+1}},
 } & \mbox{for} & 
{\displaystyle
 0 < \kappa_D < \kappa_D^{\rm MT},
 } \\[5mm]
{\displaystyle 
 \sigma
} & \mbox{for} & \kappa_D = 0.
 \end{array}\right. \label{M_lin}
\end{equation}
For $\kappa_D > \kappa_D^{\rm crit}$ we do not obtain a handy formula.

\section{Power expansion solution}
\label{sec-appendix3}

We summarize results on the power expansion solution for
$\kappa_D < \kappa_D^{\rm crit}$.
The asymptotic behavior of the mass function is given by
\begin{equation}
 \dfrac{\Sigma(x)}{\Sigma_0} = \left\{
  \begin{array}{l@{\qquad}c@{\qquad}l}
  {\displaystyle
    c_1 \left( \dfrac{x}{\Sigma_0^2} \right)^{-\frac{1}{2}\nu(1-\omega)}
  + d_1 \left( \dfrac{x}{\Sigma_0^2} \right)^{-\frac{1}{2}\nu(1+\omega)},
  }
  & \mbox{for} & \kappa_D^{\rm PE} < \kappa_D < \kappa_D^{\rm crit}, \\[5mm]
  {\displaystyle
    c_1 \left( \dfrac{x}{\Sigma_0^2} \right)^{-\frac{1}{4}(\nu-1)}
   -\frac{(\nu-1)(3\nu+1)}{8(\nu+1)}\,c_1^3 
  \left(\dfrac{x}{\Sigma_0^2}\right)^{-\frac{3\nu+1}{4}}
  \ln \frac{x}{\Sigma_0^2},
  }
  & \mbox{for} & \kappa_D = \kappa_D^{\rm PE}, \\[5mm]
  {\displaystyle
    c_1 \left( \dfrac{x}{\Sigma_0^2} \right)^{-\frac{1}{2}\nu(1-\omega)}
   +c_2 \left( \dfrac{x}{\Sigma_0^2} \right)^{-\frac{3}{2}\nu(1-\omega)-1}
  }
  & \mbox{for} & 0 < \kappa_D < \kappa_D^{\rm PE}, 
  \end{array}\right.
\label{sol_pow}
\end{equation}
with
\begin{equation}
    c_2 = -\frac{\nu^2 (1-\omega^2)}
              {4(\nu+1-\nu\omega)(2\nu\omega-\nu-1)} c_1^3,
\end{equation}
where $c_1$ and $d_1$ cannot be determined in the power expansion method.

Substituting Eq.~(\ref{sol_pow}) for the UVBC (\ref{UV-BC2}) with $m_0=0$,
we obtain the gap equation 
\begin{equation}
 \left\{
  \begin{array}{l@{\qquad}c@{\qquad}c}
  {\displaystyle
   \frac{g^{\rm crit}}{g}=1+\frac{4\omega d_1}{1-\omega^2}
   \frac{\left(\dfrac{\Sigma_0}{\Lambda}\right)^{2\nu\omega}}
        {c_1+\frac{1-\omega}{1+\omega}d_1
         \left(\dfrac{\Sigma_0}{\Lambda}\right)^{2\nu\omega}},
  } & \mbox{for} & 
 {\displaystyle
  \kappa_D^{\rm PE} < 
  \kappa_D < \kappa_D^{\rm crit}, 
  } \\[12mm]
 {\displaystyle 
   \frac{g^{\rm crit}}{g}=1-\frac{2\nu^2 c_1^2 g^{\rm crit}}{\nu+1}\left(
   1-\frac{(\nu-1)^2}{16\nu^2 g}\right)
   \left(\frac{\Sigma_0}{\Lambda}\right)^{\nu+1}
   \ln \frac{\Lambda^2}{\Sigma_0^2},
  } & \mbox{for} & 
 {\displaystyle
  \kappa_D = \kappa_D^{\rm PE},
  } \\[7mm]
 {\displaystyle 
   \frac{g^{\rm crit}}{g}=1+\frac{4(\nu+1-\nu\omega)c_2}{\nu(1-\omega^2)}
   \dfrac{\left(\dfrac{\Sigma_0}{\Lambda}\right)^{2\nu(1-\omega)+2}}
         {c_1 + \frac{\nu(3\omega-1)-2}{\nu(1+\omega)}c_2
         \left(\dfrac{\Sigma_0}{\Lambda}\right)^{2\nu(1-\omega)+2}},
  } & \mbox{for} & 
 {\displaystyle
  0 < \kappa_D < \kappa_D^{\rm PE}.
  }
  \end{array}\right.
\end{equation}
The scaling relation is then found as
\begin{equation}
  \frac{\Sigma_0}{\Lambda} = \left\{
  \begin{array}{l@{\qquad}c@{\qquad}c}
  {\displaystyle
  \left[\,\frac{c_1}{-d_1}\frac{1-\omega^2}{4\omega}
   \left(1-\frac{g^{\rm crit}}{g}\right)\,\right]^{\frac{1}{2\nu\omega}} 
  } & \mbox{for} & 
 {\displaystyle
  \kappa_D^{\rm PE} < 
  \kappa_D < \kappa_D^{\rm crit}, 
  } \\[7mm]
 {\displaystyle 
  \left[\,\dfrac{2\nu\omega-\nu-1}{\nu c_1^2}
         \left(1-\frac{g^{\rm crit}}{g}\right)
  \,\right]^{\frac{1}{2\nu(1-\omega)+2}}
  } & \mbox{for} & 
 {\displaystyle
  0 < \kappa_D < \kappa_D^{\rm PE},
  }
  \end{array}\right.
\end{equation}
where we used $g \simeq g^{\rm crit}$.
For $\kappa_D=\kappa_D^{\rm PE}$
the scaling relation cannot be written by elementary functions:
\begin{equation}
\left(\frac{\Sigma_0}{\Lambda}\right)^{\nu+1}
 \ln \frac{\Lambda^2}{\Sigma_0^2} = 
 \frac{\nu+1}{2\nu^2 c_1^2 g^{\rm crit}}
  \frac{g-\frac{(3\nu+1)^2}{16\nu}}{g-\frac{(\nu-1)^2}{16\nu^2}}.
\end{equation}

The chiral condensation is calculated from Eq.~(\ref{UV-BC}): 
\begin{equation}
  \frac{\sigma}{\Sigma_0} = \left\{
  \begin{array}{l@{\qquad}c@{\qquad}c}
  {\displaystyle
  \frac{(1+\omega)}{2} c_1 
   \left(\dfrac{\Sigma_0}{\Lambda}\right)^{\nu (1-\omega)} +
  \frac{(1-\omega)}{2} d_1 
   \left(\frac{\Sigma_0}{\Lambda}\right)^{\nu (1+\omega)} ,
  } & \mbox{for} & 
 {\displaystyle
  \kappa_D^{\rm PE} < 
  \kappa_D < \kappa_D^{\rm crit}, 
  } \\[5mm]
 {\displaystyle 
  \frac{(3\nu+1)}{4\nu} c_1 
  \left(\dfrac{\Sigma_0}{\Lambda}\right)^{\frac{\nu-1}{2}}
 -\frac{(\nu-1)^2 (3\nu+1)}{32\nu(\nu+1)} c_1^3 
  \left(\dfrac{\Sigma_0}{\Lambda}\right)^{\frac{3\nu+1}{2}}
          \ln \frac{\Lambda^2}{\Sigma_0^2},
  } & \mbox{for} & 
 {\displaystyle
  \kappa_D = \kappa_D^{\rm PE},
  } \\[5mm]
 {\displaystyle 
  \frac{(1+\omega)}{2} c_1 
   \left(\dfrac{\Sigma_0}{\Lambda}\right)^{\nu (1-\omega)}
  +\left(\frac{3\omega-1}{2}-\frac{1}{\nu}\right) c_2
   \left(\dfrac{\Sigma_0}{\Lambda}\right)^{3\nu (1-\omega)+3},
  } & \mbox{for} & 
 {\displaystyle
  0 < \kappa_D < \kappa_D^{\rm PE}.
  }
  \end{array}\right.
 \label{sig_pow}
\end{equation}
Eq.~(\ref{sig_pow}) reads
\begin{equation}
  \Sigma(\Lambda^2)
 = \left\{
 \begin{array}{l@{\qquad}c@{\qquad}c}
 {\displaystyle
\frac{2}{1+\omega} \sigma 
+\frac{2\omega}{1+\omega}d_1 \Lambda
 \left(\frac{2c_1^{-1}}{1+\omega}
       \frac{\sigma}{\Lambda}\right)^{1+
       \frac{2\nu\omega}{\nu(1-\omega)+1}},
 } & \mbox{for} & 
{\displaystyle \kappa_D^{\rm PE} < \kappa_D < \kappa_D^{\rm crit}, 
 } \\[5mm]
{\displaystyle 
\frac{4\nu}{3\nu+1}\sigma
 -\frac{\nu-1}{\nu+1}\,\Lambda
  \left(\frac{4\nu}{3\nu+1}\frac{\sigma}{\Lambda}\right)^3
  \ln \frac{\Lambda}{\sigma},
 } & \mbox{for} & 
{\displaystyle
 \kappa_D = \kappa_D^{\rm PE},
 } \\[5mm]
{\displaystyle 
\frac{2}{1+\omega}\sigma
 -\frac{\nu(1-\omega)}{2(2\nu\omega-\nu-1)}\, \Lambda\,
  \left(\frac{2}{1+\omega}\frac{\sigma}{\Lambda}\right)^3,
 } & \mbox{for} & 
{\displaystyle
 0 < \kappa_D < \kappa_D^{\rm PE}.
 }
 \end{array}\right. \label{M_pow}
\end{equation}
\end{widetext} 
These formulae are useful for calculation of the effective potential.

\section{More on nonrenormalizable region}
\label{app-D}

In Sec.~\ref{sec-RG}, we studied the renormalization of 
the effective potential, keeping the kinetic term of 
the auxiliary field finite.
It should be so in the viewpoint of renormalization of 
the effective action.
For $0 \leq \kappa_D \leq \kappa_D^{\rm MT}$, however,
the theory cannot get into the ``broken phase'' 
$g_R > g^{\rm crit}$, because the ``renormalized'' potential 
(\ref{eq:renormalized_pot2}) becomes unbounded from below.
In this appendix, we attempt another procedure in which 
the VEV of $\sigma_R$ remains finite
independently of the value of $\kappa_D$.
For such a purpose we may admit to incorporate any required operators.
Can be found a necessary and sufficient operator set?
Before details, we jump to conclusions:
It depends on approximation methods of the ladder SD equation.

The dynamical mass should be finite, so that 
Eq.~(\ref{UV-BC}) yields the wave function renormalization constant 
of $\sigma$ as
\begin{equation}
  Z_\sigma^{-1} = \left(\dfrac{\Lambda}{\mu}\right)^{2\nu(1-\omega)}, 
  \quad (0 \leq \kappa_D < \kappa_D^{\rm crit})
\label{Z-sig}
\end{equation}
for the whole region.
(Compare Eq.~(\ref{Z-sig}) with Eq.~(\ref{eq:Z_wav}).)
By using Eq.~(\ref{Z-sig}), we obtain the renormalized Yukawa vertex
\begin{equation}
    \Gamma_s^{(R)}(-q^2) 
    \propto
    \left(\frac{-q^2}{\mu^2}\right)^{-\frac{\nu}{2}(1-\omega)},
    \quad (0 \leq \kappa_D < \kappa_D^{\rm crit}),
\end{equation}
in the continuum limit $\Lambda\rightarrow\infty$.
(See also Eq.~(\ref{eq:ren_yukawa}).)
On the other hand, 
the ``renormalized'' propagator of $\sigma$
behaves as
\begin{equation}
  iD_{\sigma(R)}^{-1}(p)-iD_{\sigma(R)}^{-1}(0) \propto
  \left(\frac{\Lambda}{\mu}\right)^{2(\nu\omega-1)} \,
  \mu^{2\nu}\,\left(\dfrac{x}{\mu^2}\right)
\end{equation}
for $0 \leq \kappa_D < \kappa_D^{\rm MT}$ and
\begin{equation}
  iD_{\sigma(R)}^{-1}(p)-iD_{\sigma(R)}^{-1}(0) \propto
  \mu^{2\nu}\,\left(\dfrac{x}{\mu^2}\right)
  \ln\left(\dfrac{\Lambda^2}{x}\right)
\end{equation}
for $\kappa_D = \kappa_D^{\rm MT}$.
Thus the kinetic term of $\sigma$ diverges in the continuum limit
$\Lambda \to \infty$
for $0 \leq \kappa_D \leq \kappa_D^{\rm MT}$.
We need to add the operator 
$\partial_M (\bar\psi\psi)\partial^M (\bar\psi\psi)$ 
in the effective action
in order to absorb the divergence.
Moreover, the decay constant is also divergent in the region, so that 
the interactions of the NG boson vanishes in $\Lambda \to \infty$.

We investigate the effective potential 
for $0 \leq \kappa_D < \kappa_D^{\rm MT}$ separately in two approximations.

\subsection{Linearized approximation}

The scaling relation 
for $0 \leq \kappa_D < \kappa_D^{\rm MT}$ is given by
\begin{equation}
  \Sigma_0 \propto \Lambda \sqrt{1-\frac{g^{\rm crit}}{g}}.
\end{equation}
(See Eq.~(\ref{gapeq-linear}).)
The four-fermion coupling $g$ is thus renormalized as
\begin{equation}
  \left(\,\dfrac{1}{g(\Lambda)}-\dfrac{1}{g^{\rm crit}}\,\right)\Lambda^2
 = \left(\,\dfrac{1}{g_R(\mu)}-\dfrac{1}{g^{\rm crit}}\,\right) \mu^2,
\end{equation}
so as to make the dynamical mass $\Sigma_0$ finite.
This is the very same as Eq.~(\ref{ren_g2}).

The bare effective potential has been already obtained 
in Eq.~(\ref{eq:linearized_pot2}).
We rewrite Eq.~(\ref{eq:linearized_pot2}) in terms of $\sigma_R$ and $g_R$:
\begin{eqnarray}
 V(\sigma_R) &\propto& \mu^{D}\left[\,
 -\frac{1}{g}\frac{m_0 \sigma_R}{\mu^2}
  \left(\frac{\Lambda}{\mu}\right)^{\nu(1+\omega)}
  \right. \nonumber\\
&&\qquad
 +\frac{1}{2}
  \left(\frac{1}{g_R}-\frac{1}{g^{\rm crit}}\right)
  \dfrac{\sigma_R^2}{\mu^2}
  \left(\frac{\Lambda}{\mu}\right)^{2(\nu\omega-1)} \nonumber \\
&&\qquad \left.
 +A_2
 \left(\frac{\sigma_R}{\mu}\right)^{2+\frac{2}{\nu(1-\omega)+1}}
  \left(\frac{\Lambda}{\mu}\right)^{2(\nu\omega-1)} 
 \!\!\!\!\!\! +\cdots \,\right] . \nonumber \\
\label{bare_pot1}
\end{eqnarray}
We renormalize the bare mass term, i.e., the first line of 
Eq.~(\ref{bare_pot1}) as follows: 
\begin{equation}
  \dfrac{m_0(\Lambda)}{g(\Lambda)} \Lambda^{\nu(1+\omega)} =
  \dfrac{m_R(\mu)}{g_R(\mu)} \mu^{\nu(1+\omega)},
\end{equation}
which is the same as Eq.~(\ref{ren_m1}). 
The beta function and anomalous dimensions are then found,
\begin{equation}
  \beta (g_R) = 
   2 g_R \left( 1 - \dfrac{g_R}{g^{\rm crit}} \right),
\end{equation}
and 
\begin{eqnarray}
  \gamma_m (g_R) &=& 
  -\dfrac{\beta(g_R)}{g_R} + \nu ( 1 + \omega) , \nonumber \\
 &=&
  \nu (1+\omega) -2 + 2 \dfrac{g_R}{g^{\rm crit}},
\end{eqnarray}
respectively.
At the UVFP of $g_R$ the anomalous dimension is
$\gamma_m=\nu(1+\omega)$.
Therefore the mismatch of the $\Lambda$ dependence between $m_0$ and 
$\VEV{(\bar{\psi}\psi)_0}$ is resolved.
(See also Eq.~(\ref{chi-cond2}).)

Troublesome are the $\sigma_R^2$ term and interaction term ($A_2$ term)
in Eq.~(\ref{bare_pot1}).
The divergence still remains after the ``renormalization'' of 
the four-fermion term.
In addition, the operator having the fractional power of 
$(\bar{\psi}\psi)$ is required.

\subsection{Power expansion method}

The scaling relation is given by
\begin{equation}
  \Sigma_0 \sim \left\{
    \begin{array}{l@{\quad}c}
      {\displaystyle
        \Lambda\, \left(
          1 - \dfrac{g^{\rm crit}}{g}
        \right)^{\frac{1}{2\nu\omega}}
      }, & 
      {\displaystyle
        (\kappa_D^{\rm PE} < \kappa_D < \kappa_D^{\rm crit}),
      } \\[4ex]
      {\displaystyle
        \Lambda\, \left(
          1 - \dfrac{g^{\rm crit}}{g}
        \right)^{\frac{1}{2\nu(1-\omega)+2}}
      },
      & 
      {\displaystyle
        ( 0 < \kappa_D < \kappa_D^{\rm PE} ).
      }
    \end{array}
  \right. 
\end{equation}
(See Eq.~(\ref{gapeq-pow}).)
In order to keep $\Sigma_0$ finite,
we renormalize the gap equation,
\begin{equation}
  \left(\,\dfrac{1}{g(\Lambda)}-\dfrac{1}{g^{\rm crit}}\,\right)
  \Lambda^{2\nu\omega}
 = \left(\,\dfrac{1}{g_R(\mu)}-\dfrac{1}{g^{\rm crit}}\,\right)
   \mu^{2\nu\omega},
\end{equation}
for $\kappa_D^{\rm PE} < \kappa_D < \kappa_D^{\rm crit}$
and
\begin{eqnarray}
\lefteqn{\left(\,\dfrac{1}{g(\Lambda)}-\dfrac{1}{g^{\rm crit}}\,\right)
  \Lambda^{2\nu(1-\omega)+2}
}\nonumber\\
  & & = \left(\,\dfrac{1}{g_R(\mu)}-\dfrac{1}{g^{\rm crit}}\,\right)
  \mu^{2\nu(1-\omega)+2},
\end{eqnarray}
for $0 < \kappa_D < \kappa_D^{\rm PE}$.
By rewriting the bare effective potential in Eqs.~(\ref{eff-pot-pow1}) and
(\ref{eff_pot_pw2}) in terms of $\sigma_R$ and $g_R$,
we obtain
\begin{eqnarray}
 V(\sigma_R) &\propto& \mu^{D}\left[\,
 -\frac{1}{g}\frac{m_0 \sigma_R}{\mu^2}
  \left(\frac{\Lambda}{\mu}\right)^{\nu(1+\omega)}
  \right. \nonumber\\
&&\qquad
 +\frac{1}{2}
  \left(\frac{1}{g_R}-\frac{1}{g^{\rm crit}}\right)
  \dfrac{\sigma_R^2}{\mu^2}
 \nonumber \\
&&\qquad \left.
 +\tilde{A}_1
 \left(\frac{\sigma_R}{\mu}\right)^{2+\frac{2\nu\omega}{\nu(1-\omega)+1}}
 \!\!\!\!\!\! +\cdots \,\right] , \nonumber \\
\label{bare_pot2}
\end{eqnarray}
for $\kappa_D^{\rm PE} < \kappa_D < \kappa_D^{\rm crit}$, and 
\begin{eqnarray}
 V(\sigma_R) &\propto& \mu^{D}\left[\,
 -\frac{1}{g}\frac{m_0 \sigma_R}{\mu^2}
  \left(\frac{\Lambda}{\mu}\right)^{\nu(1+\omega)}
  \right. \nonumber\\
&&\qquad
 +\frac{1}{2}
  \left(\frac{1}{g_R}-\frac{1}{g^{\rm crit}}\right)
  \dfrac{\sigma_R^2}{\mu^2}
  \left(\frac{\Lambda}{\mu}\right)^{2(2\nu\omega-\nu-1)} \nonumber \\
&&\qquad \left.
 +\tilde{A}_2
 \left(\frac{\sigma_R}{\mu}\right)^{4}
  \left(\frac{\Lambda}{\mu}\right)^{2(2\nu\omega-\nu-1)} 
 \!\!\!\!\!\! +\cdots \,\right] , \nonumber \\
\label{bare_pot3}
\end{eqnarray}
for $0 < \kappa_D < \kappa_D^{\rm PE}$. 
The bare mass term is then renormalized
\begin{equation}
  \dfrac{m_0(\Lambda)}{g(\Lambda)} \Lambda^{\nu(1+\omega)} =
  \dfrac{m_R(\mu)}{g_R(\mu)} \mu^{\nu(1+\omega)},
\end{equation}
for $0 < \kappa_D < \kappa_D^{\rm crit}$. 
The beta function and anomalous dimensions are 
\begin{widetext}
\begin{eqnarray}
  \beta (g_R) = 
  \left\{\begin{array}{l@{\qquad}c}
   2 \nu\omega g_R \left( 1 - \dfrac{g_R}{g^{\rm crit}} \right),&
   (\kappa_D^{\rm PE} < \kappa_D < \kappa_D^{\rm crit}) \\[3mm]
   2 [\nu(1-\omega)+1]\, g_R \left( 1 - \dfrac{g_R}{g^{\rm crit}} \right),&
   (0 < \kappa_D < \kappa_D^{\rm PE})  
  \end{array}\right. 
\end{eqnarray}
and 
\begin{eqnarray}
  \gamma_m (g_R) = -\dfrac{\beta(g_R)}{g_R} + \nu ( 1 + \omega)
  =
  \left\{\begin{array}{l@{\qquad}c}
  \nu \left[\,1-\omega + 2 \omega \dfrac{g_R}{g^{\rm crit}}\,\right],&
   (\kappa_D^{\rm PE} < \kappa_D < \kappa_D^{\rm crit}) \\[3mm]
  3\nu\omega-\nu-2+2 [\nu(1-\omega)+1] \dfrac{g_R}{g^{\rm crit}},&
   (0 < \kappa_D < \kappa_D^{\rm PE})  \end{array}\right. 
\end{eqnarray}
\end{widetext}
Similarly to the linearizing approximation,
this $\gamma_m$ at the UVFP of $g_R$ 
is consistent with the renormalization of $\bar{\psi}\psi$.

After the renormalization, the effective potential for 
$\kappa_D^{\rm PE} < \kappa_D < \kappa_D^{\rm crit}$
has no divergent term.
For $0 < \kappa_D < \kappa_D^{\rm PE}$,
on the other hand,
the four-fermion ($\sigma_R^2$) and eight-fermion ($\sigma_R^4$)
terms diverge in the continuum limit.
Therefore the corresponding counter terms are required.

In passing, the bifurcation solution leads to the same results
as above ones for $\kappa_D^{\rm PE} < \kappa_D < \kappa_D^{\rm crit}$
in the whole region.
Thus only the counter term of the auxiliary field propagator
is needed in the nonrenormalizable region 
$0 \leq \kappa_D \leq \kappa_D^{\rm MT}$.

\end{document}